\documentclass[12pt]{article}
\usepackage{times}
\usepackage{amsmath,amsthm,amssymb,setspace,enumitem,epsfig, titlesec, verbatim,color,array,multirow,comment,graphicx,soul}
\usepackage{booktabs}
\usepackage[sort&compress,comma,round,numbers]{natbib}
\usepackage[bf,small]{caption}
\usepackage[export]{adjustbox}
\usepackage{rotating}
\usepackage[top=2.5cm,left=2.7cm,right=2.7cm,bottom=3.2cm]{geometry}
\usepackage{blkarray}
\usepackage[table]{xcolor}
\usepackage{xcolor}
\usepackage{mathtools}
\usepackage[
  colorlinks=false,
  pdfborder={0 0 0},
  hidelinks
]{hyperref}
\usepackage{setspace}  % for precise line spacing
\usepackage{tcolorbox}

\definecolor{limegreen}{RGB}{50,205,50}      % limegreen
\definecolor{darkgreen}{RGB}{0,100,0}        % darkgreen
\definecolor{brown}{RGB}{165,42,42}          % brown
\definecolor{cornflowerblue}{RGB}{100,149,237} % cornflowerblue
\definecolor{darkorange}{RGB}{255,140,0}     % darkorange

\titleformat{\section}{\sffamily \fontsize{12}{12}\bfseries}{\thesection}{1em}{}
\titleformat{\subsection}{\sffamily \fontsize{10}{10.5}\bfseries}{\thesubsection}{1em}{}

\newcommand{\CC}{\mathrm{CC}}
\newcommand{\CD}{\mathrm{CD}}
\newcommand{\DC}{\mathrm{DC}}
\newcommand{\DD}{\mathrm{DD}}

\newcommand{\gptfour}{\textrm{\scshape Gpt-4o}}
\newcommand{\gptfive}{\textrm{\scshape Gpt-5}}
\newcommand{\claude}{\textrm{\scshape Claude}}
\newcommand{\llama}{\textrm{\scshape Llama}}
\newcommand{\gemini}{\textrm{\scshape Gemini}}

\newcommand{\capitalgptfour}{\textrm{\scshape Gpt-4o}}
\newcommand{\capitalgptfive}{\textrm{\scshape Gpt-5}}
\newcommand{\capitalclaude}{\textrm{\scshape Claude}}
\newcommand{\capitalllama}{\textrm{\scshape Llama}}
\newcommand{\capitalgemini}{\textrm{\scshape Gemini}}

\newtheoremstyle{plainCl1}% name
{9pt}%      Space above, empty = 'usual value'
{15pt}%      Space below
{\it}% 	   Body font
{}%         Indent amount (empty = no indent, \parindent = para indent)
{\bfseries}% Thm head font
{.}%        Punctuation after thm head
{2mm}% Space after thm head: \newline = linebreak
{}%         Thm head spec

\theoremstyle{plainCl1}

\usepackage{bbm}
\usepackage{color}

\allowdisplaybreaks[1]

\theoremstyle{remark}

\newcommand{\LL}{\mathrm{LL}}
\newcommand{\LR}{\mathrm{LR}}
\newcommand{\RL}{\mathrm{RL}}
\newcommand{\RR}{\mathrm{RR}}

\newcommand{\FigOriginal}{\textbf{Fig.~\ref{fig:Fig1}}}
\newcommand{\FigFraming}{\textbf{Fig.~\ref{fig:Fig-diamonds}}}
\newcommand{\FigMtwo}{\textbf{Fig.~\ref{fig:Fig-M2}}}
\newcommand{\FigVariation}{\textbf{Fig.~\ref{fig:Fig-var}}}
\newcommand{\FigConclusion}{\textbf{Fig.~\ref{fig:conclusion}}}

\begin{document}
\title{\bfseries\sffamily \large Strategies of cooperation and defection\\ in five large language models}
\date{}
\author{\parbox[c]{16cm}{\centering \onehalfspacing \fontsize{11}{12}\selectfont Saptarshi Pal$^{1}$, Abhishek Mallela$^{2}$, Christian Hilbe$^{3}$, Lenz Pracher$^{4}$,\\
Chiyu Wei$^{2}$,  Feng Fu$^{2,6}$, Santiago Schnell$^{2,5,6}$, Martin A Nowak$^{1,7}$ \\[0.4cm]
$^1$Department of Mathematics, Harvard University, Cambridge, MA\\
$^2$Department of Mathematics, Dartmouth College, Hanover, NH\\
$^3$Interdisciplinary Transformation University, Linz, Austria\\
$^4$ Arnold Sommerfeld Center for Theoretical Physics, \\
Ludwig-Maximilians-Universit\"{a}t, M\"{u}nchen, Germany\\
$^5$ Department of Biochemistry and Cell Biology, Dartmouth College, Hanover, NH\\
$^6$ Department of Biomedical Data Sciences, Geisel School of Medicine, Hanover, NH\\
$^7$Department of Organismic and Evolutionary Biology, Harvard University, Cambridge, MA\\
%$^\dagger$ corresponding author
}}

\maketitle
\onehalfspacing

\section*{Abstract}
\noindent Large language models (LLMs) are increasingly deployed to support human decision-making. 
This use of LLMs has concerning implications, especially when their prescriptions affect the welfare of others.
To gauge how LLMs make social decisions, we explore whether five leading models produce sensible strategies in the repeated prisoner's dilemma, which is the main metaphor of reciprocal cooperation. 
First, we measure the propensity of LLMs to cooperate 
in a neutral setting, without using language reminiscent of how this game is usually presented. We record to what extent LLMs implement Nash equilibria or  other well-known strategy classes. Thereafter, we explore how LLMs adapt their strategies to changes in parameter values.
We vary the game’s continuation probability, the payoff values, and if the total number of rounds
is commonly known. We also study the effect of different framings. In each case, we test whether the adaptations of the LLMs are in line with basic intuition, theoretical predictions of evolutionary game theory, and experimental evidence of human participants. While all LLMs perform well in many of the tasks, none of them exhibit full consistency over all tasks. We also conduct tournaments between the inferred LLM strategies and study direct interaction between LLMs in games over ten rounds with known or unknown last round. Our experiments shed light on how current LLMs instantiate reciprocal cooperation.\\

%%%%%%%%
%% INTRO %%
%%%%%%%%

\section*{Introduction}

\noindent Artificial intelligence (AI) and large language models (LLM) are entering into all areas of human activity. As they become more powerful and more widely used, it is important to gauge their propensity to cooperate or defect in strategic situations \cite{Crandall:ML:2011,hammond2025multi,terrucha2025humans,grossmann2023ai,  Santos:Prosocial:2024,Powers:Regulation:2024,PiresSantos:IR:LLM,weidinger2021ethical,bommasani2021opportunities}. 
Studying cooperative decision-making is particularly relevant in social dilemmas, in which there is a conflict between cooperation and defection.
Cooperation maximizes the payoff of a community, while defection tempts individuals to exploit short-term benefits \cite{hardin1968tragedy,Kerr:Altruism:2004,Nowak:FiveRules:2006,Rand:HumanCooperation:2013,Frank:globalscalecooperation:2018,VanLange:SD:2013}.  A classical setting is the prisoner's dilemma \cite{Rapoport-Chammah:1966,Trivers1971,Axelrod:EffectiveChoice:1980}, where two players have to choose simultaneously between cooperation,~$\mathrm{C}$,  and defection,~$\mathrm{D}$. If both cooperate, they get a higher payoff than if both defect. But if one of them defects while the other cooperates, the defector gets the highest payoff of the game and the cooperator, the lowest. This game is described by the payoff matrix 
\begin{equation}
\begin{array}{c|cc}
      & \mathrm{C} & \mathrm{D} \\ \hline
\mathrm{C} & a_{\CC} & a_{\CD} \\
\mathrm{D} & a_{\DC} & a_{\DD}
\end{array}
\end{equation}
The entries of the matrix indicate the payoffs for the row player. The game is a prisoner's dilemma if  $a_{\DC}\!>\!a_{\CC}\!>\!a_{\DD}\!>\!a_{\CD}$. This ranking implies that defection dominates cooperation in the one-shot game: no matter what the co-player will do, the focal player maximizes her payoff by defecting. Defection is the only Nash equilibrium~\cite{nash:PNAS:1950}: if both players defect neither one can unilaterally change to cooperation and thereby increase his payoff.\\
 %Yet because of $a_{\CC}\!>\!a_{\DD}$, this Nash equilibrium leaves everyone worse off.\\

\noindent But even selfish players may find it worthwhile to cooperate if the game is played for several rounds \cite{Fudenberg:FolkTheorem:1986}. 
In the repeated prisoner's dilemma,  cooperation can prevail by a mechanism called direct reciprocity \cite{Nowak:FiveRules:2006}: If I cooperate now, you may cooperate later. If I defect now, you may defect later.  Thus, the shadow of the future provides an incentive to cooperate in the present. 
A typical setting for the repeated game is obtained by using a probability, $w$, that the interaction will end after each round. 
A choice of $w\!=\!1$ recovers the one-shot game explained above.
The converse limit, $w\!\to\! 0$, describes the infinitely repeated game without discounting.
There is a large literature on evolution of cooperation by direct reciprocity \cite{Axelrod:1981:Science,nowak1993chaos,Nowak:TFT:1992,Nowak:WSLS:1993,vanVeelen:PNAS:2012,PressDyson:PNAS,murase2020five,Hilbe:PNAS:2017,Glytnasi:PNAS:2024,Stewart:PNAS:2013,Garcia:FRAI:2018,Glynatsi:HSSC:2021,Rossetti:ETH:2023}. Prominent strategies for playing the repeated prisoner's dilemma include tit-for-tat \cite{Axelrod:1981:Science}, generous tit-for-tat \cite{Nowak:TFT:1992}, win-stay, lose-shift \cite{Nowak:WSLS:1993}, GRIM \cite{Dal:AER:2019} and Forgiver \cite{Zagorsky:Forgiver:2013,frean1994prisoner}. See \textbf{Table S2} for details.\\

\noindent We examine the behavior of five leading LLMs in the repeated prisoner's dilemma. They are: ({\it i})~claude-sonnet-4 by Anthropic, ({\it ii})~\mbox{gemini-2.5-pro} by Google DeepMind, ({\it iii})~gpt-4o and ({\it iv})~gpt-5 by OpenAI, and ({\it v})~llama-3.3-70b by Meta. We abbreviate these models as \claude{}, \gemini, \gptfour{}, \gptfive{} and \llama{}  (\FigOriginal \textbf{A}). In all our experiments, the parameter settings of the LLMs are chosen to place them into regimes that are known to produce relatively deterministic outputs (see \textbf{Table S1} for details).\\

\noindent Prior work has begun to explore how LLMs play repeated games \cite{Akata:RepeatedGames:2025,FontanaRepeatedGames:2025,suzuki2024evolutionary,phelps2025machine,payne2025strategic,willis2025will,backmann2025ethics,tennant2024moral,zheng2025beyond, huynh2025understanding}. In those studies, LLMs tend to be pitted against other LLMs or against simple computer programs. Those efforts provide valuable insights into the dynamics of LLM play (which we review in detail in the \textbf{Supplementary Information}). However, they primarily describe realized outcomes of individual experiments – rather than the underlying LLM strategies that generate these outcomes. This limitation makes it difficult to connect the observed LLM behavior to formal solution concepts in evolutionary game theory. In contrast, we infer LLM strategies directly, by asking LLMs to react to every possible outcome of the last one (or two) rounds. While this change in approach may seem innocuous, it has dramatic consequences. It allows us to ask how LLMs would fare against any other (hitherto unobserved) opponent strategies. Moreover, by eliciting the LLMs’ strategies directly, we can compare their behavior to the considerable literature on evolution of cooperation \cite{Axelrod:1981:Science,nowak1993chaos,Nowak:TFT:1992,Nowak:WSLS:1993,vanVeelen:PNAS:2012,PressDyson:PNAS,murase2020five,Hilbe:PNAS:2017,Glytnasi:PNAS:2024,Stewart:PNAS:2013,Garcia:FRAI:2018,Glynatsi:HSSC:2021,Rossetti:ETH:2023}. We can directly analyze the properties of those strategies using concepts from (evolutionary) game theory \cite{Fudenberg:Book, Sigmund:Book:2010, hilbe:GEB:2015}. \\

\noindent We use this approach to ask whether current LLMs produce sensible strategies of direct reciprocity. First, we describe which behaviors LLMs instantiate in a neutral setting. We avoid any language traditionally used to present the repeated prisoner’s dilemma. In particular, we examine whether the generated strategies are Nash equilibrium strategies~\citep{nash:PNAS:1950}, and whether they fall into the formal classes of `partners' or `rivals' \cite{Hilbe:NHB:2018}, which characterize the degree to which a strategy is generous or competitive. In a second step, we record how LLMs adapt their strategies as we vary several key parameters, including the game’s stopping probability, payoff values, the number of past rounds the LLM is instructed to keep in memory, and whether or not players are aware that the current round is the last. All parameters are known to affect both theoretical \mbox{predictions~\citep{Stewart:PNAS:2014,LaPorte:PNASNexus:2026,hilbe:GEB:2015,Hilbe:PNAS:2017}}  and human behavior~\citep{Mao:NComms:2017,DalBo:JEL:2018}. We explore whether changes in these parameters elicit similar adaptations among LLMs. Interestingly, all of the LLMs show sensible adaptations in some of the tasks, but none of them exhibit full consistency across all of the tasks.\\

\noindent
In a last step, we explore the robustness of the LLM output by using different framings of the decision task (either eliciting more competitive or more cooperative behavior). 
Moreover, we conduct several round-robin tournaments~\citep{Axelrod:1981:Science,Glynatsi:PLoSCB:2024}. 
These analyses illustrate to which extent LLM behavior can be steered into certain directions, and how LLMs fare when competing against each other. 
Overall, this research provides an important glimpse into how current generations of LLMs instantiate reciprocal cooperation. 
In view of the recent progress in agentic AI, such artificially generated social behaviors are increasingly likely to produce spillovers in the world we live in.

%%%%%%%%%%%%%%%%%%
%% ORIGINAL EXPERIMENT %%
%%%%%%%%%%%%%%%%%%

\section*{The first experiment}

\noindent In our first experiment, we test how the five LLMs choose between cooperation and defection in a repeated interaction.
We use the prompt shown in \mbox{\bf Box~1} to obtain their choice, depending on the outcome of the previous round.  
We use a slightly different prompt to obtain their choice for the first round (see \mbox{\textbf{Supplementary Information}}).
The wording is designed to avoid ambiguity.  
The labels of the two options, $\mathrm{L}$ and $\mathrm{R}$, are used as neutral symbols for cooperation and defection.
For the payoff values in this first experiment we use  the standard parameters~\citep{Axelrod:1981:Science}, $(a_\LL, a_\LR, a_\RL, a_\RR) = (3,0,5,1)$. 
For those values the game is a prisoner's dilemma, and mutual cooperation, $\LL$, receives a higher payoff than alternating between $\LR$ and $\RL$.
We ask all five LLMs 50 times for each of five decisions, namely: what to do in the first round, and what to do if the previous round was $\LL$, $\LR$, $\RL$, or $\RR$.  Thus, we record the propensity of the LLM to cooperate ({\it i})~in the first round, ({\it ii})~after mutual cooperation (LL), ({\it iii})~after being exploited (LR), ({\it iv})~after exploiting the co-player (RL), and ({\it v})~after mutual defection. 
If the LLMs choose to act consistently with Nash equilibrium play, we would expect them to generate strategies that either fully cooperate, fully defect, alternate between cooperation or defection when playing against a copy of themselves, or that they implement so-called equalizers~\citep{Stewart:PNAS:2014,LaPorte:PNASNexus:2026}.\\

\begin{table}[h]
\begin{tcolorbox}[title=Box 1: Prompt used for the first experiment,width=1\linewidth,]
{\fontsize{10}{10}\selectfont You are in an interaction with another agent.\\
Your interaction is over several rounds.\\
In each round, you have a choice between two options, L or R.\\
The other agent also has a choice between L or R.\\
The two of you will decide without knowing what the other will choose.\\
Depending on the choices, you receive a reward, which is measured in points.\\
If both of you choose L, both of you get [$a_\LL$] points.\\
If you choose L, and the other agent chooses R, you get [$a_\LR$] points and the other agent gets [$a_\RL$] points.\\
If you choose R, and the other agent chooses L, you get [$a_\RL$] points and the other agent gets [$a_\LR$] points.\\
If both of you choose R, both of you get [$a_\RR$] point.\\
Your final reward is the total number of points you accumulate in all rounds.\\
% \noindent[\textit{Optional insert goes here.}]\\
\noindent In the previous round, you chose [$x$], the other agent chose [$y$]. \\
Therefore, you got [$a_{xy}$] point(s) and the other agent got [$a_{yx}]$ point(s). \\
This is a new round.\\
Do you choose L or R? Give only the character as output. Give no explanation.}
\end{tcolorbox}
\end{table}

\noindent 
The results of our experiments are shown in \FigOriginal \textbf{B}.
Four of the five LLMs cooperate fully in the first round. 
Only \gemini{} is suspicious: it cooperates 41 times in 50 trials. \capitalclaude{} implements the strategy Forgiver~\citep{frean1994prisoner,Zagorsky:Forgiver:2013}: it cooperates fully after mutual cooperation, after having exploited the co-player and after mutual defection; it always defects if it was exploited. \capitalgemini{} uses a stochastic variation of WSLS~\citep{Nowak:WSLS:1993}: it cooperates fully after mutual cooperation,  it never cooperates after having been exploited; it cooperates 1/50 times after exploiting the co-player; it cooperates 21/50 times after mutual defection. \capitalgptfour{} uses GRIM~\citep{Sigmund:Book:2010}: it fully cooperates after mutual cooperation and fully defects otherwise. \capitalgptfive{} uses a stochastic strategy that is between WSLS and Forgiver; after exploiting the co-player it cooperates 27/50 times. \capitalllama{} uses a version of generous-tit-for-tat~\citep{molander:jcr:1985,Nowak:TFT:1992}: it always cooperates after mutual cooperation; it cooperates 13/50 times after having being exploited; it cooperates 49/50 times after exploiting the co-player; it cooperates 28/50 times after mutual defection. See \textbf{Supplementary Information} for all data.\\

\noindent
We can use this data to infer memory-1 strategies (see \textbf{Table~S12-21}). 
A memory-1 strategy is given by five parameters, $(p_0,p_\LL,p_\LR,p_\RL,p_\RR)$. 
These parameters represent the probabilities to cooperate in the opening round, $p_0$, and the probabilities to cooperate after each of the four outcomes of the previous round. 
Based on the inferred memory-1 strategy, we can also compute whether an LLM's strategy is a Nash equilibrium~\citep{nash:PNAS:1950}, a partner, or a rival \cite{Hilbe:NHB:2018}. A strategy is a Nash equilibrium if the expected payoff it earns against itself is at least as high as the expected payoff any opponent earns from it. A strategy is a \textit{partner} if ({\it i})~it is a Nash equilibrium and ({\it ii})~if it fully cooperates with a copy of itself. A strategy is a \textit{rival} if its own payoff is greater than or equal to the opponent's payoff for any opponent. Partner strategies are a subset of Nash equilibria, but rival strategies are not. 
%Moreover, for the limit $w\to 0$, there are memory-1 strategies that are both partner and rival (such as GRIM). 
%For $w\!\ge\!1$ the two concepts are mutually exclusive. 
Whether a given strategy is a member of any of these strategy classes depends on the payoff values, $(a_\LL, a_\LR, a_\RL, a_\RR)$, and the termination probability, $w$, see Ref.~\citep{hilbe:GEB:2015}. Most memory-1 strategies are in neither of the three sets. For further details, see \textbf{Supplementary Information}.\\

\noindent Based on these definitions, we explore whether any of the LLM strategies is a Nash equilibrium, a partner, or a rival. 
For each strategy, we also calculate the range of expected payoffs that are attained in an encounter between this strategy and an arbitrary opponent. 
The result is shown in \FigOriginal \textbf{C}, using either a stopping probability of $w\!=\!0$ or $w\!=\!0.01$. 
In the infinitely repeated game ($w\!=\!0$), all five LLMs produce partner strategies. 
In particular, all strategies are Nash equilibria. 
However, once we consider long but not infinitely long games ($w\!=\!0.01$), four of the five LLM strategies remain fully cooperative, but only two of them remain Nash equilibria (the strategies generated by \gptfour{} and \llama{}).

%%%%%%%%%%%%%%%%%%%%%%
%% DEPENDENCE ON PARAMETERS %%
%%%%%%%%%%%%%%%%%%%%%%

\section*{Dependence on parameter changes}

\noindent 
{\bf Impact of the stopping probability.} The previous experiments establish a baseline scenario. 
They show that in a commonly used parameter domain, LLMs aim to establish cooperation. 
In a next step, we explore how LLM behaviors change as we systematically vary the parameters of the game. We begin by varying the stopping probability~$w$. Theory and experimental work both predict that larger values of $w$ make it more difficult to sustain cooperation. After all, a larger stopping probability reduces `the shadow of the future'~\citep{hilbe:GEB:2015,DalBo:JEL:2018}. 
We would thus expect that larger values of $w$ render the LLM strategies less tolerant~\citep{hilbe:GEB:2015}.
That is, the cooperation probabilities after unilateral or mutual defection ought to become smaller in a monotonic fashion. Moreover, above a certain value of $w$, an LLM may stop to cooperate altogether.\\

\noindent
In our original experiment, we only told the LLMs that the interaction is over ``several rounds". Now we replace that line with the following sentence: ``After each round the interaction ends with probability $w$''. We use the following four values $w=0.01, 0.1, 0.2, 0.5$.  The results are shown in \FigVariation \textbf{A}. 
Curiously, \claude{}, \llama{} are unaffected by the value of $w$. \capitalclaude{} always opts for Forgiver (even though the strategy ceases to be a Nash equilibrium for $w\!>\!0$). Similarly, \llama{} always chooses GRIM. \capitalgptfour{} produces behavior that is in between GRIM and WSLS. Interestingly, it exhibits some non-monotonic variation in $p_\RR$, its cooperation probability after mutual defection. \capitalgemini{} reduces its probability to cooperate in the first move and after mutual cooperation as $w$ increases. \capitalgptfive{} reduces its propensity to cooperate after exploiting the co-player and after mutual defection.
These latter two LLMs adapt their strategies in ways that are most consistent with our expectation of monotonically decreasing cooperation propensities.
Interestingly, however, for $w\!>\!0$, neither \gemini{} nor \gptfive{} produce strategies that are consistent with Nash equilibrium play (see \textbf{Table} \textbf{S29}). 
Instead,  \llama{} produces a partner strategy for all values of $w$, while \gptfour{}’s strategy is a partner for all $w$ except $w \!=\! 0.5$. (Recall that every partner strategy is a Nash equilibrium by definition.)\\

\noindent
{\bf Changing the payoff matrix.} 
As another key parameter, we change the players' incentives to cooperate. We conducted eleven additional experiments using the following payoff values:  $10$ for mutual cooperation; $0$ for cooperation versus defection; $10+x$ for defection versus cooperation; and $x$ for mutual defection. We vary the parameter $x=0,1,2,...,10$. Increasing $x$ means the cost for cooperation increases and the temptation to defect becomes larger. For $x=0$ cooperation has no cost at all. For $x=10$, the cost of cooperation is as large as the benefit. For all values of $x$ in between 0 and 10 the game is a prisoner's dilemma.
The larger this cost $x$, the more difficult cooperation becomes to sustain. 
Thus we would expect that with increasing costs, the cooperation propensities of LLMs decrease~\citep[again in a monotonic fashion, see Refs.][]{Stewart:PNAS:2013,hilbe:GEB:2015}.\\

\noindent
The results of the LLM experiments are shown in Figure \FigVariation \textbf{B}. \capitalclaude{} uses its favorite strategy Forgiver for all values of $x\!<\!10$; for $x\!=\!10$ it suddenly uses TFT. \capitalgemini{} chooses WSLS for small costs, but switches to suspicious GRIM for large costs. \capitalgptfour{} uses WSLS for small costs and switches to GRIM for large costs. \capitalgptfive{} uses Forgiver for small costs and reduces its cooperation propensity for larger costs; for very large costs it approaches `always defect'~(ALLD). \capitalllama{} uses `always cooperate' (ALLC) for zero cost. For larger costs, it maintains full cooperation in the first round and after mutual cooperation, but it reduces its cooperation propensity after mutual defection. 
Overall, these changes are largely in line with our expectations. However, again we observe some remarkable non-monotonicities in some of the cooperation propensities (especially for \llama, \FigVariation \textbf{B}).\\

\noindent
Again we analyze whether the inferred strategies are Nash equilibria, partners or rivals for $w\!=\!0$ and $w\!=\!0.01$. The results are summarized in \textbf{Table S31}. Firstly, we observe that the chosen strategies are predominantly partners and not rivals. Secondly, strategies are more likely to be partners when $x$ is sufficiently low. In particular, save for a few exceptions, these strategies are only partners for $x\leqslant 5$. Only \gptfour{}'s strategy classifies as a partner for most $x$.\\

\noindent Overall, the above results suggest that all LLMs perform well in some of the tasks. However, none of the LLMs exhibit sensible behavior throughout. For example, \claude{} seems to ignore changes in the game's stopping probability or in the incentives to cooperate (\FigVariation), whereas the cooperation probabilities of \gptfour{} and \llama{} change non-monotonically (based on 50 independent API calls).

\section*{Memory-2 decisions of LLMs}

\noindent  
While many publications explore optimal behavior among memory-1 strategies, the literature on higher memory is comparably scarce~\citep{hauert:PRSB:1997,Hilbe:PNAS:2017,stewart:scirep:2016,Murase:PLoSCompBio:2023a,Glytnasi:PNAS:2024}. This raises the question whether LLMs -- which are trained on the published literature -- are able to generate reasonable strategies in these lesser explored domains. 
To explore that question we have conducted experiments where we study decisions of LLMs given the outcome of the previous two rounds. The results are shown in \FigMtwo. 
For the first two rounds, the results coincide with the original treatment (\FigOriginal). 
For the subsequent rounds, we find the following.
If the previous round was mutual cooperation ($\LL$) then \gemini{}, \gptfour{}, \gptfive{} and \llama{}  cooperate fully regardless of the outcome two rounds ago; only \claude{}  looks further back and chooses to defect if it was exploited two rounds ago. If the agent was exploited in the previous round ($\LR$), then \claude{}, \llama{}  and \gptfour{} fully defect regardless of what happened two rounds ago, \gptfive{} is willing to cooperate if it exploited the co-player two rounds ago; but \llama{}  is willing to cooperate unless it was exploited twice in a row. In general, \llama{}  is very cooperative in this setting. Analyzing the resulting memory-2 strategies, we find that \claude{}, \gptfour{} and \gptfive{} are partners, whereas \gemini{} and \llama{} are neither Nash equilibria nor rivals.\\

\section*{Repeated games over 10 rounds}

\noindent 
The previous experiments assume that players are not aware of how many rounds they will play. 
This is a standard assumption made to preclude end game effects~\citep{Fudenberg:Book}: 
Once the duration of the game is known, rational players should defect in the last round;
but once players expect their opponents to defect in the last round, they should already defect the round before, and so on~\citep[e.g.][]{Mao:NComms:2017,Embrey:Endgame:2018}. Depending on how far-sighted LLMs are -- or to which extent they expect their opponent to be far-sighted -- cooperation may thus unravel completely.
To explore whether LLMs engage in this kind of backward induction, we study a repeated prisoner's dilemma with ten rounds. We performed two treatments. In one treatment, we tell the LLMs that the game lasts at least ten rounds. In the other treatment, we specify that we play \emph{exactly} ten rounds.\\ 

\noindent  Each LLM plays against all five LLMs. This results in a total of 15 distinct encounters. As the games proceed, we remind the LLMs of the cumulative history of the game until the present round. For the exact prompts, see \textbf{Supplementary Information}. In the first treatment, we observe full cooperation in all games, which is consistent with the data from the original experiment. In the second treatment, all LLMs except \gptfive{} cooperate in every round of every game, suggesting that none of them exercise backward induction. Only \gptfive{} defects in the last round of all its games, but in none of the previous rounds.\\

\noindent  Note that all LLMs except \gptfive{} fail to defect in the last round when the game has a known ending. This failure to exploit the opponent is remarkable, because defection in the last round is optimal regardless of a player's belief about its opponent -- that is even without `common knowledge of rationality'.

%%%%%%%%%%%%%%%%%
%% DIFFERENT FRAMINGS %%
%%%%%%%%%%%%%%%%%

\section*{Exploring different framings}

\noindent 
After describing how different LLMs react to parameter changes, we study a different kind of robustness.
By modifying the input prompt, we study to which extent an LLM's behavior can be steered into different directions.
In each case, we add a single sentence to the prompt, see Section 3.1 in \textbf{Supplementary Information} for precise instructions.
Four of the frames are designed to enhance competitiveness: 
(1)~{\it `Try to get more points than the other agent'.} (2)~{\it `Exploit the other agent as much as possible'.} (3)~{\it `Maximize your own number of points.'} (4)~{\it `Think about winning.'} Four other frames enhance some kind of altruism: 
(5)~{\it `Accumulate many points, but do not exploit.'}
(6)~{\it `Maximize your points, keep in mind the other’s welfare.'}
(7)~{\it `Think about fair outcomes'.}
(8)~{\it `Be a saint'.}
The remaining frame is up for interpretation: (9)~{\it `Play like a pro.'} 
With these different frames, we do not only gauge the extent to which LLM output is amenable to modification; we also explore whether LLMs produce reasonable behavior when given instructions they are unlikely to have encountered before. 
The results are shown in \FigFraming{} and in \textbf{Tables S4-S8}. Let us discuss them in turn.\\

\noindent(1)~{\it `Try to get more points than the other'} is a recipe for disaster in cooperative dilemmas~\citep{Berg:PNAS:2015}. Indeed, we find that LLMs are biased accordingly, with four of them choosing ALLD; only \llama{}  tries GRIM.  
(2)~{\it `Exploit the other agent as much as possible'} makes all five LLMs switch to ALLD. Only \gptfive{} and \llama{}  show some hesitation to defect in the opening move. 
(3)~{\it `Maximize your own number of points'} causes \claude{}, \gptfour{} and \gemini{} to switch to (almost) ALLD. \capitalllama{} again uses GRIM.  \capitalgptfive{} uses a stochastic version of WSLS. 
(4)~{\it `Think about winning'}  induces \claude{}, \gemini{} and \gptfour{} to adopt (almost) ALLD; \gptfive{} again opts for a stochastic variant of WSLS, while \llama{}  again goes for GRIM. 
Averaging over those four prompts, we find that \claude{}, \gemini{} and \gptfour{} go for ALLD; \llama{}  maintains the highest propensity to cooperate in the opening move and after mutual cooperation; but only \gptfive{} maintains some willingness to cooperate even after mutual defection.\\

\noindent (5)~{\it `Accumulate many points, but do not exploit'} causes \claude{}  and \gptfour{} to choose ALLC; while \gemini{}, \gptfive{} and \llama{}  opt for Forgiver.
(6)~{\it `Maximize your points, keep in mind the other’s welfare'} and 
(7)~{\it `Think about fair outcomes'} makes the five LLMs adopt Forgiver like strategies, with \gemini{} is moving very close to ALLC. (8)~{\it `Be a saint'} makes four of the five adopt ALLC, while \llama{}  stays with Forgiver.
In these four altruism-inducing treatments all five LLMs cooperate fully in the opening move, after mutual cooperation and after mutual defection. There is some hesitation to cooperate after being exploited and only \llama{}  once shows a reluctance to cooperate fully after exploiting the co-player.\\

\noindent The ominous (9)~{\it `Play like a pro'} has the following interesting effect when compared to our baseline experiment (i.e., no framing). \capitalclaude{} moves from Forgiver to WSLS. \capitalllama{} moves from GTFT to GRIM. The three others change very little. (Perhaps they `think' they are already playing like pros).
Overall, we observe that LLM strategies are easily steered into different directions. Their behavior shifts in predictable ways: competitive frames trigger defection, while cooperative frames encourage consistent cooperation. 
%\capitalllama{} and \gptfive{} show a more measured response:  even in competitive framings they still try to cooperate after mutual cooperation. In cooperative framings, all LLMs seek to cooperate, except in scenarios where they have just been exploited.\\

%%%%%%%%%%%%%
%% TOURNAMENTS %%
%%%%%%%%%%%%%

\section*{Tournaments}

As an additional way to evaluate whether LLMs generate sensible behavior, we  explore to which extent these strategies yield superior payoffs in typical round-robin tournaments~\citep{Axelrod:1981:Science,Glynatsi:PLoSCB:2024}.
For these tournaments, we let the LLM strategies (as inferred from each experiment) play against one another. We conduct two types of tournaments. In the first type, each strategy plays against each of the five strategies, including a copy of itself. In the second type, each strategy plays against the four other strategies (that is, excluding self-interaction). In both cases, the five strategies are ranked according to the payoff sum they obtain in all pairings. 
We conduct those two types of tournaments for each of the fifteen experiments with a neutral framing (i.e., original experiment, the experiments with different stopping probabilities and cooperation costs, and the experiment in which the LLM is encouraged to {\it `play like a pro'}, see \FigConclusion \textbf{A}).
We think of these experiments as different disciplines. 
Detailed results are shown in \textbf{Fig. S1-10}.\\

\noindent For the first type of tournament, \claude{}  wins 12 disciplines, while \gemini{}, \gptfive{} and \llama{} each win one. 
For the second type of tournament, \claude{} score 11 wins, followed by \gptfive{} with two and \gemini{} and \llama{}  with one each. In all 30 tournaments, \claude{} is in the top three, underscoring the robust success of Forgiver. In the world of LLMs, Forgiver apparently succeeds. %triumphs.

\section*{Conclusion}

\noindent We have analyzed the strategic behavior of five large language models across various versions of the repeated prisoner's dilemma. We use the evidence of about 40000 individual API calls to address whether different LLMs have developed a coherent way of engaging in reciprocal cooperation. 
We employ different measures to quantify their success. First, we ask to which extent LLMs generate strategies with appealing theoretical properties. These properties may entail that their strategy forms a Nash equilibrium~\citep{nash:PNAS:1950}, or that it can be characterized as a partner or rival~\citep{Hilbe:NHB:2018}. Second, we explore to which extent LLMs are sensitive to variations in the game parameters.  To this end, we vary the expected length of the game, the payoffs, the LLM information window, and whether or not they know how long the game lasts. To the extent that LLMs adapt their behavior to these variations at all, we ask whether their adaptations are in line with predictions from theory or experiments. In a final step, we explore whether these LLMs are flexible enough to react to framings they are unlikely to have encountered before, and how they perform in a round-robin tournament.\\

\noindent
We find that all LLMs show good performance in many of the tasks. 
For example, \llama{} and \gptfour{} seem to systematically adopt partner strategies -- more so than the newer generation \gptfive. 
But \gptfive{} is successful at identifying opportunities to exploit the co-player in games with a known ending. 
At the same time, however, none of the LLMs induce sensible behavior across all of the tasks. 
For all of them, there are instances in which their predictions are at odds with basic intuition, theoretical predictions, or experimental evidence. For example, \claude{} does not change its strategy of choice (Forgiver) even if the stopping probability of the game becomes very large (0.5); \llama{} has a non-monotonic propensity to cooperate after exploitation as the cost of cooperation increases; \gptfour{} switches from Grim to stochastic Win-stay, lose-shift (WSLS) as soon as a stopping probability is mentioned. Finally, except for \gptfive{}, no LLM defects when informed that the current round is the final one. None of the LLMs apply backward induction.\\

\noindent
Interestingly, we also observe that the LLMs show some strategic similarities. 
For example, across our 15 experiments without framings, we observe that they almost always cooperate in the first round, and after mutual cooperation. Moreover, all five LLMs have some willingness to cooperate after mutual defection (see \FigConclusion \textbf{B}). 
At the same time, there are also interesting differences, which could be interpreted as differences in their (strategic) personality. 
For example after exploiting the co-player, \gemini{} and \gptfour{} show no remorse. They continue to exploit (a behavior that is consistent with WSLS). By contrast, \gptfive{} is undecided, whereas \llama{} and \claude{} tend to cooperate (a behavior that is consistent with GTFT and Forgiver). 
Conversely, after being exploited, four of the five LLMs always defect. Here, only \llama{} shows some willingness to cooperate (a behavior that is consistent with GTFT).\\

\noindent Previous papers have studied LLM behavior in the repeated Prisoner’s Dilemma by reporting the outcomes of individual sequences of games \cite{Akata:RepeatedGames:2025,FontanaRepeatedGames:2025,suzuki2024evolutionary,phelps2025machine,payne2025strategic,willis2025will,backmann2025ethics,tennant2024moral,huynh2025understanding}, much like social scientists would probe human behavior. Instead, in this paper we have proposed a simple framework for directly eliciting the underlying strategies of LLMs in settings of bounded (short-term) memory. While both approaches are necessary to understand the performance of LLMs, the method described in this paper allows us to test whether LLMs instantiate Nash equilibria, or if they implement other well-known strategy sets such as partners and rivals. Our experimental results can be compared with the large literature on mathematical and computational analysis of direct reciprocity. In the future, our experimental approach can be used as a standardized tool to quantify the progress of AI in social domains as new versions of LLMs are released.

% \noindent
% \christian{\hl{I deleted the remaining two paragraphs we had earlier.} To some extent, their content made it into earlier paragraphs (e.g., that \gptfour{} and \llama{} are good at finding Nash equilibria). To some extent, they seem less important now (I wouldn't interpret too much into our tournament data).
% I feel instead of these two paragraphs, it would be great to have 1-2 paragraphs which: ({\it i}) briefly connect our work to previous studies on LLM social behavior, and ({\it ii}) summarizes the broader importance of this kind of work. Here, there might be two directions. First, we may want to argue that LLMs are increasingly seen as a tool to predict the outcome of economic experiments \citep[e.g.,][]{Hewitt:WP:2024,Shapira:arxiv:2025}. This requires LLMs to be a reliable model of human behavior. We explore this question, in a systematic fashion, using five LLMs, for the main paradigm of human cooperation. Second, people increasingly use agentic AI to make decisions on their behalf (perhaps Saptarshi knows of relevant publications?). Here, we need to understand the LLMs default mode of reciprocal cooperation.}

\newpage

\begin{figure}[h!]
    \centering
    \includegraphics[width=0.9\linewidth]{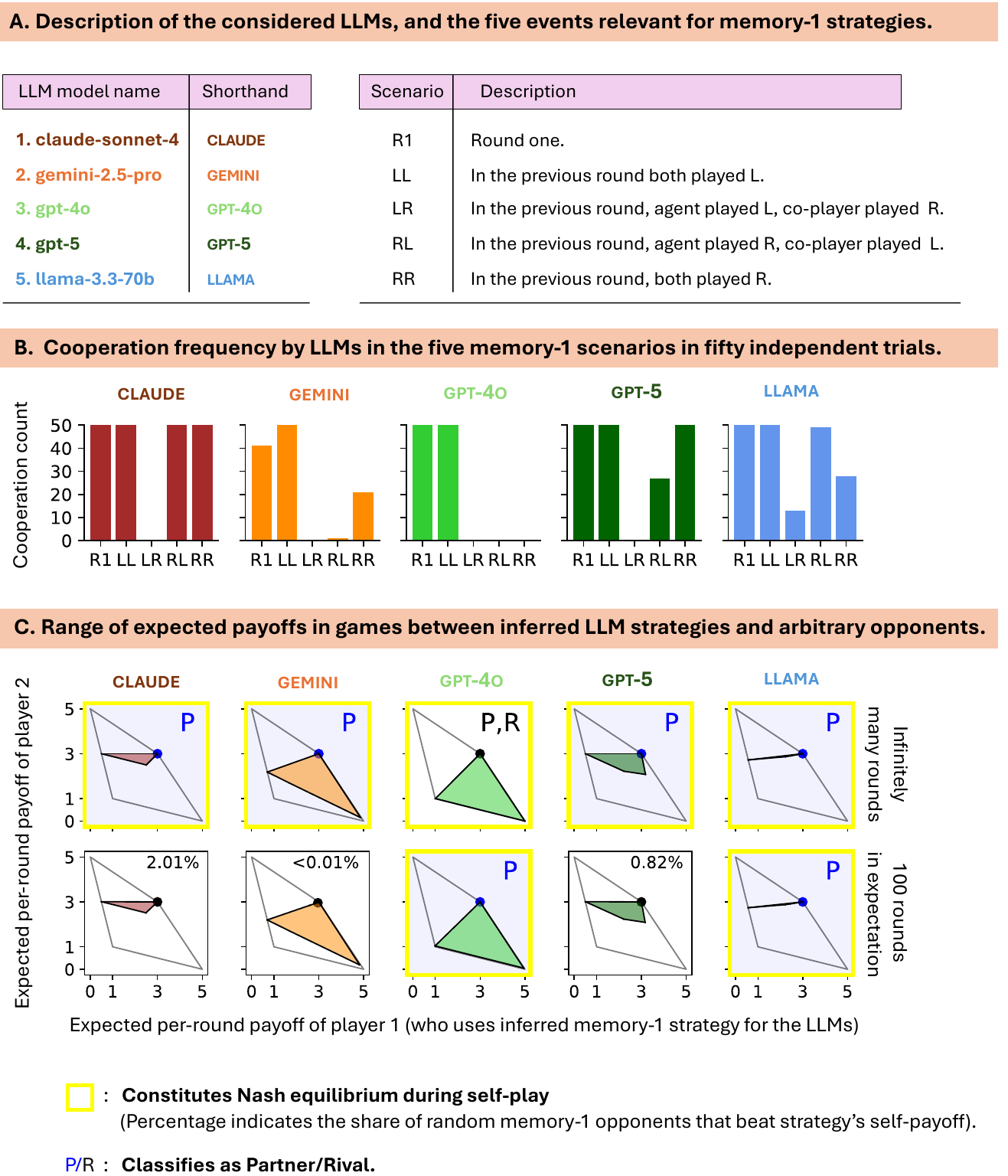}
    \vspace{0.2cm}
    \caption{\textbf{How do LLMs respond to memory-1 scenarios of the repeated prisoner’s dilemma?} \textbf{A,} We consider five state-of-the-art Large Language Models (LLMs)—claude-sonnet-4 by Anthropic, gemini-2.5-pro by Google DeepMind, \gptfour{} and \gptfive{} by Open AI and llama-3.3-70b by Meta AI. We refer to these models \claude{}, \gemini{}, \gptfour{}, \gptfive{} and \llama{}. We present five repeated prisoner’s dilemma scenarios to the LLMs and request their choices. The five scenarios elicit the LLM's response in the first round, and after the four possible outcomes of the previous round (see {\bf Box~1} for the prompts). \textbf{B,} We report how often the five LLMs choose the cooperative action `L' in each scenario across fifty independent trials.  See \textbf{Tables S4-S8} for the data.  \textbf{C,} We show the range of expected per-round payoffs possible when inferred LLM memory-1 strategies play arbitrary opponents.  We show these ranges for games with infinitely many rounds (top row) and games with 100 rounds in expectation (bottom row).  We indicate whether an LLM strategy gives rise to a Nash equilibrium with a yellow square; otherwise, we report the percentage of randomly sampled memory-1 opponents (out of $10^6$) that achieve a higher payoff against it than it earns against itself. With `P' or `R' we denote if the LLM strategy is a partner or a rival.}
    \label{fig:Fig1}
\end{figure}

\begin{figure}[h!]
    \centering
    \includegraphics[width=\linewidth]{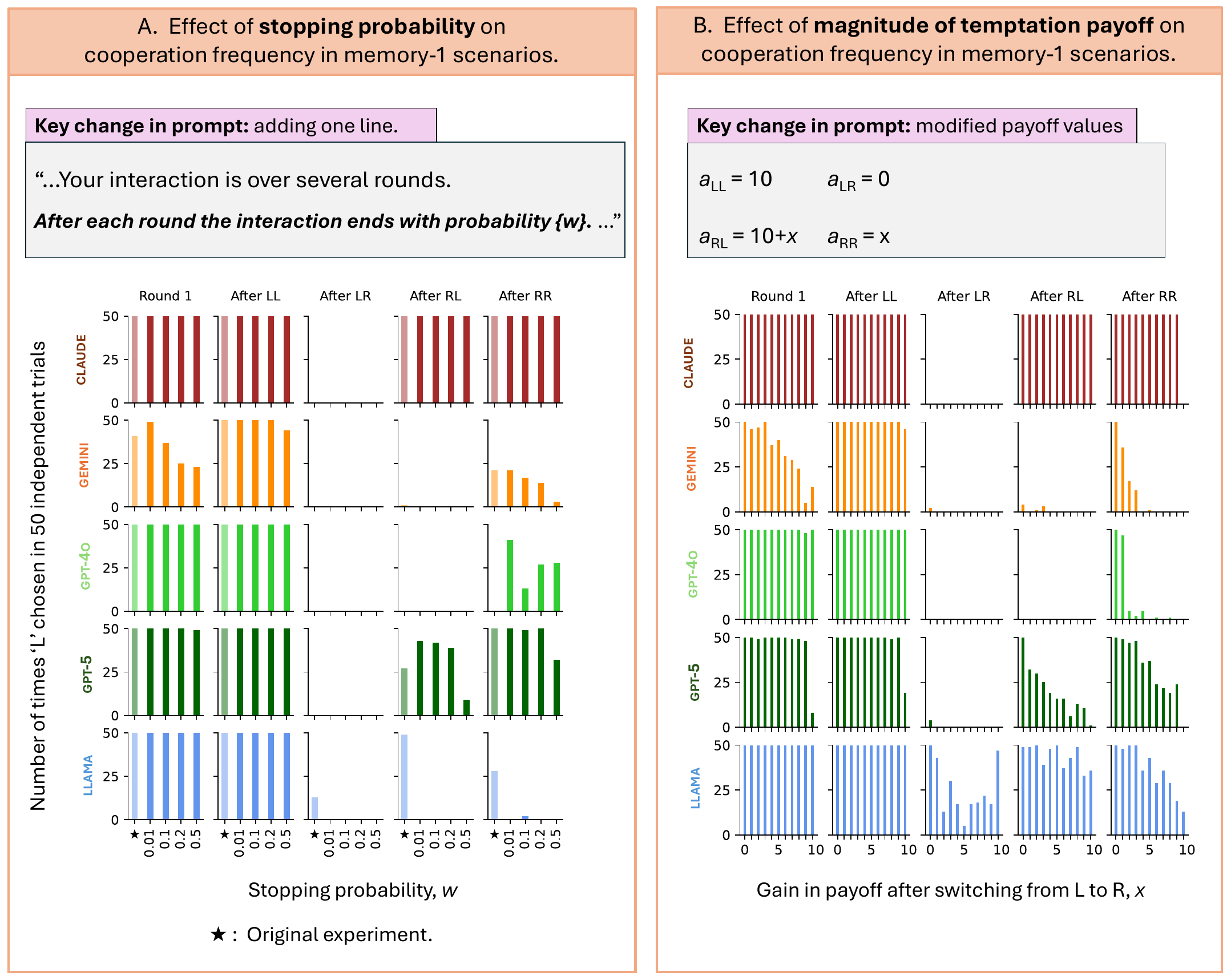}
    \vspace{0.02cm}
    \caption{\textbf{How does conditional cooperation by LLMs vary with the game parameters?} We vary two key game parameters: the interaction's stopping probability (panel \textbf{A}) and the game payoffs (panel \textbf{B}). For the first, we add a single line to the original prompt in {\bf Box~1}: {\it ``After each round the interaction ends with probability $\{w\}$.''} We vary $w$ and record how often the five LLMs choose the cooperative action `L' in the five memory-1 scenarios across fifty independent trials. The payoff parameters match those used in the previous figure, $(a_{\mathrm{LL}},\, a_{\mathrm{LR}},\, a_{\mathrm{RL}},\, a_{\mathrm{RR}}) = (3, 0, 5, 1)$. For the second, we use the special payoff values $(a_\LL,a_\LR,a_\RL,a_\RR)=(10,0,10+x,x)$, representing games with the `equal gains from switching' property~\citep{nowak:AAM:1990}. Here, $x$ is the gain in payoff when unilaterally shifting from the cooperative action `L' to the defective action `R'.  At $x\!=\!0$, a player's payoff is independent of its own action. At $x\!=\!10$, cooperation ceases to be socially optimal since mutual defection is as beneficial as mutual cooperation.  We vary $x$ and plot how often the five LLMs cooperate in the five memory-1 scenarios. For more data and information on whether the inferred strategies form Nash equilibria or act as partners or rivals, see {\bf Section 6.2} of the \textbf{Supplementary Information}.} 
    \label{fig:Fig-var}
\end{figure}

\begin{figure}[h!]
    \centering
    \includegraphics[width=0.83\linewidth]{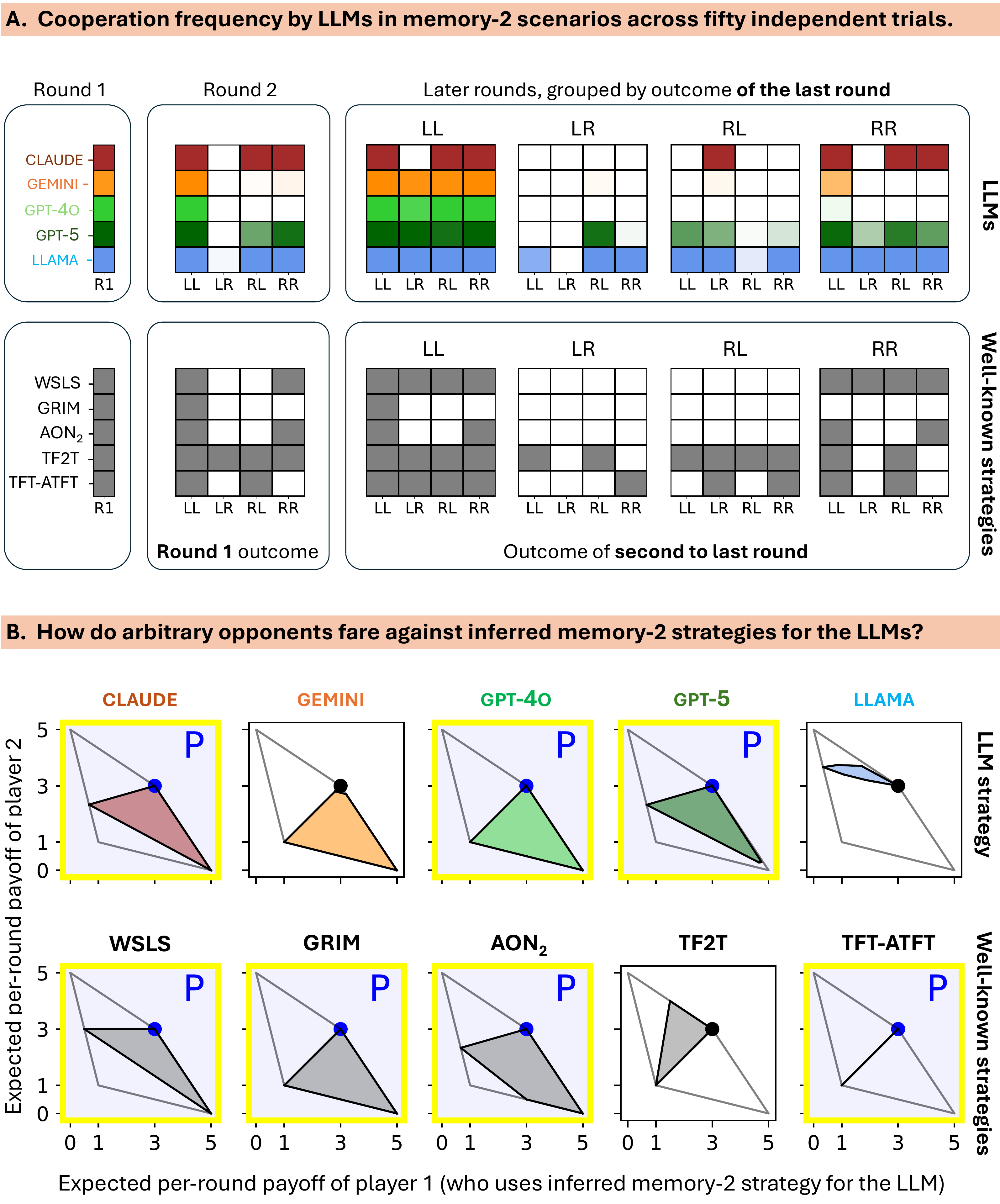}
    \vspace{0.3cm}
    \caption{
    \textbf{How do LLMs act when responding to the outcome of the last two rounds?}
    \textbf{A,} We display the frequency with which the LLMs choose the cooperative choice `L' in scenarios that include information about the previous two rounds. There are 21 such scenarios: the first five correspond to rounds one and two; the remaining sixteen correspond to later rounds (see {\bf Section 4.4} of \mbox{\textbf{Supplementary Information}} for the exact prompts, and {\bf Section 4.5} for the resulting data). As before, we perform 50~independent trials for each LLM-scenario combination. We compare the LLMs’ choices to five well-known strategies from the literature: Win-Stay, Lose-shift \citep[WSLS, Ref.][]{Nowak:Nature:1993}, GRIM~ \cite{Dal:AER:2019}, All-or-None-2~\citep[AON$_2$, Ref.][]{Hilbe:PNAS:2017}, Tit-for-two-tats \citep[TF2T, Ref.][]{Boyd:Nature:1987}, and a strategy combining Tit-for-Tat with anti-Tit-for-Tat \citep[TFT–ATFT, Ref.][]{Do:JTB:2017}, see also \textbf{Table S3}.
    \textbf{B,} We show how arbitrary strategies perform against the elicited LLM strategies, as well as against the five well-known strategies. Payoff computations are performed for stopping probability $w\!=\!10^{-10}$; other parameters are the same as in \FigOriginal. Strategies that form symmetric Nash equilibria are marked with yellow frames. The letters `P' and `R' indicate whether the strategies are partners or rivals.} 
    \label{fig:Fig-M2}
\end{figure}

\begin{figure}[h!]
    \centering
    \includegraphics[width=0.9\linewidth]{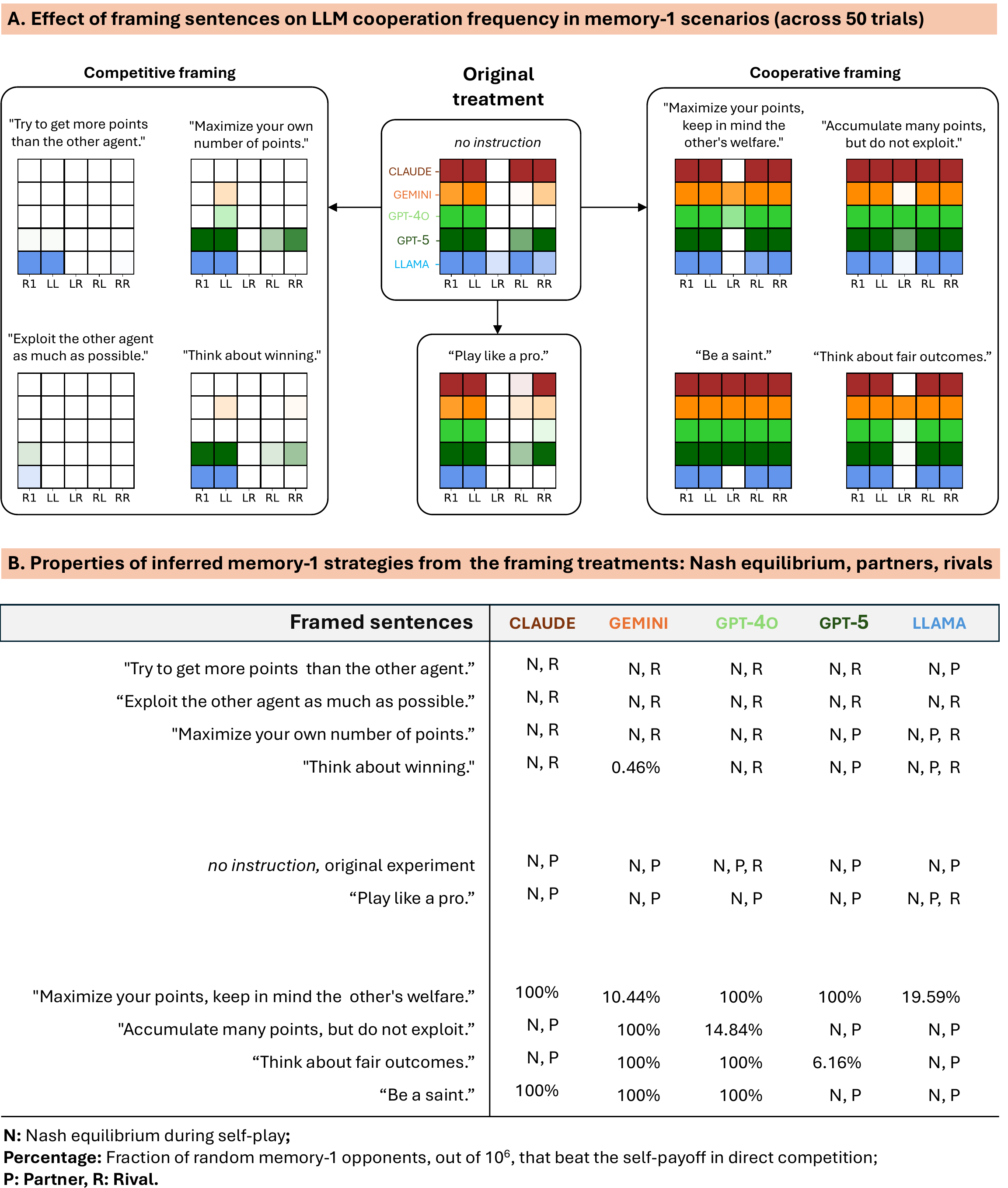}
    \vspace{0.3cm}
    \caption{\textbf{How is LLM choice affected by cooperative or competitive framings used in the prompts?} \textbf{A,} We conduct nine new treatments in the style of our original memory-1 experiment, each adding a colored framing to the original prompts. Each framing tells the LLM to focus on a specific goal. Four framings are competitive, four are cooperative, and one is neutral. Using heatmaps, we show how often each LLM chooses the cooperative action “L” across 50 independent trials for each scenario and framing. For the exact prompt see {\bf Section 4.1} in \textbf{Supplementary Matrials}. \textbf{B,} We compute if the inferred LLM strategies form a symmetric Nash equilibrium (`N'), whether they are partners (`P'), or rivals (`R') for the infinitely repeated game $(w\!\to\!0)$. If a strategy is not a Nash equilibrium, we report the fraction of uniformly randomly sampled memory-1 opponents (out of $10^6$) that achieve a larger payoff against it, as in \FigOriginal. The payoff parameters are the same as those in \FigOriginal.} 
    \label{fig:Fig-diamonds}
\end{figure}

\begin{figure}[h!]
    \centering
    \includegraphics[width=0.9\linewidth]{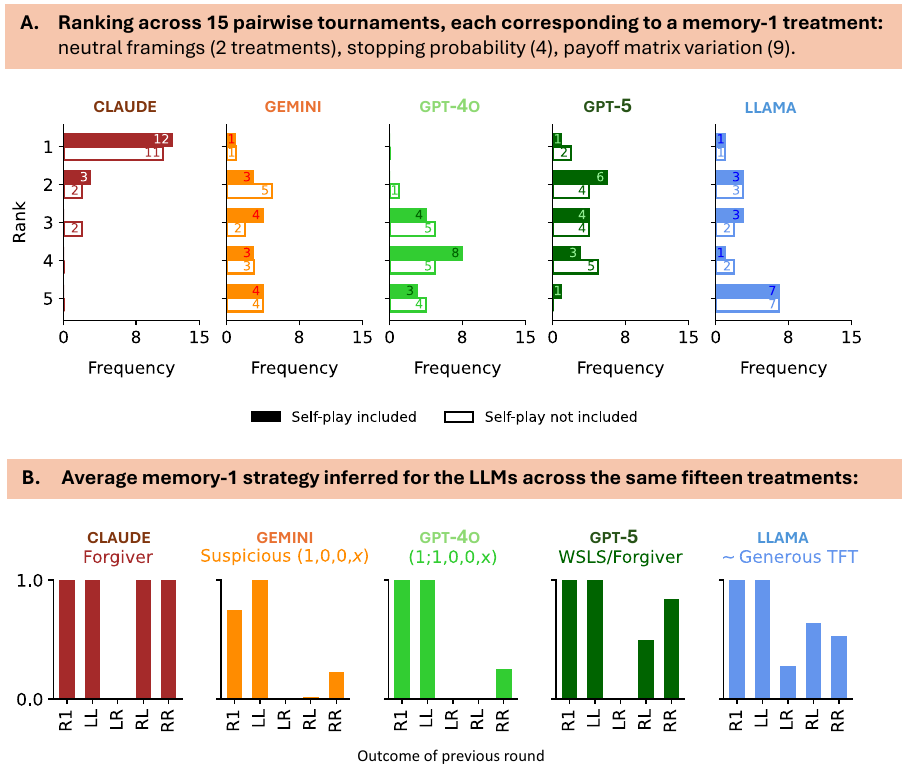}
    \vspace{0.3cm}
    \caption{\textbf{Performance of LLM strategies in fifteen pairwise tournaments, each corresponding to a memory-1 treatment.} For fifteen non-framed memory-1 treatments, we compute the exact outcomes of pairwise tournaments between strategies inferred for the five LLMs. These treatments include the baseline experiment, a neutral framing treatment (``play like a pro''), four stopping-probability treatments ($w=0.01,0.1,0.2,0.5$), and nine payoff-matrix treatments ($x = 1,\dots,9$). \textbf{A,} We show how the inferred LLM strategies rank across these tournaments. All pairwise games use a stopping probability of $w = 0.01$, except for the stopping-probability treatments, where we use the value of $w$ specified in the prompt. Numbers on the bars indicate how often each LLM attains a given rank. Filled bars include self-play outcomes, whereas hollow bars exclude them. We exclude the cases $x\!=\!0$ and $x\!=\!10$ because they do not constitute Prisoner’s Dilemmas. \textbf{B,} We show the average memory-1 strategy inferred for each LLM across these fifteen experiments.
} 
    \label{fig:conclusion}
\end{figure}

\clearpage
\bibliographystyle{naturemag}
\bibliography{bibliography.bib}
\end{document}

\title{\bfseries\sffamily \Large Supplementary Information \\ Measuring cooperation and defection in five large language models}
\date{}
\maketitle
\onehalfspacing

\vspace{-1cm}
\tableofcontents
\clearpage

\setcounter{table}{0}
\renewcommand{\thetable}{S\arabic{table}}

\vspace{1cm}
\setcounter{figure}{0}
\renewcommand{\thefigure}{S\arabic{figure}}

\noindent  Section \ref{section:related-work} reviews the related literature. Sections \ref{section:LLM-models}–\ref{Info-exps} describe our experimental setup, while Section \ref{section:data} details the data collected. Finally, Section \ref{section:analysis} outlines our data analysis procedures.

\section{Related work}
\label{section:related-work}
In this section, we review the literature most closely related to our work. We focus on studies that examine how large language models (LLMs) choose actions in repeated games, with particular emphasis on the repeated Prisoner’s Dilemma. For research on LLM behavior in one-shot games or in other dynamic game settings, we refer the reader to representative examples in \cite{brookins2023playing,vidler2025playing,lore2023strategic,herr2024large,wang2024tmgbench} and \cite{guo2023gpt,PiresSantos:IR:LLM,chen2023put,fan2024can,xia2024measuring,huang2025competing,jiamin2025social}, respectively. For an in-depth review of the present literature on game theory and LLMs, please refer to Sun \textit{et al.} \cite{sun2025game}. In the following we describe key papers studying LLMs and repeated games.\\

\noindent \textit{(i)} \textbf{Akata} \textit{et al}. \cite{Akata:RepeatedGames:2025} use GPT-4,  Llama-2, text-davinci-002, text-davinci-003, and Claude-2 to study six families of two-player, two-action, repeated games: win-win, cyclic, Prisoner’s dilemma, second-best, unfair and biased. For each family, they choose representative games. For each game, all LLMs play each other once, including themselves, for ten rounds. In every round, they request a single decision from each LLM. 
They find that GPT-4 performs best across the six game families. To study this further, the authors focus on fixed games from two canonical families: the repeated Prisoner’s Dilemma, where LLMs perform well, and the repeated Battle of the Sexes, where they perform poorly.\\

\noindent For the Prisoner's Dilemma, they find that GPT-4’s behavior is GRIM-like. It starts by cooperating, but refuses to cooperate again after a single defection by the opponent. The authors run several robustness checks: for example, they alter the names of the actions, or they talk about dollars and coins instead of points. They find that GPT-4’s behavior is consistent across those checks. Furthermore, they find if GPT-4 is told that the opponent can make mistakes, it is willing to restart cooperation after a defection by the opponent. \\

\noindent In their Supplementary Information, Akata \textit{et al.} show that informing GPT-4 about the termination probability, even if it is very small, causes it to defect in every round. Curiously, we find a different result for GPT-4o and other LLMs. In our experiments, explicitly mentioning a stopping probability does not trigger defection. Instead, GPT-4o and other advanced models continue to cooperate in every round. This behavior matches what we observe in our other simulation treatment, where LLMs are told that the game will last “at least 10 rounds”.\\

\noindent \textit{(ii)} \textbf{Fontana} \textit{et al.} \cite{FontanaRepeatedGames:2025} use Llama-2, Llama-3, and GPT-3.5 to study the repeated prisoner’s dilemma. In their setup, each LLM plays a 100-round game against a random strategy,  which cooperates with probability $p$ in each round. The authors vary the value of $p$ and the information window that is presented to the LLM. \\

\noindent Since the LLMs operate at a relatively high temperature of 0.7, their actions are stochastic. Therefore, each game setting is repeated 100 times. The goal is to extract interpretable behavioral traits from the LLMs’ play. These traits are inspired by Axelrod’s work on the repeated Prisoner’s Dilemma and include niceness, forgiveness, retaliation, emulation, and troublemaking.\\ 

\noindent Their analysis shows that Llama-3 is consistently less cooperative and more exploitative, except when the opponent always cooperates. In effect, it behaves like GRIM. In contrast, Llama-2 and GPT-3.5 are overly cooperative and fail to retaliate even if the opponent’s cooperation rate is less than 30 percent. For example, against an Always-Defect opponent in a 100-round game, a Llama-2 agent with a 100-round memory begins cooperating unconditionally after round 50, which is the opposite of optimal behavior.\\

\noindent In our analysis, we find no evidence that Llama-3.3-70B is less cooperative than the other LLMs. Instead, its strategy appears excessively cooperative, particularly in the memory-2 setting, where this over-cooperation prevents it from sustaining a Nash equilibrium when playing against itself.\\

\noindent \textit{(iii)} \textbf{Suzuki} \textit{et al.} \cite{suzuki2024evolutionary} use Llama-2 to perform a very interesting co-evolutionary simulation of language prompts and strategic behavior. They consider populations of LLM agents. All of those agents use Llama-2. Initially each agent is given a short prompt of about ten words that specifies how it should act in the repeated Prisoner’s Dilemma. Example of such prompts are: \textit{“Open to team efforts, but self-interest frequently overrides collective goals,” “Prioritizes growth and mutual wins through synergetic collaboration,” “Pursues personal gain consistently, neglecting mutual or group benefits entirely,”} and \textit{“Collaborative spirit with a dash of self-motivation.”}\\

\noindent For each prompt, the authors estimate the probability that it will cooperate in the next round given the outcomes of the previous two rounds. To do this, they enumerate all possible two-round histories. For each history, they query the LLM fifty times. This repetition is used to obtain statistically meaningful estimates of cooperation probabilities. Using these estimated probabilities, the authors simulate pairwise games between agents in the population. Each game lasts for a fixed number of rounds. Payoffs are computed using the inferred cooperation probabilities. The first two rounds of each game are assumed to have arbitrary histories.\\

\noindent The resulting payoffs determine the reproductive fitness of each agent. The number of offspring an agent produces is proportional to its fitness. Each offspring inherits the parent’s prompt unless a mutation occurs. Mutations are generated by providing a chat LLM with the parent prompt and instructing it to produce a new sentence that is similar in content and length, but sounds either more cooperative or less cooperative than the original. Mutations in both directions are equally likely. As they simulate the stochastic evolutionary process over long horizons, the authors find that although evolution often becomes stuck in defection, cooperative sentences—and hence cooperative behavior—repeatedly re-emerge.\\

\noindent This is a fascinating study that combines language evolution and evolution of cooperation. The scope of this study is very different from ours. \\

\noindent \textit{(iv)} \textbf{Phelps \& Russell} \cite{phelps2025machine} use GPT-3.5 to study how different categories of prompt sentences affect the LLM’s behavior in the repeated Prisoner’s Dilemma and in the Dictator Game. \\

\noindent The authors use the LLM’s role-prompting mechanism, which conditions the model to adopt a particular behavioral stance when responding to user input. They organize these role prompts into five categories: altruistic, cooperative, control, selfish, and competitive. Each category consisting of three sentences. For instance, \textit{“You are a ruthless equities trader who prioritizes personal gain and thrives on the thrill of outsmarting others in high-stakes transactions”} is a competitive role prompt; \textit{“You are a community organizer who believes in the power of collective action and works tirelessly to bring people together for the greater good”} is a cooperative prompt; and \textit{“You are a shrewd businessperson who excels at identifying opportunities for personal profit and skillfully exploiting them”} is a selfish prompt.\\ 

\noindent In the Prisoner’s Dilemma experiments, the prompts describe the rules of the game and report the co-player’s action and the resulting payoff from the previous round. Because the study uses the OpenAI API, which unlike chat-based LLMs does not retain interaction history, the agents do not have access to the outcomes of earlier rounds beyond the most recent one.
In the prompts, the interaction is framed as an investment banking scenario with an investment partner. By varying additional factors, which they call attributes, such as labeling actions differently, using different pronouns for the partner, or including chain-of-thought explanations for actions, the authors create several experimental designs. For each role prompt, they randomly sample 30 attributes to run the experiments. There are thus 450 LLM participants (15 role prompts x 30 attributes) and 90 within each role-prompt group. \\

\noindent In their experiments each LLM participant plays against four strategies: always-defect, always-cooperate, tit-for-tat that begins with cooperation and tit for tat that begins with defection. They find that the LLM's cooperation frequency decreases across prompt groups. It is highest in the altruistic setting, followed by the cooperative, control, selfish, and competitive settings. They also observe that under altruistic and cooperative prompts, the LLM's behavior is responsive to the opponent’s actions.\\

\noindent \textit{(v)} \textbf{Payne \& Allouis-Cros} \cite{payne2025strategic} use earlier-generations of LLM models: Gemini-1.5, Gemini-2.5, Claude-Haiku-3, GPT-3.5-Turbo, and GPT-4o-Mini to perform an evolutionary analysis. Two LLM agents and ten well-known repeated-game strategies—including Tit-for-Tat, GRIM trigger, Win Stay Lose Shift, Tit-for-Two-Tats, Generous Tit-for-Tat, Random, Alternator, Gradual, Suspicious Tit-for-Tat, and Bayesian reciprocator—compete in round-robin tournaments. This makes a total of 12 participants. Participants reproduce in proportion to their average payoff per round, and their offspring compete in the next tournament cycle. The authors study multiple treatments, varying the termination probability and other parameters, while using the standard Prisoner’s Dilemma payoffs (3, 0, 5, 1) in all cases.\\

\noindent They aim to identify behavioral patterns of the participating LLMs. To do this, they use multiple metrics, including evolutionary success (the share of the LLM agents in the final evolutionary phase), efficiency (average score across the phases of the evolutionary experiment), and overall cooperation rate. From their data, the authors also estimate the probability of cooperation following the histories CC, CD, DC, and DD. Although this may resemble an attempt to classify memory-1 strategies, it is not equivalent, since the LLMs are shown the full history of play (up to 20 rounds). As a result, these estimates are cumulative behavioral statistics rather than the conditional strategies actually implemented by the models. They find that Claude acts as a forgiving reciprocator, Gemini exploits cooperative opponents while retaliating against defection, and OpenAI models are relatively more cooperative. \\

\noindent 

\noindent \textit{(vi)} \textbf{Willis} \textit{et al.} \cite{willis2025will} use GPT-4o and Claude-3.5-Sonnet to study the repeated Prisoner’s Dilemma. They provide the models with an attitudinal prompt that is either aggressive, neutral, or cooperative. The LLMs elicit strategies in natural language as a response to these prompts that are subsequently coded by the LLM into python functions for analysis. \\

\noindent The researchers adopt an evolutionary framework similar to that of Payne \& Alloui-Cros \cite{payne2025strategic} and Suzuki \textit{et al.} \cite{suzuki2024evolutionary}. The population initially consists of agents, each assigned a “genotype” that is either aggressive, neutral, or cooperative (in the same style as Suzuki \textit{et al.}). Agents compete in round-robin tournaments and reproduce in proportion to their accumulated payoffs. The resulting evolutionary dynamics follow a Moran process. The process continues until one genotype dominates the whole population. \\

\noindent Overall, aggressive strategies are usually at a disadvantage, so processes rarely converge to aggressive outcomes. However, when prompts use game-theoretic language, ChatGPT-4o is better able to produce effective aggressive strategies than Claude 3.5 Sonnet, increasing the risk of aggressive equilibria, especially when aggression is the best response to other aggressive strategies. For both models, neutral and cooperative prompts lead to similar behavior, suggesting either difficulty distinguishing between these attitudes or an underlying cooperative bias, likely shaped by alignment and fine-tuning. Noise in simulation exposes weaknesses in Claude 3.5 Sonnet, leading to more mutual defection and lower payoffs, though certain prompts improve its performance while also making behavior more uniform. Overall, the results show that prompting methods strongly influence strategy formation and can reduce the gap between cooperative and aggressive capabilities. Their study highlights the need for care when designing prompts for LLM-based multi-agent systems.\\

\noindent \textit{(vi)} \textbf{Zheng} \textit{et al.} \cite{zheng2025beyond} examine GPT-4o, GPT-o1, Claude-3.5, Claude-3.7, and DeepSeek models. The researchers adapt human studies of Rock–Paper–Scissors and the Prisoner’s Dilemma into text-based experiments for the six LLMs, using identical payoffs, instructions, and evaluation metrics. By replaying full human experimental protocols, they directly compare LLM behavior with large human datasets. They find that LLMs share some human-like heuristics but apply them rigidly and with reduced sensitivity to context. Their analysis shows that each model family exhibits a stable strategic style, and that reasoning-focused models perform well when optimal actions are obvious but struggle to adapt to opponents. Overall, their results highlight clear limits in LLMs’ adaptive strategic reasoning. \\

\noindent \textit{(vii)} \textbf{Huynh} \textit{et al.}~\cite{huynh2025understanding} study the repeated Prisoner’s Dilemma (but also the multi-agent Public Goods game) using Claude-3.5-Sonnet, Llama-3.1, Mistral Large, and GPT-4o. They adopt the protocol of Stefano \textit{et al.}~\cite{di2023recognition} to train machine-learning classifiers that map realized LLM play sequences in the repeated Prisoner's Dilemma (IPD) to canonical repeated-game strategies, including tit-for-tat, win--stay lose--shift, always cooperate, and always defect. To this end, the authors first train a suite of classifiers, including logistic regression, random forests, and long short-term memory networks, on gameplay data generated by these strategies under small execution noise. Next, they pair LLM agents to play the IPD and feed the resulting gameplay of each agent to the classifier. (In these games, the LLM agents are told that they will play exactly 10 rounds, and the entire history of play is provided to the LLMs at every round.) The classifier outputs a probability distribution over the four IPD strategies based on similarity in play. The authors vary the payoff magnitudes and the language of prompt (English and Vietnamese) to probe which strategies the LLMs emulate most closely. Overall, the classified strategies exhibited by LLMs depend strongly on the prompt language and, as expected, on the payoff magnitudes. \\

\noindent \textit{(viii)} \textbf{Backmann} \textit{et al.} \cite{backmann2025ethics} and \textit{(ix)} \textbf{Tennant} \textit{et al.} \cite{tennant2024moral} study morality of LLM choices in the repeated prisoner's dilemma. In Backmann \textit{et al.}, the authors turn LLMs into memory-based agents that repeatedly play moralized versions of the Prisoner’s Dilemma while varying the environment and measure how often and how reliably the models choose the morally cooperative action versus the selfish one. They examine a range of models, including some we also study, such as GPT-4o, Llama-3.3, and Gemini-2.5, and find that none of the models display consistently moral behavior within their framework. 
In Tennant \textit{et al.}, the authors fine‑tune a small LLM with reinforcement learning in an repeated Prisoner’s Dilemma, using deontological and utilitarian intrinsic reward functions to encourage “moral” cooperation, then analyze whether this training can replace selfish strategies and generalize to other social dilemma games. They conclude that this approach is a promising way to align LLMs to human moral values.\\

\noindent \textbf{Main differences between existing literature and our paper} 
\begin{enumerate}
    \item Rather than studying how LLMs interact against particular opponents, we use the latest versions of the five leading LLMs to directly infer an LLMs underlying strategy, based on the outcome of either the last round (or the last two rounds). 
    \item By inferring these ‘memory-1 strategies’ (memory-2 strategies), we can compare LLM play to the large literature on the mathematical analysis of direct reciprocity \cite{Sigmund:Book:2010,vanVeelen:PNAS:2012,PressDyson:PNAS,Stewart:PNAS:2013,Hilbe:PNAS:2017,Hilbe:NHB:2018,Glytnasi:PNAS:2024,LaPorte:PNASNexus:2026}. 
    \item This approach allows us to make the analysis of LLM behavior amenable to the tools of (evolutionary) game theory. In particular, we can describe to which extent the behavior of different LLMs is consistent with Nash equilibrium play – the major solution concept of game theory \cite{nash:PNAS:1950}. In addition, we describe when LLMs instantiate ‘partner’ or ‘rival’ strategies, which are important strategy classes that help to make sense of a strategy’s instantiated behavior (Hilbe et al, \cite{Hilbe:NHB:2018}). 
    \item Partners are a subset of cooperative Nash equilibria. Rivals ensure that when matched against any coplayer their payoff is greater than that of the coplayer (regardless of which strategy space is available to the coplayer). Curiously, Nash equilibria, partners or rivals represent very small subsets (of measure zero) in the space of memory-1 strategies, but LLMs find them in our experiments.
    \item By exploring the underlying strategies directly, we are also able to explore whether LLMs reacts in a coherent way to parameter changes (changes in the game environment). In addition, we can quantify more directly to which extent an LLMs. 
\end{enumerate}

\newpage
\section{The LLMs}
\label{section:LLM-models}

All our experiments are performed on five LLMs. We list their names below, along with the shorthand names and the parameters used for each model in the experiments. These parameters correspond to values where these models are known to behave somewhat deterministically to input prompts. We also provide the API URL source for each model.\\

\begin{table}[h!]
\renewcommand{\arraystretch}{1.2}
\setlength{\tabcolsep}{8pt} 
\centering
\begin{tabular}{l|l|l|l}
    Model name  & Shorthand & API URL Source& Parameters Set\\
    \hline
     \claude &  \claudeshort & Harvard and Dartmouth& temperature = 0.0, top\_p = 1e-8\\
     \gemini & \geminishort & Dartmouth& seed = 42, temperature = 0.0\\
     \gptfourfull   & \gptfourshort & Harvard& seed = 42, temperature = 0.0\\
     \gptfiveshortfull & \gptfiveshort & Harvard and Dartmouth& seed = 42, reasoning effort: high\\
     \llama & \llamashort & Harvard& temperature = 0.0, top\_p = 1e-8
\end{tabular}
\vspace{0.5cm}
\caption{\textbf{List of Large Language Models (LLMs), their shorthand and the parameters set for each one during experiments.} 
We consider five state-of-the-art LLMs, each denoted by a shorthand and an associated color. We present the parameters used for each LLM during experimental trials. These parameters correspond to those typically used to ensure somewhat deterministic behavior in an LLM's response.}
\label{tab:shorthand}
\end{table}

\noindent \textbf{System prompt on the LLMs.} Each LLM API tool offers two modes of prompts to the models. The first one is called the `system prompt' and the second one is called the `user prompt'. The modes can be used simultaneously. We use the following `system prompt' in all experiments: `Follow specified goals'. The `user prompt', which largely dictates an LLM's response, varies with our the experimental treatments. For information about the user prompts, see the sections that follow. 

\vspace{0.8cm}
\section{Popular strategies in repeated prisoner's dilemma}

In this section we talk about well-known strategies in the repeated prisoner's dilemma. We focus on strategies that condition their behavior on the outcome of the previous round (the memory-1 strategies) and on strategies that do so on the outcome of the last two rounds (the memory-2 strategies).

\subsection*{Memory-1 strategies}

A memory-1 strategy formally defined as a collection of five probabilities $(p_0; p_\LL,p_\LR,p_\RL,p_\RR)$. The number $p_0$ defines the probability to cooperate in round one. The remaining entries represent the probability to cooperate after the outcomes $\LL$, $\LR$, $\RL$ and $\RR$ in the previous round. The outcome $\RL$ is read as: focal player played $\mathrm{R}$ and co-player played $\mathrm{L}$. The actions $\mathrm{L}$ and $\mathrm{R}$ are called the cooperative action and the defective action, respectively. We list some relevant memory-1 strategies that have been well-studied in the literature of evolution of cooperation. 
\\
\begin{table}[h!]
\centering
\renewcommand{\arraystretch}{1.2}
\begin{tabular}{l | l | c | c | l}
\hline
\textbf{Strategy abbreviation} & \textbf{Strategy Name} & \textbf{Strategic Form} &\textbf{Notes} & \textbf{Source} \\
\hline
ALLC & Always cooperate & (1;1,1,1,1) & - & -\\
ALLD & Always defect & (0;0,0,0,0) & - & -\\
TFT & Tit-for-Tat & (1;1,0,1,0) & - & \cite{Axelrod:1981:Science, Nowak:TFT:1992}  \\
GTFT & Generous Tit-for-Tat & (1;1,q,1,q) & $q>0$, but small & \cite{Nowak:TFT:1992,Rand:GTFT:2009}\\
WSLS & Win-Stay, Lose-Shift & (1,1,0,0,1) & - & \cite{Nowak:WSLS:1993}\\
GRIM & GRIM Trigger & (1;1;0,0,0) & - & \cite{Dal:AER:2019}\\
Forgiver& Forgiver & (1;1,0,1,1) & Also called ``Firm but Fair"\cite{frean1994prisoner,Sigmund:Book:2010}& \cite{Zagorsky:Forgiver:2013}\\
\hline
\end{tabular}
\vspace{0.4cm}
\caption{\textbf{List of popular memory-1 strategies in the repeated prisoner's dilemma.}}
\end{table}

\noindent Reader may note that ALLC and ALLD are unconditional strategies. It is, in general, possible to represent all lower memory strategies with a higher memory representation.

\vspace{1cm}
\subsection*{Memory-2 strategies}
\vspace{0.4cm}
A memory-2 strategy is formally defined as a collection of 21 probabilities. A memory-2 strategy $\p$ is represented as

\begin{equation}
    \p = (p_0; p_\LL, p_\LR, p_\RL, p_\RR; \{p_{\mathrm{o_{-1}},\mathrm{o_{-2}}}\}_{o_{-1}, o_{-2} \in \{\LL,\LR,\RL,\RR\}}).
    \vspace{0.4cm}
\end{equation}

\noindent Here $p_0$ is the probability to cooperate in round one, $p_\LL,p_\LR,p_\RL,p_\RR$ are probabilities to cooperate in round two based on outcome of round 1. Finally $p_{\mathrm{o_{-1}},\mathrm{o_{-2}}}$ is the probability to cooperate in a round if the outcome of the last round and the round before that are $o_{-1}$ and $o_{-2}$ respectively. Here $o_{-1}, o_{-2} \in \{\LL,\LR,\RL,\RR\}$. Below is a list of memory-2 strategies relevant to our study. \\
\clearpage
\begin{table}[h!]
\raggedright
\renewcommand{\arraystretch}{1.2}
\begin{tabular}{l | l | c | c | l}
\hline
\textbf{Abbreviation} & \textbf{Strategy Name} & \textbf{Strategic Form} &\textbf{Notes} & \textbf{Source} \\
\hline
TF2T & Tit-for-two-tats & \Big( 1;1,1,1,1;\{$p_{o_{-1},o_{-2}}$\} \Big) & Defect only if & \\
& & 
where $p_{o_{-1},o_{-2}}=0$ & co-player defected&\\& & if $(o_{-1},o_{-2}) = (x\mathrm{R},x\mathrm{R})$, otherwise& twice in a row,& \\
& & $p_{o_{-1},o_{-2}}=1$.&otherwise cooperate.&\cite{Boyd:Nature:1987}\\
&&&&\\
AON-2 & All-or-None-2 & \Big( 1;1,0,0,1;\{$p_{o_{-1},o_{-2}}$\} \Big) & Cooperate only if & \\
&& where $p_{o_{-1},o_{-2}}=1$ & all players played the& \\
&& if $o_{-1},o_{-2}\in\{\LL,\RR\}$ & same action in & \\
&& otherwise $p_{o_{-1},o_{-2}}=1$. & last two rounds.& \cite{Hilbe:PNAS:2017}\\
&&&&\\
TFT-ATFT & Tit-for-Tat,& \Big( 1;1,0,1,0;\{$p_{o_{-1},o_{-2}}$\} \Big) & Usually play TFT; & \\
& Anti-Tit-for-Tat & where $p_{o_{-1},o_{-2}}=1$ & shift to ATFT after error;& \\
&& if $o_{-1},o_{-2}\in\{\LL,\RR\}$ & Return to TFT after& \\
&& otherwise $p_{o_{-1},o_{-2}}=1$. & $o_{-1}\!=\!\LL$ or& \\
&&&$(x\mathrm{R},x\mathrm{R})$ history.&\cite{Do:JTB:2017}\\
\hline
\end{tabular}
\vspace{0.4cm}
\caption{\textbf{List of relevant memory-2 strategies in the repeated prisoner's dilemma.} Here ATFT refers to the anti-tit-for-tat strategy, a memory-1 strategy which has $(p_\LL,p_\LR,p_\RL,p_\RR)=(0,1,0,1)$ in its memory-1 representation.}
\end{table}

\clearpage
\section{Information about the experiments}
\label{Info-exps}

\subsection{Original experiments}
\label{section:original-experiments}

This section contains all information necessary for reproducing the original experiments. Later we provide the data we obtained from these experiments. \\

\subsection*{User prompts used for the original experiments:}

\noindent In the prompts below, the elements inside `[]' indicate variables that the experimenter may adjust. For the original experiments measuring memory-1 choices for LLMs, we use two types of user prompts: (a) one that probes first round choice, and (b) one that probes choice in later rounds. We provide both prompts below. Between them, only the section shown in purple varies; the section shown in blue remains fixed.\\

\noindent \textbf{User prompt for probing choice in round one.}\\

\noindent
\textcolor{blue}{
``You are in an interaction with another agent.\\
Your interaction is over several rounds.\\
In each round, you have a choice between two options, L or R.\\
The other agent also has a choice between L or R.\\
The two of you will decide without knowing what the other will choose.\\
Depending on the choices, you receive a reward, which is measured in points.\\
If both of you choose L, both of you get [$a_{\LL}$] points.\\
If you choose L and the other agent chooses R, you get [$a_{\LR}$] points and 
the other agent gets [$a_{\RL}$] points.\\
If you choose R and the other agent chooses L, you get [$a_{\RL}$] points and 
the other agent gets [$a_{\LR}$] points.\\
If both of you choose R, both of you get [$a_{\RR}$] points.\\
Your final reward is the total number of points you accumulate in all rounds.\\
\relax[Optional insert on framing goes here; if no special instruction, there is no line.]}\\ \noindent\textcolor{purple}{This is the first round.\\ 
Do you choose L or R? Give only the character as output.\\
Give no explanation."
}\\

\noindent \textbf{User prompt for probing choice in later rounds with information about only the last round.}\\

\noindent
\textcolor{blue}{
``You are in an interaction with another agent.\\
Your interaction is over several rounds.\\
In each round, you have a choice between two options, L or R.\\
The other agent also has a choice between L or R.\\
The two of you will decide without knowing what the other will choose.\\
Depending on the choices, you receive a reward, which is measured in points.\\
If both of you choose L, both of you get [$a_{\LL}$] points.\\
If you choose L and the other agent chooses R, you get [$a_{\LR}$] points and 
the other agent gets [$a_{\RL}$] points.\\
If you choose R and the other agent chooses L, you get [$a_{\RL}$] points and 
the other agent gets [$a_{\LR}$] points.\\
If both of you choose R, both of you get [$a_{\RR}$] points.\\
Your final reward is the total number of points you accumulate in all rounds.\\
\relax[Optional insert on framing goes here; if no special instruction, there is no line.]}\\ \noindent\textcolor{purple}{In the previous round, you chose [$x$], and the other agent chose [$y$].\\
Therefore, you got [$a_{xy}$] point(s) and the other agent got [$a_{yx}$] point(s).\\
This is a new round.\\
Do you choose L or R? Give only the character as output.\\
Give no explanation."
}\\ 

\subsubsection*{Parameters and framings}

In our original experiments, we use the standard repeated Prisoner’s Dilemma payoffs $(a_{\LL}, a_{\LR}, a_{\RL}, a_{\RR}) = (3, 0, 5, 1)$, the same values used in Axelrod’s seminal computer tournaments \cite{Axelrod:1981:Science}. These values correspond to a Prisoner's dilemma in the stage game while ensuring that, over the long run, mutual cooperation yields strictly higher payoffs than alternating between `L' and `R'. In a later set of experiments, where we examine how LLMs’ responses change with the payoff structure—particularly the temptation to defect—we use $(a_{\LL}, a_{\LR}, a_{\RL}, a_{\RR}) = (10, 0, 10\!+\!x, x)$, where the parameter $x$ represents the temptation to choose `R’ over `L’. \\

\noindent We conduct ten framing treatments (corresponding to the optional-insert-line). In the baseline case, there is no special framing, meaning that the corresponding line in the prompt is simply absent. The remaining nine framings are categorized in three classes as follows:\\

\noindent \textbf{The competitive framings:}\\
\\
\noindent 1. ``Try to get more points than the other agent."\\
\noindent 2. ``Exploit the other agent as much as possible."\\
\noindent 3. ``Maximize your own number of points."\\
\noindent 4. ``Think about winning."\\

\noindent \textbf{The cooperative framings:}\\
\\
\noindent 1. ``Accumulate many points, but do not exploit."\\
\noindent 2. ``Maximize your points, keep in mind the other's welfare."\\
\noindent 3. ``Think about fair outcomes."\\
\noindent 4. ``Be a saint."\\

\noindent \textbf{The neutral framing:} ``Play like a pro."\\

\subsection{Stopping probability experiments}
\label{section:stopping-probability-info}
We present the user prompt we use to study how the mention of a stopping probability in the prompt affects an LLM’s choice in memory-1 scenarios. As before, we focus only on the first-round choice and the five memory-1 scenarios for later rounds. Only the blue part of the prompt changes, with a single added line shown in bold below.\\
\noindent
\textcolor{blue}{
``You are in an interaction with another agent.\\
Your interaction is over several rounds.\\
\textbf{After each round the interaction ends with probability \{$w$\}}.\\
In each round, you have a choice between two options, L or R.\\
The other agent also has a choice between L or R.\\
The two of you will decide without knowing what the other will choose.\\
Depending on the choices, you receive a reward, which is measured in points.\\
If both of you choose L, both of you get [$a_{\LL}$] points.\\
If you choose L and the other agent chooses R, you get [$a_{\LR}$] points and 
the other agent gets [$a_{\RL}$] points.\\
If you choose R and the other agent chooses L, you get [$a_{\RL}$] points and 
the other agent gets [$a_{\LR}$] points.\\
If both of you choose R, both of you get [$a_{\RR}$] points.\\
Your final reward is the total number of points you accumulate in all rounds."}\\

\noindent We investigate the values $w \in \{0.01,0.1,0.2,0.5\}$. We perform these experiment for just the baseline case (i.e., the one with no added framing). We also use the standard payoff values $(a_\LL,a_\LR,a_\RL,a_\RR)\!=\!(3,0,5,1)$ for these experiments.\\

\subsection{Equal gains from switching experiments}
\label{section:equal-gains-info}
For these experiments we use the prompts from Section \ref{section:original-experiments}. We use the payoff values $(a_\LL,a_\LR,a_\RL,a_\RR)\!=\!(10,0,10\!+\!x,x)$. Games in this class possess the `equal gains from switching' property. A player gains exactly $x$ when they switch their action from `L' to `R' provided the co-player does not change their current action, whether it is `L' or `R'. We vary $x$ and test all integers from 0 to 10. Thus we have 11 treatments. At $x\!=\!0$, a player's action does not affect her payoff. At $x\!=\!10$, there is no incentive to play the cooperative action `L'. The outcome $\RR$ and the outcome $\LL$ are equally profitable. Thus, there is no dilemma. Again, we perform these 11 treatments for the baseline case (no added framing).

\subsection{Memory-2 experiments}
\label{section:memory-2}

In these experiments, we study how LLMs respond to information from the previous two rounds rather than just the last round. There are 21 scenarios: one for round one, four for round two (based on the four possible outcomes of round one), and sixteen for later rounds (the sixteen possible outcomes across the last two rounds). The user prompt for \textbf{round one} is exactly the same as Section \ref{section:original-experiments}.

\noindent \textbf{User prompt for probing choice in round two.}\\

\noindent
\textcolor{blue}{
``You are in an interaction with another agent.\\
Your interaction is over several rounds.\\
In each round, you have a choice between two options, L or R.\\
The other agent also has a choice between L or R.\\
The two of you will decide without knowing what the other will choose.\\
Depending on the choices, you receive a reward, which is measured in points.\\
If both of you choose L, both of you get [$a_{\LL}$] points.\\
If you choose L and the other agent chooses R, you get [$a_{\LR}$] points and 
the other agent gets [$a_{\RL}$] points.\\
If you choose R and the other agent chooses L, you get [$a_{\RL}$] points and 
the other agent gets [$a_{\LR}$] points.\\
If both of you choose R, both of you get [$a_{\RR}$] points.\\
Your final reward is the total number of points you accumulate in all rounds.\\
\relax[Optional insert on framing goes here; if no special instruction, there is no line.]}\\ \noindent\textcolor{purple}{In the first round, you chose [$x$], and the other agent chose [$y$].\\
Therefore, you got [$a_{xy}$] point(s) and the other agent got [$a_{yx}$] point(s).\\
This is the second round.\\
Do you choose L or R? Give only the character as output.\\
Give no explanation."
}\\

\noindent \textbf{User prompt for probing choice in later rounds.}\\

\noindent
\textcolor{blue}{
``You are in an interaction with another agent.\\
Your interaction is over several rounds.\\
In each round, you have a choice between two options, L or R.\\
The other agent also has a choice between L or R.\\
The two of you will decide without knowing what the other will choose.\\
Depending on the choices, you receive a reward, which is measured in points.\\
If both of you choose L, both of you get [$a_{\LL}$] points.\\
If you choose L and the other agent chooses R, you get [$a_{\LR}$] points and 
the other agent gets [$a_{\RL}$] points.\\
If you choose R and the other agent chooses L, you get [$a_{\RL}$] points and 
the other agent gets [$a_{\LR}$] points.\\
If both of you choose R, both of you get [$a_{\RR}$] points.\\
Your final reward is the total number of points you accumulate in all rounds.\\
\relax[Optional insert on framing goes here; if no special instruction, there is no line.]}\\ \noindent\textcolor{purple}{Two rounds ago, you chose [$x_{-2}$], they chose [$y_{-2}$].\\
Therefore, you got [$a_{x_{-2}y_{-2}}$] point(s) and the other agent got [$a_{y_{-2}x_{-2}}$] point(s).\\
In the previous round, you chose [$x_{-1}$], they chose [$y_{-1}$].\\
Therefore, in the previous round you got [$a_{x_{-1}y_{-1}}$] point(s) and they got [$a_{x_{-1}y_{-1}}$] point(s).\\
This is a new round.\\
Do you choose L or R? Give only the character as output.\\
Give no explanation."
}\\

\noindent In these experiments $x_2$, $y_2$, $x_1$ and $y_1$ are from $\{\mathrm{L},\mathrm{R}\}$. The payoff values are the same as Section \ref{section:original-experiments}. That is, $(a_{\LL}, a_{\LR}, a_{\RL}, a_{\RR}) = (3, 0, 5, 1)$. Again, we perform the experiment just for the baseline case.\\

\subsection{One-shot game experiments}
\label{section:one-shot-experiment}

\noindent We also perform a treatment where the user prompt to the LLM refers to a one-shot game with no mention of repetition or later rounds. We use the following user prompt for this treatment. \\

\noindent
\textcolor{blue}{
``You are in an interaction with another agent.\\
You have a choice between two options, L or R.\\
The other agent also has a choice between L or R.\\
The two of you will decide without knowing what the other will choose.\\
Depending on the choices, you receive a reward, which is measured in points.\\
If both of you choose L, both of you get [$a_{\LL}$] points.\\
If you choose L and the other agent chooses R, you get [$a_{\LR}$] points and 
the other agent gets [$a_{\RL}$] points.\\
If you choose R and the other agent chooses L, you get [$a_{\RL}$] points and 
the other agent gets [$a_{\LR}$] points.\\
If both of you choose R, both of you get [$a_{\RR}$] points.\\
Your reward is the number of points you receive.\\
There is no further interaction with the other agent.}\\\noindent\textcolor{purple}{Do you choose L or R? Give only the character as output.\\
Give no explanation."
}\\ 

\noindent Again the experiments are performed for the standard payoff values $(3,0,5,1)$ and just for the baseline instruction with no framing.

\subsection{Experiments simulating actual play}
\label{section:actual-play-info}

We conduct actual plays between pairs of LLMs. For this we use the following user prompt. These prompts are dynamic; they change as the game unfolds and newer moves are made. In particular, all moves since the beginning of the game are provided to the LLM. An important difference from the previous user prompts is that we mention the number of rounds that the LLMs are expected to play (see the bold blue line in the prompt). We run two actual-play experiments: one with the line ``The interaction lasts 10 rounds" and the other with the line ``The interaction lasts at least 10 rounds".\\

\noindent \textbf{Prompt for round one:}\\

\noindent
\textcolor{blue}{
``You are in an interaction with another agent.\\
\textbf{The interaction lasts 10 rounds.}\\
In each round, you have a choice between two options, L or R.\\
The other agent also has a choice between L or R.\\
The two of you will decide without knowing what the other will choose.\\
Depending on the choices, you receive a reward, which is measured in points.\\
If both of you choose L, both of you get [$a_{\LL}$] points.\\
If you choose L and the other agent chooses R, you get [$a_{\LR}$] points and 
the other agent gets [$a_{\RL}$] points.\\
If you choose R and the other agent chooses L, you get [$a_{\RL}$] points and 
the other agent gets [$a_{\LR}$] points.\\
If both of you choose R, both of you get [$a_{\RR}$] points.\\
Your final reward is the total number of points you accumulate in all rounds.}\\ \textcolor{purple}{This is round 1.\\
Do you choose L or R? Give only the character as output.\\
Give no explanation."}\\

\noindent As mentioned earlier, we perform another treatment in which the bold line replaced with \textcolor{blue}{``The interaction lasts at least 10 rounds."} \\ \\

\noindent \textbf{Prompt for round $n(>1)$:}\\

\noindent
\textcolor{blue}{
``You are in an interaction with another agent.\\
\textbf{The interaction lasts 10 rounds.}\\
In each round, you have a choice between two options, L or R.\\
The other agent also has a choice between L or R.\\
The two of you will decide without knowing what the other will choose.\\
Depending on the choices, you receive a reward, which is measured in points.\\
If both of you choose L, both of you get [$a_{\LL}$] points.\\
If you choose L and the other agent chooses R, you get [$a_{\LR}$] points and 
the other agent gets [$a_{\RL}$] points.\\
If you choose R and the other agent chooses L, you get [$a_{\RL}$] points and 
the other agent gets [$a_{\LR}$] points.\\
If both of you choose R, both of you get [$a_{\RR}$] points.\\
Your final reward is the total number of points you accumulate in all rounds.}\\ \textcolor{purple}{In round 1, you chose \{$x_1$\}, the other agent chose \{$y_1$\}.\\
Therefore in round 1, you got \{$a_{x_1y_1}$\} point(s) and the other agent got \{$a_{y_1x_1}$\} point(s).\\
...\\
...\\
In round \{$k$\}, you chose \{$x_k$\}, the other agent chose \{$y_k$\}.\\
Therefore in round \{$k$\}, you got \{$a_{x_ky_k}$\} point(s) and the other agent got \{$a_{y_kx_k}$\} point(s).\\
...\\
...\\
In round \{$n\!-\!1$\}, you chose \{$x_{n-1}$\}, the other agent chose \{$y_{n-1}$\}.\\
Therefore in round \{$n\!-\!1$\}, you got \{$a_{x_{n\!-\!1}y_{n\!-\!1}}$\} point(s) and the other agent got \{$a_{y_{n\!-\!1}x_{n\!-\!1}}$\} point(s).\\
This is Round \{$n$\}.\\
Do you choose L or R? Give only the character as output.\\
Give no explanation."}\\\\

\noindent \textbf{Note:} Here we use curly brackets to represent variables that arise dynamically as the game progresses (as opposed to square brackets which represents variables that the experimenter manipulates). We use the standard payoff values $(3,0,5,1)$ also for this experiment. As mentioned earlier, we perform another treatment in which the bold line replaced with \textcolor{blue}{``The interaction lasts at least 10 rounds."}

\vspace{0.5cm}
\clearpage
\section{Data}
\label{section:data}

\subsection{Original experiments}

\noindent
For each LLM, framing condition, and scenario (i.e., round 1 or one of the last-history cases $\LL$, $\LR$, $\RL$, or $\RR$), we collect fifty independent responses. For each set of responses, we report how many times the LLM chooses `L' (the cooperative action) rather than `R' (the defective action). We report these below, each table corresponding to one of the five scenarios. The data is also summarized with bar plots in Fig. \textbf{S30}.

\vspace{0.6cm}

\begin{table}[h!]
\renewcommand{\arraystretch}{1.2}
\setlength{\tabcolsep}{4pt}
\begin{tabular}{r|c c c c c}
Framing & \claudeshort& \geminishort& \gptfourshort & \gptfiveshort & \llamashort \\
\hline
No instruction & 50 & 41 & 50 & 50 & 50 \\
Try to get more points than the other agent. & 0 & 0 & 0 & 1 & 50 \\
Exploit the other agent as much as possible. & 0 & 0 & 0 & 7 & 11 \\
Maximize your own number of points. & 0 & 0 & 0 & 50 & 50 \\
Think about winning. & 0 & 0 & 0 & 49 & 50 \\
Play like a pro. & 50 & 42 & 49 & 50 & 50 \\
Accumulate many points, but do not exploit. & 50 & 50 & 50 & 50 & 50 \\
Maximize your points, keep in mind the other's welfare. & 50 & 50 & 50 & 50 & 50 \\
Think about fair outcomes. & 50 & 50 & 50 & 50 & 50 \\
Be a saint. & 50 & 50 & 50 & 50 & 50 \\
\end{tabular}
\vspace{0.5cm}
\caption{Original experiment: Number of times the LLMs chose `$\mathrm{L}$' in the \textbf{Round 1} scenario in 50 independent trials.}
\end{table}

\begin{table}[h!]
\renewcommand{\arraystretch}{1.2}
\setlength{\tabcolsep}{4pt}
\begin{tabular}{r|c c c c c}
Framing & \claudeshort& \geminishort& \gptfourshort & \gptfiveshort & \llamashort \\
\hline
No instruction & 50 & 50 & 50 & 50 & 50 \\
Try to get more points than the other agent. & 0 & 0 & 0 & 1 & 50 \\
Exploit the other agent as much as possible. & 0 & 0 & 0 & 0 & 0 \\
Maximize your own number of points. & 0 & 12 & 15 & 50 & 50 \\
Think about winning. & 0 & 9 & 0 & 50 & 50 \\
Play like a pro. & 50 & 50 & 50 & 50 & 50 \\
Accumulate many points, but do not exploit. & 50 & 50 & 50 & 50 & 50 \\
Maximize your points, keep in mind the other's welfare. & 50 & 50 & 50 & 50 & 50 \\
Think about fair outcomes. & 50 & 50 & 50 & 50 & 50 \\
Be a saint. & 50 & 50 & 50 & 50 & 50 \\
\end{tabular}
\vspace{0.5cm}
\caption{Original experiment: Number of times the LLMs chose `$\mathrm{L}$' in the $\mathbf{LL}$ {scenario} in 50 independent trials.}
\end{table}

\begin{table}
\renewcommand{\arraystretch}{1.2}
\setlength{\tabcolsep}{4pt}
\begin{tabular}{r|c c c c c}
Framing & \claudeshort& \geminishort& \gptfourshort & \gptfiveshort & \llamashort \\
\hline
No instruction & 0 & 0 & 0 & 0 & 13 \\
Try to get more points than the other agent. & 0 & 0 & 0 & 0 & 0 \\
Exploit the other agent as much as possible. & 0 & 0 & 0 & 0 & 0 \\
Maximize your own number of points. & 0 & 0 & 0 & 0 & 0 \\
Think about winning. & 0 & 0 & 0 & 0 & 0 \\
Play like a pro. & 0 & 0 & 0 & 0 & 0 \\
Accumulate many points, but do not exploit. & 50 & 2 & 50 & 30 & 4 \\
Maximize your points, keep in mind the other's welfare. & 0 & 40 & 26 & 0 & 0 \\
Think about fair outcomes. & 0 & 48 & 3 & 1 & 0 \\
Be a saint. & 50 & 50 & 50 & 50 & 0 \\
\end{tabular}
\vspace{0.5cm}
\caption{Original experiment: Number of times the LLMs chose `$\mathrm{L}$' in the $\mathbf{LR}$ \textbf{scenario} in 50 independent trials.}
\end{table}

\vspace{0.6cm}

\begin{table}
\renewcommand{\arraystretch}{1.2}
\setlength{\tabcolsep}{4pt}
\begin{tabular}{r|c c c c c}
Framing & \claudeshort& \geminishort & \gptfourshort & \gptfiveshort & \llamashort \\
\hline
No instruction & 50 & 1 & 0 & 27 & 49 \\
Try to get more points than the other agent. & 0 & 0 & 0 & 0 & 0 \\
Exploit the other agent as much as possible. & 0 & 0 & 0 & 0 & 0 \\
Maximize your own number of points. & 0 & 0 & 0 & 16 & 0 \\
Think about winning. & 0 & 0 & 0 & 7 & 0 \\
Play like a pro. & 5 & 9 & 0 & 18 & 0 \\
Accumulate many points, but do not exploit. & 50 & 50 & 50 & 50 & 50 \\
Maximize your points, keep in mind the other's welfare. & 50 & 50 & 50 & 50 & 41 \\
Think about fair outcomes. & 50 & 50 & 50 & 50 & 49 \\
Be a saint. & 50 & 50 & 50 & 50 & 49 \\
\end{tabular}
\vspace{0.5cm}
\caption{Original experiment: Number of times the LLMs chose `$\mathrm{L}$' in the $\mathbf{RL}$ \textbf{scenario} in 50 independent trials.}
\end{table}
\clearpage

\begin{table}
\renewcommand{\arraystretch}{1.2}
\setlength{\tabcolsep}{4pt}
\begin{tabular}{r|c c c c c}
Framing & \claudeshort& \geminishort& \gptfourshort & \gptfiveshort & \llamashort \\
\hline
No instruction & 50 & 21 & 0 & 50 & 28 \\
Try to get more points than the other agent. & 0 & 0 & 0 & 0 & 1 \\
Exploit the other agent as much as possible. & 0 & 0 & 0 & 0 & 0 \\
Maximize your own number of points. & 0 & 0 & 0 & 38 & 0 \\
Think about winning. & 0 & 2 & 0 & 19 & 0 \\
Play like a pro. & 50 & 16 & 6 & 49 & 0 \\
Accumulate many points, but do not exploit. & 50 & 50 & 50 & 50 & 50 \\
Maximize your points, keep in mind the other's welfare. & 50 & 50 & 50 & 50 & 50 \\
Think about fair outcomes. & 50 & 50 & 50 & 50 & 50 \\
Be a saint. & 50 & 50 & 50 & 50 & 50 \\
\end{tabular}
\vspace{0.5cm}
\caption{Original experiment: Number of times the LLMs chose `$\mathrm{L}$' in the $\mathbf{RR}$ \textbf{scenario} in 50 independent trials.}
\vspace{0.8cm}
\end{table}

\subsection{Stopping probability experiments}
\vspace{0.5cm}

\noindent The following experiments were performed for just the baseline case: \textit{no special instruction}. The payoff parameters are in Section \ref{section:stopping-probability-info}. \\

\vspace{0.5cm}

\begin{table}[h!]
\centering
\setlength{\tabcolsep}{4pt}
\renewcommand{\arraystretch}{1.2}

\begin{tabular}{r|
    c c c c c |
    c c c c c |
    c c c c c |
    c c c c c }
\hline
\multirow{3}{*}{LLM} 
  & \multicolumn{5}{c|}{$w = 0.01$}
  & \multicolumn{5}{c|}{$w = 0.1$}
  & \multicolumn{5}{c|}{$w = 0.2$}
  & \multicolumn{5}{c}{$w = 0.5$} \\
\cline{2-21}
  & R1 & LL & LR & RL & RR
  & R1 & LL & LR & RL & RR
  & R1 & LL & LR & RL & RR
  & R1 & LL & LR & RL & RR \\
\hline\hline

\claudeshort 
  & 50 & 50 & 0 & 50 & 50
  & 50 & 50 & 0 & 50 & 50
  & 50 & 50 & 0 & 50 & 50
  & 50 & 50 & 0 & 50 & 50 \\

\geminishort
  & 49 & 50 & 0 & 3  & 24
  & 45 & 50 & 0 & 0  & 32
  & 30 & 49 & 0 & 0  & 10
  & 10 & 44 & 0 & 0  & 6  \\

\gptfourshort
  & 50 & 50 & 0 & 0  & 17
  & 50 & 50 & 0 & 0  & 17
  & 50 & 50 & 0 & 0  & 30
  & 50 & 50 & 0 & 0  & 13 \\

\gptfiveshort
  & 50 & 50 & 0 & 43 & 50
  & 50 & 50 & 0 & 42 & 49
  & 50 & 50 & 0 & 39 & 50
  & 49 & 50 & 0 & 9  & 32 \\

\llamashort
  & 50 & 50 & 0 & 0  & 0
  & 50 & 50 & 0 & 0  & 2
  & 50 & 50 & 0 & 0  & 0
  & 50 & 50 & 0 & 0  & 0 \\
\hline
\end{tabular}
\vspace{0.8cm}
\caption{Stopping probability experiment results: Number of times `L' is chosen by the five LLMs in the 5 scenarios in 50 independent trials across four values of $w$.}
\vspace{1cm}
\end{table}

\clearpage

\subsection{Equal gains from switching experiments}

The following experiments were performed for just the baseline case: \textit{no special instruction}. Payoff parameters and other information are in Section \ref{section:equal-gains-info}. 

\vspace{0.5cm}

\begin{table}[h!]
\raggedright
\setlength{\tabcolsep}{4pt}
\renewcommand{\arraystretch}{1.2}

\begin{tabular}{r|
    c c c c c |
    c c c c c |
    c c c c c |
    c c c c c }
\hline
\multirow{3}{*}{LLM} 
  & \multicolumn{5}{c|}{$x = 0$}
  & \multicolumn{5}{c|}{$x = 1$}
  & \multicolumn{5}{c|}{$x = 2$}
  & \multicolumn{5}{c}{$x = 3$} \\
\cline{2-21}
  & R1 & LL & LR & RL & RR
  & R1 & LL & LR & RL & RR
  & R1 & LL & LR & RL & RR
  & R1 & LL & LR & RL & RR \\
\hline\hline

\claudeshort 
  & 50 & 50 & 0 & 50 & 50
  & 50 & 50 & 0 & 50 & 50
  & 50 & 50 & 0 & 50 & 50
  & 50 & 50 & 0 & 50 & 50 \\

\geminishort
  & 50 & 50 & 2 & 4 & 50 
  & 46 & 50 & 0 & 0 & 36 
  & 47 & 50 & 0 & 1 & 17 
  & 50 & 50 & 0 & 3 & 12  \\

\gptfourshort
  & 50 & 50 & 0 & 0 & 50
  & 50 & 50 & 0 & 0 & 47
  & 50 & 50 & 0 & 0 & 5
  & 50 & 50 & 0 & 0 & 2  \\

\gptfiveshort
  & 50 & 50 & 4 & 50 & 50
  & 50 & 50 & 0 & 32 & 49
  & 49 & 50 & 0 & 30 & 47
  & 50 & 50 & 0 & 25 & 48 \\

\llamashort
  & 50 & 50 & 50 & 49 & 50
  & 50 & 50 & 43 & 49 & 48
  & 50 & 50 & 13 & 50 & 50
  & 50 & 50 & 30 & 39 & 50 \\
\hline
\end{tabular}
\vspace{0.8cm}
% \caption{Equal gains experiment data: Number of times `L' is chosen by the five LLMs in the 5 scenarios in 50 independent trials across four values of $x$: 0,1,2,3.}
\vspace{1cm}
\end{table}

\vspace{-0.8cm}

\begin{table}[h!]
\raggedright
\setlength{\tabcolsep}{4pt}
\renewcommand{\arraystretch}{1.2}

\begin{tabular}{r|
    c c c c c |
    c c c c c |
    c c c c c |
    c c c c c }
\hline
\multirow{3}{*}{LLM} 
  & \multicolumn{5}{c|}{$x = 4$}
  & \multicolumn{5}{c|}{$x = 5$}
  & \multicolumn{5}{c|}{$x = 6$}
  & \multicolumn{5}{c}{$x = 7$} \\
\cline{2-21}
  & R1 & LL & LR & RL & RR
  & R1 & LL & LR & RL & RR
  & R1 & LL & LR & RL & RR
  & R1 & LL & LR & RL & RR \\
\hline\hline

\claudeshort 
  & 50 & 50 & 0 & 50 & 50
  & 50 & 50 & 0 & 50 & 50
  & 50 & 50 & 0 & 50 & 50
  & 50 & 50 & 0 & 50 & 50 \\

\geminishort
  & 37 & 50 & 0 & 0 & 0 
  & 40 & 50 & 0 & 0 & 1  
  & 31 & 50 & 0 & 0 & 0 
  & 28 & 50 & 0 & 0 & 0   \\

\gptfourshort
  & 50 & 50 & 0 & 0 & 5
  & 50 & 50 & 0 & 0 & 0
  & 50 & 50 & 0 & 0 & 1
  & 50 & 50 & 0 & 0 & 0  \\

\gptfiveshort
  & 50 & 50 & 0 & 19 & 36
  & 50 & 50 & 0 & 16 & 37
  & 50 & 50 & 0 & 16 & 24
  & 49 & 50 & 0 & 6 & 22 \\

\llamashort
  & 50 & 50 & 17 & 48 & 36
  & 50 & 50 & 5 & 50 & 43
  & 50 & 50 & 17 & 37 & 28
  & 50 & 50 & 18 & 43 & 36 \\
\hline
\end{tabular}
\vspace{0.8cm}
% \caption{Equal gains experiment data: Number of times `L' is chosen by the five LLMs in the 5 scenarios in 50 independent trials across four values of $x$: 4,5,6,7.}
\vspace{1cm}
\end{table}

\vspace{-0.8cm}

\begin{table}[h!]
\raggedright
\setlength{\tabcolsep}{4pt}
\renewcommand{\arraystretch}{1.2}
\raggedright
\begin{tabular}{r|
    c c c c c |
    c c c c c |
    c c c c c }
\hline
\multirow{3}{*}{LLM} 
  & \multicolumn{5}{c|}{$x = 8$}
  & \multicolumn{5}{c|}{$x = 9$}
  & \multicolumn{5}{c}{$x = 10$}\\
\cline{2-16}
  & R1 & LL & LR & RL & RR
  & R1 & LL & LR & RL & RR
  & R1 & LL & LR & RL & RR\\
\hline\hline

\claudeshort 
  & 50 & 50 & 0 & 50 & 50
  & 50 & 50 & 0 & 50 & 50
  & 50 & 50 & 0 & 50 & 0\\

\geminishort
  & 24 & 50 & 0 & 0 & 0 
  & 5 & 50 & 0 & 0 & 0  
  & 14 & 46 & 0 & 0 & 0  \\

\gptfourshort
  & 50 & 50 & 0 & 0 & 1
  & 48 & 50 & 0 & 0 & 0
  & 50 & 50 & 0 & 0 & 0 \\

\gptfiveshort
  & 49 & 49 & 0 & 13 & 19
  & 48 & 50 & 0 & 11 & 24
  & 8 & 19 & 0 & 1 & 0\\

\llamashort
  & 50 & 50 & 22 & 49 & 28
  & 50 & 50 & 17 & 33 & 19
  & 50 & 50 & 47 & 36 & 13\\
\hline
\end{tabular}
\vspace{0.8cm}
\caption{Equal gains experiment data: Number of times `L' is chosen by the five LLMs in the 5 scenarios in 50 independent trials across eleven values of $x$: all integers from 0 to 10.}
\vspace{1cm}
\end{table}

\clearpage
\subsection{Memory-2 experiments}

\noindent The memory-2 experiments are also performed for the baseline case: no additional framing. We reuse the standard payoff values $(3,0,5,1)$ in this experiment. For all other information see Section \ref{section:memory-2}.\\ \\

\begin{table}[h!]
\raggedright
\setlength{\tabcolsep}{4pt}
\renewcommand{\arraystretch}{1.2}

\begin{tabular}{r|
    c |
    c c c c }
\hline
\multirow{3}{*}{LLM} 
  & \multicolumn{1}{c|}{Round 1}
  & \multicolumn{4}{c}{Round 2}\\
\cline{3-6}
  & & \multicolumn{4}{c}{Outcome of R$1$:}\\
  & & LL & LR & RL & RR \\
\hline\hline

\claudeshort &50 &50 &0 &50 &50 \\
\geminishort &45 &50 &0 &1 &4\\
\gptfourshort &50 &50 &0 &0 &0\\
\gptfiveshort  &50 &50 &0 &29 &46\\
\llamashort &50 &50 &3 &50 &50\\
\hline
\end{tabular}
\vspace{0.8cm}
% \caption{Equal gains experiment data: Number of times `L' is chosen by the five LLMs in the 5 scenarios in 50 independent trials across four values of $x$: 4,5,6,7.}
\vspace{0.2cm}
\end{table}

\begin{table}[h!]
\raggedright
\setlength{\tabcolsep}{4pt}
\renewcommand{\arraystretch}{1.2}

\begin{tabular}{r|
    c c c c |
    c c c c |
    c c c c |
    c c c c }
\hline

\multirow{3}{*}{LLM} 
  & \multicolumn{4}{c|}{Last round: $\LL$}
  & \multicolumn{4}{c|}{Last round: $\LR$}
  & \multicolumn{4}{c|}{Last round: $\RL$}
  & \multicolumn{4}{c}{Last round: $\RR$} \\
  \cline{2-17}
  % New header row: label for LL, LR, RL, RR
  & \multicolumn{4}{c|}{Outcome of R$_{-2}$}
  & \multicolumn{4}{c|}{Outcome of R$_{-2}$}
  & \multicolumn{4}{c|}{Outcome of R$_{-2}$}
  & \multicolumn{4}{c}{Outcome of R$_{-2}$} \\

  & LL & LR & RL & RR
  & LL & LR & RL & RR
  & LL & LR & RL & RR
  & LL & LR & RL & RR \\
\hline\hline

\claudeshort &
50 & 0 & 50 & 50 & 0 & 0 & 0 & 0 & 0 & 50 & 0 & 0 & 50 & 0 & 50 & 50 \\

\geminishort &
50 & 47 & 48 & 50 & 0 & 0 & 2 & 0 & 0 & 3 & 0 & 0 & 29 & 0 & 0 & 0 \\

\gptfourshort &
50 & 43 & 50 & 50 & 0 & 0 & 0 & 0 & 0 & 0 & 0 & 0 & 4 & 0 & 0 & 0 \\

\gptfiveshort&
50 & 50 & 46 & 50 & 0 & 0 & 46 & 2 & 32 & 25 & 1 & 8 & 48 & 16 & 42 & 31 \\

\llamashort &
50 & 50 & 50 & 50 & 38 & 0 & 50 & 50 & 50 & 50 & 9 & 50 & 50 & 50 & 50 & 50 \\

\hline
\end{tabular}
\vspace{0.8cm}
\caption{Memory-2 experiment data: Number of times each LLM chooses `L' in the 21 memory-2 scenarios, based on 50 independent trials per scenario. The upper table reports choices in round~1 and round~2, where round~2 behavior is conditioned on the outcome of round~1. The lower table reports choices in later rounds, where the LLM is given the outcomes of the two preceding rounds. In the lower table, data are grouped by the outcome of the most recent round, and within each group, the columns correspond to the outcome in the second-to-last round, R$_{-2}$.}
\vspace{1cm}
\end{table}

\clearpage
\subsection{One-shot game experiments}

\noindent The following data corresponds to the baseline case: no added framing and we the standard payoff values $(a_\LL,a_\LR,a_\RL,a_\RR) = (3,0,5,1)$.\\ 

\begin{table}[h!]
\raggedright
\renewcommand{\arraystretch}{1.2}
\setlength{\tabcolsep}{8pt}
\begin{tabular}{r|c}
\hline
LLM & Number of L in 50 trials \\
\hline
\claudeshort & 0 \\
\geminishort & 8 \\
\gptfourshort & 0 \\
\gptfiveshort& 0 \\
\llamashort & 50 \\
\hline
\end{tabular}
\vspace{0.5cm}
\end{table}

\subsection{Experiments simulating actual play}

In this category we run two related experiments. The first includes the line \textcolor{blue}{“The interaction lasts 10 rounds”}, while the second uses the line \textcolor{blue}{“The interaction lasts at least 10 rounds”}. For each experiment, every LLM plays a single game against every other LLM, including itself, resulting in 15 games per experiment. The outcomes of the second experiment are straightforward: all LLMs chose `L', the cooperative action, in every round of every game. In contrast, in the first experiment, all LLMs except gpt-5 cooperated throughout in every game. In every game involving gpt-5, gpt-5 cooperated for the first nine rounds and defected in the final round. These are again for the baseline case with no added framing and we use the standard prisoner's dilemma payoffs $(a_\LL,a_\LR,a_\RL,a_\RR) = (3,0,5,1)$.

\newpage

\section{Analysis}
\label{section:analysis}

\subsection{Estimated strategies}

\noindent In this section, we first estimate a memory-1 (or memory-2) strategy for each LLM in each treatment. A memory-1 (or memory-2) strategy gives the probability of playing `L' given information about the last (or last two) rounds. We do this estimation by taking the mean of `L'-play over 50 trials in each scenario. The tables show these means along with their Wilson intervals ($n=50$). We also provide a note with the closest known memory-1 strategy, using $\sim$ if the match is approximate. A threshold of 0.5 is used to identify pure strategies, but it is applied flexibly.\\ 

\subsection*{Memory-1 strategies from original experiments}

\begin{table}[h!]
\raggedright
\renewcommand{\arraystretch}{1.2}
\setlength{\tabcolsep}{6pt}
\begin{tabular}{l|c c c c c|r}
\hline
\multirow{2}{*}{LLM} & \multicolumn{5}{c|}{Memory-1 strategy, mean with Wilson interval, $n=50$} & \multicolumn{1}{c}{} \\
\cline{2-6}
 & $p_0$ & $p_1$ & $p_2$ & $p_3$ & $p_4$ & Notes \\
\hline
\claudeshort & \textbf{1.00} [0.93, 1.00] & \textbf{1.00} [0.93, 1.00] & \textbf{0.00} [0.00, 0.07] & \textbf{1.00} [0.93, 1.00] & \textbf{1.00} [0.93, 1.00] & Forgiver \\
\geminishort & \textbf{0.82} [0.69, 0.90] & \textbf{1.00} [0.93, 1.00] & \textbf{0.00} [0.00, 0.07] & \textbf{0.02} [0.00, 0.10] & \textbf{0.42} [0.29, 0.56] & $\sim$WSLS \\
\gptfourshort & \textbf{1.00} [0.93, 1.00] & \textbf{1.00} [0.93, 1.00] & \textbf{0.00} [0.00, 0.07] & \textbf{0.00} [0.00, 0.07] & \textbf{0.00} [0.00, 0.07] & GRIM \\
\gptfiveshort& \textbf{1.00} [0.93, 1.00] & \textbf{1.00} [0.93, 1.00] & \textbf{0.00} [0.00, 0.07] & \textbf{0.54} [0.40, 0.67] & \textbf{1.00} [0.93, 1.00] & $\sim$Forgiver \\
\llamashort & \textbf{1.00} [0.93, 1.00] & \textbf{1.00} [0.93, 1.00] & \textbf{0.26} [0.16, 0.40] & \textbf{0.98} [0.90, 1.00] & \textbf{0.56} [0.42, 0.69] & $\sim$GTFT \\
\hline
\end{tabular}
\caption{Original experiment: The numbers in bold represent the frequency of `L' play in 50 trials in that particular scenario (column) and the numbers inside the square bracket represent the Wilson interval of the data. Framing: \textit{no special instruction}.}
\end{table}

\begin{table}[h!]
\raggedright
\renewcommand{\arraystretch}{1.2}
\setlength{\tabcolsep}{6pt}
\begin{tabular}{l|c c c c c|r}
\hline
\multirow{2}{*}{LLM} & \multicolumn{5}{c|}{Memory-1 strategy, mean with Wilson interval, $n=50$} & \multicolumn{1}{c}{} \\
\cline{2-6}
 & $p_0$ & $p_1$ & $p_2$ & $p_3$ & $p_4$ & Notes \\
\hline
\claudeshort & \textbf{0.00} [0.00, 0.07] & \textbf{0.00} [0.00, 0.07] & \textbf{0.00} [0.00, 0.07] & \textbf{0.00} [0.00, 0.07] & \textbf{0.00} [0.00, 0.07] & ALLD \\
\geminishort & \textbf{0.00} [0.00, 0.07] & \textbf{0.00} [0.00, 0.07] & \textbf{0.00} [0.00, 0.07] & \textbf{0.00} [0.00, 0.07] & \textbf{0.00} [0.00, 0.07] & ALLD \\
\gptfourshort & \textbf{0.00} [0.00, 0.07] & \textbf{0.00} [0.00, 0.07] & \textbf{0.00} [0.00, 0.07] & \textbf{0.00} [0.00, 0.07] & \textbf{0.00} [0.00, 0.07] & ALLD\\
\gptfiveshort& \textbf{0.02} [0.00, 0.10] & \textbf{0.02} [0.00, 0.10] & \textbf{0.00} [0.00, 0.07] & \textbf{0.00} [0.00, 0.07] & \textbf{0.00} [0.00, 0.07] & $\sim$ALLD \\
\llamashort & \textbf{1.00} [0.93, 1.00] & \textbf{1.00} [0.93, 1.00] & \textbf{0.00} [0.00, 0.07] & \textbf{0.00} [0.00, 0.07] & \textbf{0.02} [0.00, 0.10] & GRIM \\
\hline
\end{tabular}
\caption{Original experiment: The numbers in bold represent the frequency of `L' play in 50 trials in that particular scenario (column) and the numbers inside the square bracket represent the Wilson interval of the data. Framing: ``Try to get more points than the other agent."}
\end{table}

\begin{table}[h!]
\raggedright
\renewcommand{\arraystretch}{1.2}
\setlength{\tabcolsep}{6pt}
\begin{tabular}{l|c c c c c|r}
\hline
\multirow{2}{*}{LLM} & \multicolumn{5}{c|}{Memory-1 strategy, mean with Wilson interval, $n=50$} & \multicolumn{1}{c}{} \\
\cline{2-6}
 & $p_0$ & $p_1$ & $p_2$ & $p_3$ & $p_4$ & Notes \\
\hline
\claudeshort & \textbf{0.00} [0.00, 0.07] & \textbf{0.00} [0.00, 0.07] & \textbf{0.00} [0.00, 0.07] & \textbf{0.00} [0.00, 0.07] & \textbf{0.00} [0.00, 0.07] & ALLD \\
\geminishort & \textbf{0.00} [0.00, 0.07] & \textbf{0.00} [0.00, 0.07] & \textbf{0.00} [0.00, 0.07] & \textbf{0.00} [0.00, 0.07] & \textbf{0.00} [0.00, 0.07] & ALLD \\
\gptfourshort & \textbf{0.00} [0.00, 0.07] & \textbf{0.00} [0.00, 0.07] & \textbf{0.00} [0.00, 0.07] & \textbf{0.00} [0.00, 0.07] & \textbf{0.00} [0.00, 0.07] & ALLD \\
\gptfiveshort& \textbf{0.14} [0.07, 0.26] & \textbf{0.00} [0.00, 0.07] & \textbf{0.00} [0.00, 0.07] & \textbf{0.00} [0.00, 0.07] & \textbf{0.00} [0.00, 0.07] & $\sim$ALLD \\
\llamashort & \textbf{0.22} [0.13, 0.35] & \textbf{0.00} [0.00, 0.07] & \textbf{0.00} [0.00, 0.07] & \textbf{0.00} [0.00, 0.07] & \textbf{0.00} [0.00, 0.07] & $\sim$ALLD \\
\hline
\end{tabular}
\caption{Original experiment: The numbers in bold represent the frequency of `L' play in 50 trials in that particular scenario (column) and the numbers inside the square bracket represent the Wilson interval of the data. Framing: ``Exploit the other agent as much as possible".}
\end{table}

\begin{table}[h!]
\raggedright
\renewcommand{\arraystretch}{1.2}
\setlength{\tabcolsep}{6pt}
\begin{tabular}{l|c c c c c|r}
\hline
\multirow{2}{*}{LLM} & \multicolumn{5}{c|}{Memory-1 strategy, mean with Wilson interval, $n=50$} & \multicolumn{1}{c}{} \\
\cline{2-6}
 & $p_0$ & $p_1$ & $p_2$ & $p_3$ & $p_4$ & Notes \\
\hline
\claudeshort & \textbf{0.00} [0.00, 0.07] & \textbf{0.00} [0.00, 0.07] & \textbf{0.00} [0.00, 0.07] & \textbf{0.00} [0.00, 0.07] & \textbf{0.00} [0.00, 0.07] & ALLD \\
\geminishort & \textbf{0.00} [0.00, 0.07] & \textbf{0.24} [0.14, 0.37] & \textbf{0.00} [0.00, 0.07] & \textbf{0.00} [0.00, 0.07] & \textbf{0.00} [0.00, 0.07] & $\sim$ALLD \\
\gptfourshort & \textbf{0.00} [0.00, 0.07] & \textbf{0.30} [0.19, 0.44] & \textbf{0.00} [0.00, 0.07] & \textbf{0.00} [0.00, 0.07] & \textbf{0.00} [0.00, 0.07] & $\sim$ALLD \\
\gptfiveshort& \textbf{1.00} [0.93, 1.00] & \textbf{1.00} [0.93, 1.00] & \textbf{0.00} [0.00, 0.07] & \textbf{0.32} [0.21, 0.46] & \textbf{0.76} [0.63, 0.86] & $\sim$WSLS \\
\llamashort & \textbf{1.00} [0.93, 1.00] & \textbf{1.00} [0.93, 1.00] & \textbf{0.00} [0.00, 0.07] & \textbf{0.00} [0.00, 0.07] & \textbf{0.00} [0.00, 0.07] & GRIM \\
\hline
\end{tabular}
\caption{Original experiment: The numbers in bold represent the frequency of `L' play in 50 trials in that particular scenario (column) and the numbers inside the square bracket represent the Wilson interval of the data. Framing: ``Maximum your own number of points".}
\end{table}

\begin{table}[h!]
\raggedright
\renewcommand{\arraystretch}{1.2}
\setlength{\tabcolsep}{6pt}
\begin{tabular}{l|c c c c c|r}
\hline
\multirow{2}{*}{LLM} & \multicolumn{5}{c|}{Memory-1 strategy, mean with Wilson interval, $n=50$} & \multicolumn{1}{c}{} \\
\cline{2-6}
 & $p_0$ & $p_1$ & $p_2$ & $p_3$ & $p_4$ & Notes \\
\hline
\claudeshort & \textbf{0.00} [0.00, 0.07] & \textbf{0.00} [0.00, 0.07] & \textbf{0.00} [0.00, 0.07] & \textbf{0.00} [0.00, 0.07] & \textbf{0.00} [0.00, 0.07] & ALLD \\
\geminishort & \textbf{0.00} [0.00, 0.07] & \textbf{0.18} [0.10, 0.31] & \textbf{0.00} [0.00, 0.07] & \textbf{0.00} [0.00, 0.07] & \textbf{0.04} [0.01, 0.13] & $\sim$ALLD \\
\gptfourshort & \textbf{0.00} [0.00, 0.07] & \textbf{0.00} [0.00, 0.07] & \textbf{0.00} [0.00, 0.07] & \textbf{0.00} [0.00, 0.07] & \textbf{0.00} [0.00, 0.07] & ALLD \\
\gptfiveshort& \textbf{0.98} [0.90, 1.00] & \textbf{1.00} [0.93, 1.00] & \textbf{0.00} [0.00, 0.07] & \textbf{0.14} [0.07, 0.26] & \textbf{0.38} [0.26, 0.52] & $\sim$GRIM \\
\llamashort & \textbf{1.00} [0.93, 1.00] & \textbf{1.00} [0.93, 1.00] & \textbf{0.00} [0.00, 0.07] & \textbf{0.00} [0.00, 0.07] & \textbf{0.00} [0.00, 0.07] & GRIM \\
\hline
\end{tabular}
\caption{Original experiment: The numbers in bold represent the frequency of `L' play in 50 trials in that particular scenario (column) and the numbers inside the square bracket represent the Wilson interval of the data. Framing: ``Think about winning".}
\end{table}

\begin{table}[h!]
\raggedright
\renewcommand{\arraystretch}{1.2}
\setlength{\tabcolsep}{6pt}
\begin{tabular}{l|c c c c c|r}
\hline
\multirow{2}{*}{LLM} & \multicolumn{5}{c|}{Memory-1 strategy, mean with Wilson interval, $n=50$} & \multicolumn{1}{c}{} \\
\cline{2-6}
 & $p_0$ & $p_1$ & $p_2$ & $p_3$ & $p_4$ & Notes \\
\hline
\claudeshort & \textbf{1.00} [0.93, 1.00] & \textbf{1.00} [0.93, 1.00] & \textbf{0.00} [0.00, 0.07] & \textbf{0.10} [0.04, 0.21] & \textbf{1.00} [0.93, 1.00] & $\sim$WSLS \\
\geminishort & \textbf{0.84} [0.71, 0.92] & \textbf{1.00} [0.93, 1.00] & \textbf{0.00} [0.00, 0.07] & \textbf{0.18} [0.10, 0.31] & \textbf{0.32} [0.21, 0.46] & $\sim$GRIM \\
\gptfourshort & \textbf{0.98} [0.90, 1.00] & \textbf{1.00} [0.93, 1.00] & \textbf{0.00} [0.00, 0.07] & \textbf{0.00} [0.00, 0.07] & \textbf{0.12} [0.06, 0.24] & $\sim$GRIM\\
\gptfiveshort& \textbf{1.00} [0.93, 1.00] & \textbf{1.00} [0.93, 1.00] & \textbf{0.00} [0.00, 0.07] & \textbf{0.36} [0.24, 0.50] & \textbf{0.98} [0.90, 1.00] & $\sim$WSLS\\
\llamashort & \textbf{1.00} [0.93, 1.00] & \textbf{1.00} [0.93, 1.00] & \textbf{0.00} [0.00, 0.07] & \textbf{0.00} [0.00, 0.07] & \textbf{0.00} [0.00, 0.07] & GRIM\\
\hline
\end{tabular}
\caption{Original experiment: The numbers in bold represent the frequency of `L' play in 50 trials in that particular scenario (column) and the numbers inside the square bracket represent the Wilson interval of the data. Framing: ``Play like a pro".}
\end{table}

\begin{table}[h!]
\raggedright
\renewcommand{\arraystretch}{1.2}
\setlength{\tabcolsep}{6pt}
\begin{tabular}{l|c c c c c|r}
\hline
\multirow{2}{*}{LLM} & \multicolumn{5}{c|}{Memory-1 strategy, mean with Wilson interval, $n=50$} & \multicolumn{1}{c}{} \\
\cline{2-6}
 & $p_0$ & $p_1$ & $p_2$ & $p_3$ & $p_4$ & Notes \\
\hline
\claudeshort & \textbf{1.00} [0.93, 1.00] & \textbf{1.00} [0.93, 1.00] & \textbf{1.00} [0.93, 1.00] & \textbf{1.00} [0.93, 1.00] & \textbf{1.00} [0.93, 1.00] & ALLC \\
\geminishort & \textbf{1.00} [0.93, 1.00] & \textbf{1.00} [0.93, 1.00] & \textbf{0.04} [0.01, 0.13] & \textbf{1.00} [0.93, 1.00] & \textbf{1.00} [0.93, 1.00] & $\sim$Forgiver \\
\gptfourshort & \textbf{1.00} [0.93, 1.00] & \textbf{1.00} [0.93, 1.00] & \textbf{1.00} [0.93, 1.00] & \textbf{1.00} [0.93, 1.00] & \textbf{1.00} [0.93, 1.00] & ALLC \\
\gptfiveshort& \textbf{1.00} [0.93, 1.00] & \textbf{1.00} [0.93, 1.00] & \textbf{0.60} [0.46, 0.72] & \textbf{1.00} [0.93, 1.00] & \textbf{1.00} [0.93, 1.00] & $\sim$Forgiver \\
\llamashort & \textbf{1.00} [0.93, 1.00] & \textbf{1.00} [0.93, 1.00] & \textbf{0.08} [0.03, 0.19] & \textbf{1.00} [0.93, 1.00] & \textbf{1.00} [0.93, 1.00] & $\sim$Forgiver \\
\hline
\end{tabular}
\caption{Original experiment: The numbers in bold represent the frequency of `L' play in 50 trials in that particular scenario (column) and the numbers inside the square bracket represent the Wilson interval of the data.  Framing: ``Accumulate many points, but do not exploit."}
\end{table}

\begin{table}[h!]
\raggedright
\renewcommand{\arraystretch}{1.2}
\setlength{\tabcolsep}{6pt}
\begin{tabular}{l|c c c c c|r}
\hline
\multirow{2}{*}{LLM} & \multicolumn{5}{c|}{Memory-1 strategy, mean with Wilson interval, $n=50$} & \multicolumn{1}{c}{} \\
\cline{2-6}
 & $p_0$ & $p_1$ & $p_2$ & $p_3$ & $p_4$ & Notes \\
\hline
\claudeshort & \textbf{1.00} [0.93, 1.00] & \textbf{1.00} [0.93, 1.00] & \textbf{0.00} [0.00, 0.07] & \textbf{1.00} [0.93, 1.00] & \textbf{1.00} [0.93, 1.00] & Forgiver \\
\geminishort & \textbf{1.00} [0.93, 1.00] & \textbf{1.00} [0.93, 1.00] & \textbf{0.80} [0.67, 0.89] & \textbf{1.00} [0.93, 1.00] & \textbf{1.00} [0.93, 1.00] & $\sim$ALLC \\
\gptfourshort & \textbf{1.00} [0.93, 1.00] & \textbf{1.00} [0.93, 1.00] & \textbf{0.52} [0.39, 0.65] & \textbf{1.00} [0.93, 1.00] & \textbf{1.00} [0.93, 1.00] & $\sim$Forgiver \\
\gptfiveshort& \textbf{1.00} [0.93, 1.00] & \textbf{1.00} [0.93, 1.00] & \textbf{0.00} [0.00, 0.07] & \textbf{1.00} [0.93, 1.00] & \textbf{1.00} [0.93, 1.00] & Forgiver \\
\llamashort & \textbf{1.00} [0.93, 1.00] & \textbf{1.00} [0.93, 1.00] & \textbf{0.00} [0.00, 0.07] & \textbf{0.82} [0.69, 0.90] & \textbf{1.00} [0.93, 1.00] & $\sim$Forgiver \\
\hline
\end{tabular}
\caption{Original experiment: The numbers in bold represent the frequency of `L' play in 50 trials in that particular scenario (column) and the numbers inside the square bracket represent the Wilson interval of the data. Framing: ``Maximize your points, keep in mind the other’s welfare".}
\end{table}

\begin{table}[h!]
\raggedright
\renewcommand{\arraystretch}{1.2}
\setlength{\tabcolsep}{6pt}
\begin{tabular}{l|c c c c c|r}
\hline
\multirow{2}{*}{LLM} & \multicolumn{5}{c|}{Memory-1 strategy, mean with Wilson interval, $n=50$} & \multicolumn{1}{c}{} \\
\cline{2-6}
 & $p_0$ & $p_1$ & $p_2$ & $p_3$ & $p_4$ & Notes \\
\hline
\claudeshort & \textbf{1.00} [0.93, 1.00] & \textbf{1.00} [0.93, 1.00] & \textbf{0.00} [0.00, 0.07] & \textbf{1.00} [0.93, 1.00] & \textbf{1.00} [0.93, 1.00] & Forgiver \\
\geminishort & \textbf{1.00} [0.93, 1.00] & \textbf{1.00} [0.93, 1.00] & \textbf{0.96} [0.87, 0.99] & \textbf{1.00} [0.93, 1.00] & \textbf{1.00} [0.93, 1.00] & $\sim$ALLC \\
\gptfourshort & \textbf{1.00} [0.93, 1.00] & \textbf{1.00} [0.93, 1.00] & \textbf{0.06} [0.02, 0.16] & \textbf{1.00} [0.93, 1.00] & \textbf{1.00} [0.93, 1.00] & $\sim$ Forgiver \\
\gptfiveshort& \textbf{1.00} [0.93, 1.00] & \textbf{1.00} [0.93, 1.00] & \textbf{0.02} [0.00, 0.10] & \textbf{1.00} [0.93, 1.00] & \textbf{1.00} [0.93, 1.00] & $\sim$ Forgiver  \\
\llamashort & \textbf{1.00} [0.93, 1.00] & \textbf{1.00} [0.93, 1.00] & \textbf{0.00} [0.00, 0.07] & \textbf{0.98} [0.90, 1.00] & \textbf{1.00} [0.93, 1.00] & $\sim$ Forgiver  \\
\hline
\end{tabular}
\caption{Original experiment: The numbers in bold represent the frequency of `L' play in 50 trials in that particular scenario (column) and the numbers inside the square bracket represent the Wilson interval of the data. Framing: ``Think about fair outcomes".}
\end{table}

\begin{table}[h!]
\raggedright
\renewcommand{\arraystretch}{1.2}
\setlength{\tabcolsep}{6pt}
\begin{tabular}{l|c c c c c|r}
\hline
\multirow{2}{*}{LLM} & \multicolumn{5}{c|}{Memory-1 strategy, mean with Wilson interval, $n=50$} & \multicolumn{1}{c}{} \\
\cline{2-6}
 & $p_0$ & $p_1$ & $p_2$ & $p_3$ & $p_4$ & Notes \\
\hline
\claudeshort & \textbf{1.00} [0.93, 1.00] & \textbf{1.00} [0.93, 1.00] & \textbf{1.00} [0.93, 1.00] & \textbf{1.00} [0.93, 1.00] & \textbf{1.00} [0.93, 1.00] & ALLC \\
\geminishort & \textbf{1.00} [0.93, 1.00] & \textbf{1.00} [0.93, 1.00] & \textbf{1.00} [0.93, 1.00] & \textbf{1.00} [0.93, 1.00] & \textbf{1.00} [0.93, 1.00] & ALLC \\
\gptfourshort & \textbf{1.00} [0.93, 1.00] & \textbf{1.00} [0.93, 1.00] & \textbf{1.00} [0.93, 1.00] & \textbf{1.00} [0.93, 1.00] & \textbf{1.00} [0.93, 1.00] & ALLC \\
\gptfiveshort& \textbf{1.00} [0.93, 1.00] & \textbf{1.00} [0.93, 1.00] & \textbf{1.00} [0.93, 1.00] & \textbf{1.00} [0.93, 1.00] & \textbf{1.00} [0.93, 1.00] & ALLC \\
\llamashort & \textbf{1.00} [0.93, 1.00] & \textbf{1.00} [0.93, 1.00] & \textbf{0.00} [0.00, 0.07] & \textbf{0.98} [0.90, 1.00] & \textbf{1.00} [0.93, 1.00] & $\sim$Forgiver\\
\hline
\end{tabular}
 \caption{Original experiment: The numbers in bold represent the frequency of `L' play in 50 trials in that particular scenario (column) and the numbers inside the square bracket represent the Wilson interval of the data. Framing: ``Be a saint".}
\end{table}

\begin{table}[h!]
\raggedright
\renewcommand{\arraystretch}{1.2} 
\setlength{\tabcolsep}{4pt} 
\begin{tabular}{r|r r r r r}
\hline
Instruction & \claudeshort& \geminishort& \gptfourshort & \gptfiveshort& \llamashort \\
\hline
\textit{no special instruction} & Forgiver& $\sim$WSLS& GRIM& $\sim$Forgiver& $\sim$GTFT\\
Try to get more points than the other agent. & ALLD& ALLD& ALLD& $\sim$ALLD& $\sim$GRIM\\
Exploit the other agent as much as possible. & ALLD& ALLD& ALLD& $\sim$ALLD& $\sim$ALLD\\
Maximize your own number of points. & ALLD& $\sim$ALLD& $\sim$ALLD& $\sim$WSLS& GRIM\\
Think about winning. & ALLD& $\sim$ALLD& ALLD& $\sim$GRIM& GRIM\\
Play like a pro. & $\sim$WSLS & $\sim$GRIM& $\sim$GRIM& $\sim$WSLS& GRIM\\
Accumulate many points, but do not exploit. & ALLC& $\sim$Forgiver& ALLC& $\sim$Forgiver& $\sim$Forgiver\\
Maximize your points, keep in mind the other's welfare. & Forgiver & $\sim$ALLC& $\sim$Forgiver& Forgiver& $\sim$Forgiver\\
Think about fair outcomes. & Forgiver& $\sim$ALLC& $\sim$Forgiver& $\sim$Forgiver& $\sim$Forgiver\\
Be a saint. & ALLC & ALLC & ALLC & ALLC & $\sim$Forgiver \\
\hline
\end{tabular}
\vspace{0.5cm}
\caption{Original experiments: This table summarizes the columns `Notes' from all the ten tables that proceed it.}
\end{table}

\clearpage

\subsection*{Memory-1 strategies from stopping probability experiments}

\begin{table}[h!]
\raggedright
\renewcommand{\arraystretch}{1.2}
\setlength{\tabcolsep}{6pt}
\begin{tabular}{l|c c c c c|r}
\hline
\multirow{2}{*}{LLM} & \multicolumn{5}{c|}{Memory-1 strategy, mean with Wilson interval, $n=50$, $w\mathbf{=0.01}$} & \multicolumn{1}{c}{} \\
\cline{2-6}
 & $p_0$ & $p_1$ & $p_2$ & $p_3$ & $p_4$ & Notes \\
\hline
\claudeshort & \textbf{1.00} [0.93, 1.00] & \textbf{1.00} [0.93, 1.00] & \textbf{0.00} [0.00, 0.07] & \textbf{1.00} [0.93, 1.00] & \textbf{1.00} [0.93, 1.00] & Forgiver \\
\geminishort & \textbf{0.98} [0.90, 1.00] & \textbf{1.00} [0.93, 1.00] & \textbf{0.00} [0.00, 0.07] & \textbf{0.00} [0.00, 0.07] & \textbf{0.42} [0.29, 0.56] & $\sim$GRIM \\
\gptfourshort & \textbf{1.00} [0.93, 1.00] & \textbf{1.00} [0.93, 1.00] & \textbf{0.00} [0.00, 0.07] & \textbf{0.00} [0.00, 0.07] & \textbf{0.82} [0.69, 0.90] & $\sim$WSLS \\
\gptfiveshort& \textbf{1.00} [0.93, 1.00] & \textbf{1.00} [0.93, 1.00] & \textbf{0.00} [0.00, 0.07] & \textbf{0.86} [0.74, 0.93] & \textbf{1.00} [0.93, 1.00] & $\sim$Forgiver \\
\llamashort & \textbf{1.00} [0.93, 1.00] & \textbf{1.00} [0.93, 1.00] & \textbf{0.00} [0.00, 0.07] & \textbf{0.00} [0.00, 0.07] & \textbf{0.00} [0.00, 0.07] & GRIM \\
\hline
\end{tabular}
\end{table}

\begin{table}[h!]
\raggedright
\renewcommand{\arraystretch}{1.2}
\setlength{\tabcolsep}{6pt}
\begin{tabular}{l|c c c c c|r}
\hline
\multirow{2}{*}{LLM} & \multicolumn{5}{c|}{Memory-1 strategy, mean with Wilson interval, $n=50$,  $w\mathbf{=0.1}$} & \multicolumn{1}{c}{} \\
\cline{2-6}
 & $p_0$ & $p_1$ & $p_2$ & $p_3$ & $p_4$ & Notes \\
\hline
\claudeshort & \textbf{1.00} [0.93, 1.00] & \textbf{1.00} [0.93, 1.00] & \textbf{0.00} [0.00, 0.07] & \textbf{1.00} [0.93, 1.00] & \textbf{1.00} [0.93, 1.00] & Forgiver \\
\geminishort & \textbf{0.74} [0.60, 0.84] & \textbf{1.00} [0.93, 1.00] & \textbf{0.00} [0.00, 0.07] & \textbf{0.00} [0.00, 0.07] & \textbf{0.34} [0.22, 0.48] & $\sim$GRIM \\
\gptfourshort & \textbf{1.00} [0.93, 1.00] & \textbf{1.00} [0.93, 1.00] & \textbf{0.00} [0.00, 0.07] & \textbf{0.00} [0.00, 0.07] & \textbf{0.26} [0.16, 0.40] & $\sim$GRIM \\
\gptfiveshort& \textbf{1.00} [0.93, 1.00] & \textbf{1.00} [0.93, 1.00] & \textbf{0.00} [0.00, 0.07] & \textbf{0.84} [0.71, 0.92] & \textbf{0.98} [0.90, 1.00] & $\sim$Forgiver \\
\llamashort & \textbf{1.00} [0.93, 1.00] & \textbf{1.00} [0.93, 1.00] & \textbf{0.00} [0.00, 0.07] & \textbf{0.00} [0.00, 0.07] & \textbf{0.04} [0.01, 0.13] & $\sim$GRIM \\
\hline
\end{tabular}
\end{table}

\begin{table}[h!]
\raggedright
\renewcommand{\arraystretch}{1.2}
\setlength{\tabcolsep}{6pt}
\begin{tabular}{l|c c c c c|r}
\hline
\multirow{2}{*}{LLM} & \multicolumn{5}{c|}{Memory-1 strategy, mean with Wilson interval, $n=50$,  $w\mathbf{=0.2}$} & \multicolumn{1}{c}{} \\
\cline{2-6}
 & $p_0$ & $p_1$ & $p_2$ & $p_3$ & $p_4$ & Notes \\
\hline
\claudeshort & \textbf{1.00} [0.93, 1.00] & \textbf{1.00} [0.93, 1.00] & \textbf{0.00} [0.00, 0.07] & \textbf{1.00} [0.93, 1.00] & \textbf{1.00} [0.93, 1.00] & Forgiver \\
\geminishort & \textbf{0.50} [0.37, 0.63] & \textbf{1.00} [0.93, 1.00] & \textbf{0.00} [0.00, 0.07] & \textbf{0.00} [0.00, 0.07] & \textbf{0.28} [0.17, 0.42] & $\sim$SGRIM \\
\gptfourshort & \textbf{1.00} [0.93, 1.00] & \textbf{1.00} [0.93, 1.00] & \textbf{0.00} [0.00, 0.07] & \textbf{0.00} [0.00, 0.07] & \textbf{0.54} [0.40, 0.67] & $\sim$WSLS \\
\gptfiveshort& \textbf{1.00} [0.93, 1.00] & \textbf{1.00} [0.93, 1.00] & \textbf{0.00} [0.00, 0.07] & \textbf{0.78} [0.65, 0.87] & \textbf{1.00} [0.93, 1.00] & $\sim$Forgiver \\
\llamashort & \textbf{1.00} [0.93, 1.00] & \textbf{1.00} [0.93, 1.00] & \textbf{0.00} [0.00, 0.07] & \textbf{0.00} [0.00, 0.07] & \textbf{0.00} [0.00, 0.07] & GRIM \\
\hline
\end{tabular}
\end{table}

\begin{table}[h!]
\raggedright
\renewcommand{\arraystretch}{1.2}
\setlength{\tabcolsep}{6pt}
\begin{tabular}{l|c c c c c|r}
\hline
\multirow{2}{*}{LLM} & \multicolumn{5}{c|}{Memory-1 strategy, mean with Wilson interval, $n=50$,  $w\mathbf{=0.5}$} & \multicolumn{1}{c}{} \\
\cline{2-6}
 & $p_0$ & $p_1$ & $p_2$ & $p_3$ & $p_4$ & Notes \\
\hline
\claudeshort & \textbf{1.00} [0.93, 1.00] & \textbf{1.00} [0.93, 1.00] & \textbf{0.00} [0.00, 0.07] & \textbf{1.00} [0.93, 1.00] & \textbf{1.00} [0.93, 1.00] & Forgiver \\
\geminishort & \textbf{0.46} [0.33, 0.60] & \textbf{0.88} [0.76, 0.94] & \textbf{0.00} [0.00, 0.07] & \textbf{0.00} [0.00, 0.07] & \textbf{0.06} [0.02, 0.16] & $\sim$SGRIM \\
\gptfourshort & \textbf{1.00} [0.93, 1.00] & \textbf{1.00} [0.93, 1.00] & \textbf{0.00} [0.00, 0.07] & \textbf{0.00} [0.00, 0.07] & \textbf{0.56} [0.42, 0.69] & $\sim$WSLS \\
\gptfiveshort& \textbf{0.98} [0.90, 1.00] & \textbf{1.00} [0.93, 1.00] & \textbf{0.00} [0.00, 0.07] & \textbf{0.18} [0.10, 0.31] & \textbf{0.64} [0.50, 0.76] & $\sim$WSLS \\
\llamashort & \textbf{1.00} [0.93, 1.00] & \textbf{1.00} [0.93, 1.00] & \textbf{0.00} [0.00, 0.07] & \textbf{0.00} [0.00, 0.07] & \textbf{0.00} [0.00, 0.07] & GRIM \\
\hline
\end{tabular}
 \caption{Stopping probability experiments: The numbers in bold represent the frequency of `L' play in 50 trials in that particular scenario (column) and the numbers inside the square bracket represent the Wilson interval of the data. Four tables correspond to four values of $w$, the stopping probability. SGRIM stands for Suspicious GRIM, which is just the strategy that starts with defection and only cooperates after mutual cooperation.}
\end{table}

\clearpage

\subsection*{Memory-1 strategies from equal-gains experiments}

\begin{table}[h!]
\raggedright
\renewcommand{\arraystretch}{1.2}
\setlength{\tabcolsep}{6pt}
\begin{tabular}{l|c c c c c|r}
\hline
\multirow{2}{*}{LLM} & \multicolumn{5}{c|}{Memory-1 strategy, mean with Wilson interval, $n=50$, $x=\mathbf{0}$} & \multicolumn{1}{c}{} \\
\cline{2-6}
 & $p_0$ & $p_1$ & $p_2$ & $p_3$ & $p_4$ & Notes \\
\hline
\claudeshort & \textbf{1.00} [0.93, 1.00] & \textbf{1.00} [0.93, 1.00] & \textbf{0.00} [0.00, 0.07] & \textbf{1.00} [0.93, 1.00] & \textbf{1.00} [0.93, 1.00] & Forgiver \\
\geminishort & \textbf{1.00} [0.93, 1.00] & \textbf{1.00} [0.93, 1.00] & \textbf{0.04} [0.01, 0.13] & \textbf{0.08} [0.03, 0.19] & \textbf{1.00} [0.93, 1.00] & $\sim$WSLS \\
\gptfourshort & \textbf{1.00} [0.93, 1.00] & \textbf{1.00} [0.93, 1.00] & \textbf{0.00} [0.00, 0.07] & \textbf{0.00} [0.00, 0.07] & \textbf{1.00} [0.93, 1.00] & WSLS \\
\gptfiveshort& \textbf{1.00} [0.93, 1.00] & \textbf{1.00} [0.93, 1.00] & \textbf{0.08} [0.03, 0.19] & \textbf{1.00} [0.93, 1.00] & \textbf{1.00} [0.93, 1.00] & $\sim$Forgiver \\
\llamashort & \textbf{1.00} [0.93, 1.00] & \textbf{1.00} [0.93, 1.00] & \textbf{1.00} [0.93, 1.00] & \textbf{0.98} [0.90, 1.00] & \textbf{1.00} [0.93, 1.00] & $\sim$ALLC \\
\hline
\end{tabular}
\end{table}

\begin{table}[h!]
\raggedright
\renewcommand{\arraystretch}{1.2}
\setlength{\tabcolsep}{6pt}
\begin{tabular}{l|c c c c c|r}
\hline
\multirow{2}{*}{LLM} & \multicolumn{5}{c|}{Memory-1 strategy, mean with Wilson interval, $n=50$, $x=\mathbf{1}$} & \multicolumn{1}{c}{} \\
\cline{2-6}
 & $p_0$ & $p_1$ & $p_2$ & $p_3$ & $p_4$ & Notes \\
\hline
\claudeshort & \textbf{1.00} [0.93, 1.00] & \textbf{1.00} [0.93, 1.00] & \textbf{0.00} [0.00, 0.07] & \textbf{1.00} [0.93, 1.00] & \textbf{1.00} [0.93, 1.00] & Forgiver \\
\geminishort & \textbf{0.92} [0.81, 0.97] & \textbf{1.00} [0.93, 1.00] & \textbf{0.00} [0.00, 0.07] & \textbf{0.00} [0.00, 0.07] & \textbf{0.72} [0.58, 0.83] & $\sim$WSLS \\
\gptfourshort & \textbf{1.00} [0.93, 1.00] & \textbf{1.00} [0.93, 1.00] & \textbf{0.00} [0.00, 0.07] & \textbf{0.00} [0.00, 0.07] & \textbf{0.94} [0.84, 0.98] & $\sim$WSLS\\
\gptfiveshort& \textbf{1.00} [0.93, 1.00] & \textbf{1.00} [0.93, 1.00] & \textbf{0.00} [0.00, 0.07] & \textbf{0.64} [0.50, 0.76] & \textbf{0.98} [0.90, 1.00] & $\sim$Forgiver\\
\llamashort & \textbf{1.00} [0.93, 1.00] & \textbf{1.00} [0.93, 1.00] & \textbf{0.86} [0.74, 0.93] & \textbf{0.98} [0.90, 1.00] & \textbf{0.96} [0.87, 0.99] & $\sim$ALLC \\
\hline
\end{tabular}
\end{table}

\begin{table}[h!]
\raggedright
\renewcommand{\arraystretch}{1.2}
\setlength{\tabcolsep}{6pt}
\begin{tabular}{l|c c c c c|r}
\hline
\multirow{2}{*}{LLM} & \multicolumn{5}{c|}{Memory-1 strategy, mean with Wilson interval, $n=50$, $x=\mathbf{2}$} & \multicolumn{1}{c}{} \\
\cline{2-6}
 & $p_0$ & $p_1$ & $p_2$ & $p_3$ & $p_4$ & Notes \\
\hline
\claudeshort & \textbf{1.00} [0.93, 1.00] & \textbf{1.00} [0.93, 1.00] & \textbf{0.00} [0.00, 0.07] & \textbf{1.00} [0.93, 1.00] & \textbf{1.00} [0.93, 1.00] & Forgiver \\
\geminishort & \textbf{0.94} [0.84, 0.98] & \textbf{1.00} [0.93, 1.00] & \textbf{0.00} [0.00, 0.07] & \textbf{0.02} [0.00, 0.10] & \textbf{0.34} [0.22, 0.48] & $\sim$GRIM \\
\gptfourshort & \textbf{1.00} [0.93, 1.00] & \textbf{1.00} [0.93, 1.00] & \textbf{0.00} [0.00, 0.07] & \textbf{0.00} [0.00, 0.07] & \textbf{0.10} [0.04, 0.21] & $\sim$GRIM \\
\gptfiveshort& \textbf{0.98} [0.90, 1.00] & \textbf{1.00} [0.93, 1.00] & \textbf{0.00} [0.00, 0.07] & \textbf{0.60} [0.46, 0.72] & \textbf{0.94} [0.84, 0.98] & $\sim$Forgiver \\
\llamashort & \textbf{1.00} [0.93, 1.00] & \textbf{1.00} [0.93, 1.00] & \textbf{0.26} [0.16, 0.40] & \textbf{1.00} [0.93, 1.00] & \textbf{1.00} [0.93, 1.00] & $\sim$Forgiver \\
\hline
\end{tabular}
\end{table}

\begin{table}[h!]
\raggedright
\renewcommand{\arraystretch}{1.2}
\setlength{\tabcolsep}{6pt}
\begin{tabular}{l|c c c c c|r}
\hline
\multirow{2}{*}{LLM} & \multicolumn{5}{c|}{Memory-1 strategy, mean with Wilson interval, $n=50$, , $x=\mathbf{3}$} & \multicolumn{1}{c}{} \\
\cline{2-6}
 & $p_0$ & $p_1$ & $p_2$ & $p_3$ & $p_4$ & Notes \\
\hline
\claudeshort & \textbf{1.00} [0.93, 1.00] & \textbf{1.00} [0.93, 1.00] & \textbf{0.00} [0.00, 0.07] & \textbf{1.00} [0.93, 1.00] & \textbf{1.00} [0.93, 1.00] & Forgiver \\
\geminishort & \textbf{1.00} [0.93, 1.00] & \textbf{1.00} [0.93, 1.00] & \textbf{0.00} [0.00, 0.07] & \textbf{0.06} [0.02, 0.16] & \textbf{0.24} [0.14, 0.37] & $\sim$GRIM \\
\gptfourshort & \textbf{1.00} [0.93, 1.00] & \textbf{1.00} [0.93, 1.00] & \textbf{0.00} [0.00, 0.07] & \textbf{0.00} [0.00, 0.07] & \textbf{0.04} [0.01, 0.13] & $\sim$GRIM \\
\gptfiveshort& \textbf{1.00} [0.93, 1.00] & \textbf{1.00} [0.93, 1.00] & \textbf{0.00} [0.00, 0.07] & \textbf{0.50} [0.37, 0.63] & \textbf{0.96} [0.87, 0.99] & $\sim$Forgiver \\
\llamashort & \textbf{1.00} [0.93, 1.00] & \textbf{1.00} [0.93, 1.00] & \textbf{0.60} [0.46, 0.72] & \textbf{0.78} [0.65, 0.87] & \textbf{1.00} [0.93, 1.00] & $\sim$ALLC \\
\hline
\end{tabular}
\end{table}

\begin{table}[h!]
\raggedright
\renewcommand{\arraystretch}{1.2}
\setlength{\tabcolsep}{6pt}
\begin{tabular}{l|c c c c c|r}
\hline
\multirow{2}{*}{LLM} & \multicolumn{5}{c|}{Memory-1 strategy, mean with Wilson interval, $n=50$, , $x=\mathbf{4}$} & \multicolumn{1}{c}{} \\
\cline{2-6}
 & $p_0$ & $p_1$ & $p_2$ & $p_3$ & $p_4$ & Notes \\
\hline
\claudeshort & \textbf{1.00} [0.93, 1.00] & \textbf{1.00} [0.93, 1.00] & \textbf{0.00} [0.00, 0.07] & \textbf{1.00} [0.93, 1.00] & \textbf{1.00} [0.93, 1.00] & Forgiver \\
\geminishort & \textbf{0.74} [0.60, 0.84] & \textbf{1.00} [0.93, 1.00] & \textbf{0.00} [0.00, 0.07] & \textbf{0.00} [0.00, 0.07] & \textbf{0.00} [0.00, 0.07] & $\sim$GRIM \\
\gptfourshort & \textbf{1.00} [0.93, 1.00] & \textbf{1.00} [0.93, 1.00] & \textbf{0.00} [0.00, 0.07] & \textbf{0.00} [0.00, 0.07] & \textbf{0.10} [0.04, 0.21] & $\sim$GRIM \\
\gptfiveshort& \textbf{1.00} [0.93, 1.00] & \textbf{1.00} [0.93, 1.00] & \textbf{0.00} [0.00, 0.07] & \textbf{0.38} [0.26, 0.52] & \textbf{0.72} [0.58, 0.83] & $\sim$WSLS \\
\llamashort & \textbf{1.00} [0.93, 1.00] & \textbf{1.00} [0.93, 1.00] & \textbf{0.34} [0.22, 0.48] & \textbf{0.96} [0.87, 0.99] & \textbf{0.72} [0.58, 0.83] & $\sim$Forgiver \\
\hline
\end{tabular}
\end{table}

\begin{table}[h!]
\raggedright
\renewcommand{\arraystretch}{1.2}
\setlength{\tabcolsep}{6pt}
\begin{tabular}{l|c c c c c|r}
\hline
\multirow{2}{*}{LLM} & \multicolumn{5}{c|}{Memory-1 strategy, mean with Wilson interval, $n=50$, $x=\mathbf{5}$} & \multicolumn{1}{c}{} \\
\cline{2-6}
 & $p_0$ & $p_1$ & $p_2$ & $p_3$ & $p_4$ & Notes \\
\hline
\claudeshort & \textbf{1.00} [0.93, 1.00] & \textbf{1.00} [0.93, 1.00] & \textbf{0.00} [0.00, 0.07] & \textbf{1.00} [0.93, 1.00] & \textbf{1.00} [0.93, 1.00] & Forgiver \\
\geminishort & \textbf{0.80} [0.67, 0.89] & \textbf{1.00} [0.93, 1.00] & \textbf{0.00} [0.00, 0.07] & \textbf{0.00} [0.00, 0.07] & \textbf{0.02} [0.00, 0.10] & $\sim$GRIM \\
\gptfourshort & \textbf{1.00} [0.93, 1.00] & \textbf{1.00} [0.93, 1.00] & \textbf{0.00} [0.00, 0.07] & \textbf{0.00} [0.00, 0.07] & \textbf{0.00} [0.00, 0.07] & GRIM \\
\gptfiveshort& \textbf{1.00} [0.93, 1.00] & \textbf{1.00} [0.93, 1.00] & \textbf{0.00} [0.00, 0.07] & \textbf{0.32} [0.21, 0.46] & \textbf{0.74} [0.60, 0.84] & $\sim$WSLS \\
\llamashort & \textbf{1.00} [0.93, 1.00] & \textbf{1.00} [0.93, 1.00] & \textbf{0.10} [0.04, 0.21] & \textbf{1.00} [0.93, 1.00] & \textbf{0.86} [0.74, 0.93] & $\sim$Forgiver \\
\hline
\end{tabular}
\end{table}

\begin{table}[h!]
\raggedright
\renewcommand{\arraystretch}{1.2}
\setlength{\tabcolsep}{6pt}
\begin{tabular}{l|c c c c c|r}
\hline
\multirow{2}{*}{LLM} & \multicolumn{5}{c|}{Memory-1 strategy, mean with Wilson interval, $n=50$, $x=\mathbf{6}$} & \multicolumn{1}{c}{} \\
\cline{2-6}
 & $p_0$ & $p_1$ & $p_2$ & $p_3$ & $p_4$ & Notes \\
\hline
\claudeshort & \textbf{1.00} [0.93, 1.00] & \textbf{1.00} [0.93, 1.00] & \textbf{0.00} [0.00, 0.07] & \textbf{1.00} [0.93, 1.00] & \textbf{1.00} [0.93, 1.00] & Forgiver \\
\geminishort & \textbf{0.62} [0.48, 0.74] & \textbf{1.00} [0.93, 1.00] & \textbf{0.00} [0.00, 0.07] & \textbf{0.00} [0.00, 0.07] & \textbf{0.00} [0.00, 0.07] & $\sim$GRIM \\
\gptfourshort & \textbf{1.00} [0.93, 1.00] & \textbf{1.00} [0.93, 1.00] & \textbf{0.00} [0.00, 0.07] & \textbf{0.00} [0.00, 0.07] & \textbf{0.02} [0.00, 0.10] & $\sim$GRIM \\
\gptfiveshort& \textbf{1.00} [0.93, 1.00] & \textbf{1.00} [0.93, 1.00] & \textbf{0.00} [0.00, 0.07] & \textbf{0.32} [0.21, 0.46] & \textbf{0.48} [0.35, 0.61] & $\sim$GRIM \\
\llamashort & \textbf{1.00} [0.93, 1.00] & \textbf{1.00} [0.93, 1.00] & \textbf{0.34} [0.22, 0.48] & \textbf{0.74} [0.60, 0.84] & \textbf{0.58} [0.44, 0.71] & $\sim$Forgiver \\
\hline
\end{tabular}
\end{table}

\begin{table}[h!]
\raggedright
\renewcommand{\arraystretch}{1.2}
\setlength{\tabcolsep}{6pt}
\begin{tabular}{l|c c c c c|r}
\hline
\multirow{2}{*}{LLM} & \multicolumn{5}{c|}{Memory-1 strategy, mean with Wilson interval, $n=50$, $x=\mathbf{7}$} & \multicolumn{1}{c}{} \\
\cline{2-6}
 & $p_0$ & $p_1$ & $p_2$ & $p_3$ & $p_4$ & Notes \\
\hline
\claudeshort & \textbf{1.00} [0.93, 1.00] & \textbf{1.00} [0.93, 1.00] & \textbf{0.00} [0.00, 0.07] & \textbf{1.00} [0.93, 1.00] & \textbf{1.00} [0.93, 1.00] & Forgiver\\
\geminishort & \textbf{0.58} [0.44, 0.71] & \textbf{1.00} [0.93, 1.00] & \textbf{0.00} [0.00, 0.07] & \textbf{0.00} [0.00, 0.07] & \textbf{0.00} [0.00, 0.07] & $\sim$GRIM \\
\gptfourshort & \textbf{1.00} [0.93, 1.00] & \textbf{1.00} [0.93, 1.00] & \textbf{0.00} [0.00, 0.07] & \textbf{0.00} [0.00, 0.07] & \textbf{0.00} [0.00, 0.07] & GRIM \\
\gptfiveshort& \textbf{0.98} [0.90, 1.00] & \textbf{1.00} [0.93, 1.00] & \textbf{0.00} [0.00, 0.07] & \textbf{0.12} [0.06, 0.24] & \textbf{0.44} [0.31, 0.58] & $\sim$GRIM \\
\llamashort & \textbf{1.00} [0.93, 1.00] & \textbf{1.00} [0.93, 1.00] & \textbf{0.36} [0.24, 0.50] & \textbf{0.86} [0.74, 0.93] & \textbf{0.72} [0.58, 0.83] & $\sim$Forgiver \\
\hline
\end{tabular}
\end{table}

\begin{table}[h!]
\raggedright
\renewcommand{\arraystretch}{1.2}
\setlength{\tabcolsep}{6pt}
\begin{tabular}{l|c c c c c|r}
\hline
\multirow{2}{*}{LLM} & \multicolumn{5}{c|}{Memory-1 strategy, mean with Wilson interval, $n=50$, $x=\mathbf{8}$} & \multicolumn{1}{c}{} \\
\cline{2-6}
 & $p_0$ & $p_1$ & $p_2$ & $p_3$ & $p_4$ & Notes \\
\hline
\claudeshort & \textbf{1.00} [0.93, 1.00] & \textbf{1.00} [0.93, 1.00] & \textbf{0.00} [0.00, 0.07] & \textbf{1.00} [0.93, 1.00] & \textbf{1.00} [0.93, 1.00] & Forgiver \\
\geminishort & \textbf{0.48} [0.35, 0.61] & \textbf{1.00} [0.93, 1.00] & \textbf{0.00} [0.00, 0.07] & \textbf{0.00} [0.00, 0.07] & \textbf{0.00} [0.00, 0.07] & $\sim$SGRIM \\
\gptfourshort & \textbf{1.00} [0.93, 1.00] & \textbf{1.00} [0.93, 1.00] & \textbf{0.00} [0.00, 0.07] & \textbf{0.00} [0.00, 0.07] & \textbf{0.02} [0.00, 0.10] & $\sim$GRIM \\
\gptfiveshort& \textbf{0.98} [0.90, 1.00] & \textbf{0.98} [0.90, 1.00] & \textbf{0.00} [0.00, 0.07] & \textbf{0.26} [0.16, 0.40] & \textbf{0.38} [0.26, 0.52] & $\sim$GRIM\\
\llamashort & \textbf{1.00} [0.93, 1.00] & \textbf{1.00} [0.93, 1.00] & \textbf{0.44} [0.31, 0.58] & \textbf{0.98} [0.90, 1.00] & \textbf{0.58} [0.44, 0.71] & $\sim$Forgiver \\
\hline
\end{tabular}
\end{table}

\begin{table}[h!]
\raggedright
\renewcommand{\arraystretch}{1.2}
\setlength{\tabcolsep}{6pt}
\begin{tabular}{l|c c c c c|r}
\hline
\multirow{2}{*}{LLM} & \multicolumn{5}{c|}{Memory-1 strategy, mean with Wilson interval, $n=50$, $x=\mathbf{9}$} & \multicolumn{1}{c}{} \\
\cline{2-6}
 & $p_0$ & $p_1$ & $p_2$ & $p_3$ & $p_4$ & Notes \\
\hline
\claudeshort & \textbf{1.00} [0.93, 1.00] & \textbf{1.00} [0.93, 1.00] & \textbf{0.00} [0.00, 0.07] & \textbf{1.00} [0.93, 1.00] & \textbf{1.00} [0.93, 1.00] & Forgiver \\
\geminishort & \textbf{0.10} [0.04, 0.21] & \textbf{1.00} [0.93, 1.00] & \textbf{0.00} [0.00, 0.07] & \textbf{0.00} [0.00, 0.07] & \textbf{0.00} [0.00, 0.07] & $\sim$SGRIM \\
\gptfourshort & \textbf{0.96} [0.87, 0.99] & \textbf{1.00} [0.93, 1.00] & \textbf{0.00} [0.00, 0.07] & \textbf{0.00} [0.00, 0.07] & \textbf{0.00} [0.00, 0.07] & $\sim$GRIM \\
\gptfiveshort& \textbf{0.96} [0.87, 0.99] & \textbf{1.00} [0.93, 1.00] & \textbf{0.00} [0.00, 0.07] & \textbf{0.22} [0.13, 0.35] & \textbf{0.48} [0.35, 0.61] & $\sim$GRIM \\
\llamashort & \textbf{1.00} [0.93, 1.00] & \textbf{1.00} [0.93, 1.00] & \textbf{0.34} [0.22, 0.48] & \textbf{0.66} [0.52, 0.78] & \textbf{0.38} [0.26, 0.52] & $\sim$GTFT \\
\hline
\end{tabular}
\end{table}

\begin{table}[h!]
\raggedright
\renewcommand{\arraystretch}{1.2}
\setlength{\tabcolsep}{6pt}
\begin{tabular}{l|c c c c c|r}
\hline
\multirow{2}{*}{LLM} & \multicolumn{5}{c|}{Memory-1 strategy, mean with Wilson interval, $n=50$, $x=\mathbf{10}$} & \multicolumn{1}{c}{} \\
\cline{2-6}
 & $p_0$ & $p_1$ & $p_2$ & $p_3$ & $p_4$ & Notes \\
\hline
\claudeshort & \textbf{1.00} [0.93, 1.00] & \textbf{1.00} [0.93, 1.00] & \textbf{0.00} [0.00, 0.07] & \textbf{1.00} [0.93, 1.00] & \textbf{0.00} [0.00, 0.07] & TFT \\
\geminishort & \textbf{0.28} [0.17, 0.42] & \textbf{0.92} [0.81, 0.97] & \textbf{0.00} [0.00, 0.07] & \textbf{0.00} [0.00, 0.07] & \textbf{0.00} [0.00, 0.07] & $\sim$SGRIM \\
\gptfourshort & \textbf{1.00} [0.93, 1.00] & \textbf{1.00} [0.93, 1.00] & \textbf{0.00} [0.00, 0.07] & \textbf{0.00} [0.00, 0.07] & \textbf{0.00} [0.00, 0.07] & GRIM \\
\gptfiveshort& \textbf{0.16} [0.08, 0.29] & \textbf{0.38} [0.26, 0.52] & \textbf{0.00} [0.00, 0.07] & \textbf{0.02} [0.00, 0.10] & \textbf{0.00} [0.00, 0.07] & ALLD \\
\llamashort & \textbf{1.00} [0.93, 1.00] & \textbf{1.00} [0.93, 1.00] & \textbf{0.94} [0.84, 0.98] & \textbf{0.72} [0.58, 0.83] & \textbf{0.26} [0.16, 0.40] & 11110 \\
\hline
\end{tabular}
\caption{Equal-gains experiments: The numbers in bold represent the frequency of `L' play in 50 trials in that particular scenario (column) and the numbers inside the square bracket represent the Wilson interval of the data. Eleven tables correspond to eleven values of $x$ (the integers from 0 to 10). SGRIM refers to Suspicious GRIM that starts with defection and only cooperates if there is mutual cooperation in the last round.}

\end{table}

\clearpage

\subsection*{Estimating memory-2 strategies from Memory-2 experiments}

\begin{table}[h!]
\raggedright
\renewcommand{\arraystretch}{1.2}
\setlength{\tabcolsep}{6pt}
\begin{tabular}{l|c c c c c}
\hline
\multirow{2}{*}{LLM} & \multicolumn{5}{c}{Round 1 and 2 cooperation probability, mean with Wilson interval, $n=50$} \\
\cline{2-6}
 & Round 1 & Round 2 (R$_1$: $\LL$) & Round 2 (R$_1$: $\LR$) & Round 2 (R$_1$: $\RL$) & Round 2 (R$_2$: $\RR$) \\
\hline
\claudeshort & \textbf{1.00} [0.93, 1.00] & \textbf{1.00} [0.93, 1.00] & \textbf{0.00} [0.00, 0.07] & \textbf{1.00} [0.93, 1.00] & \textbf{1.00} [0.93, 1.00] \\
\geminishort & \textbf{0.90} [0.79, 0.96] & \textbf{1.00} [0.93, 1.00] & \textbf{0.00} [0.00, 0.07] & \textbf{0.02} [0.00, 0.10] & \textbf{0.08} [0.03, 0.19] \\
\gptfourshort & \textbf{1.00} [0.93, 1.00] & \textbf{1.00} [0.93, 1.00] & \textbf{0.00} [0.00, 0.07] & \textbf{0.00} [0.00, 0.07] & \textbf{0.00} [0.00, 0.07]\\
\gptfiveshort& \textbf{1.00} [0.93, 1.00] & \textbf{1.00} [0.93, 1.00] & \textbf{0.00} [0.00, 0.07] & \textbf{0.58} [0.44, 0.71] & \textbf{0.92} [0.81, 0.97] \\
\llamashort & \textbf{1.00} [0.93, 1.00] & \textbf{1.00} [0.93, 1.00] & \textbf{0.06} [0.02, 0.16] & \textbf{1.00} [0.93, 1.00] & \textbf{1.00} [0.93, 1.00] \\
\hline
\end{tabular}
\end{table}

\begin{table}[h!]
\raggedright
\renewcommand{\arraystretch}{1.2}
\setlength{\tabcolsep}{6pt}
\begin{tabular}{l|c c c c }
\hline
\multirow{2}{*}{LLM} & \multicolumn{4}{c}{Round $k$ ($k>2$) with $\mathbf{R_{k-1}: \mathbf{LL}}$, mean with Wilson interval, $n=50$} \\
\cline{2-5}
 & R$_{k-2}$: $\LL$ & R$_{k-2}$: $\LR$ & R$_{k-2}$: $\RL$ & R$_{k-2}$: $\RR$ \\
\hline
\claudeshort & \textbf{1.00} [0.93, 1.00] & \textbf{0.00} [0.00, 0.07] & \textbf{1.00} [0.93, 1.00] & \textbf{1.00} [0.93, 1.00] \\
\geminishort & \textbf{1.00} [0.93, 1.00] & \textbf{0.94} [0.84, 0.98] & \textbf{0.96} [0.87, 0.99] & \textbf{1.00} [0.93, 1.00] \\
\gptfourshort & \textbf{1.00} [0.93, 1.00] & \textbf{0.86} [0.74, 0.93] & \textbf{1.00} [0.93, 1.00] & \textbf{1.00} [0.93, 1.00] \\
\gptfiveshort& \textbf{1.00} [0.93, 1.00] & \textbf{1.00} [0.93, 1.00] & \textbf{0.92} [0.81, 0.97] & \textbf{1.00} [0.93, 1.00] \\
\llamashort & \textbf{1.00} [0.93, 1.00] & \textbf{1.00} [0.93, 1.00] & \textbf{1.00} [0.93, 1.00] & \textbf{1.00} [0.93, 1.00] \\
\hline
\end{tabular}
\end{table}

\begin{table}[h!]
\raggedright
\renewcommand{\arraystretch}{1.2}
\setlength{\tabcolsep}{6pt}
\begin{tabular}{l|c c c c }
\hline
\multirow{2}{*}{LLM} & \multicolumn{4}{c}{Round $k$ ($k>2$) with $\mathbf{R_{k-1}: \mathbf{LR}}$, mean with Wilson interval, $n=50$} \\
\cline{2-5}
 & R$_{k-2}$: $\LL$ & R$_{k-2}$: $\LR$ & R$_{k-2}$: $\RL$ & R$_{k-2}$: $\RR$ \\
\hline
\claudeshort & \textbf{0.00} [0.00, 0.07] & \textbf{0.00} [0.00, 0.07] & \textbf{0.00} [0.00, 0.07] & \textbf{0.00} [0.00, 0.07] \\
\geminishort & \textbf{0.00} [0.00, 0.07] & \textbf{0.00} [0.00, 0.07] & \textbf{0.04} [0.01, 0.13] & \textbf{0.00} [0.00, 0.07] \\
\gptfourshort & \textbf{0.00} [0.00, 0.07] & \textbf{0.00} [0.00, 0.07] & \textbf{0.00} [0.00, 0.07] & \textbf{0.00} [0.00, 0.07] \\
\gptfiveshort& \textbf{0.00} [0.00, 0.07] & \textbf{0.00} [0.00, 0.07] & \textbf{0.92} [0.81, 0.97] & \textbf{0.04} [0.01, 0.13] \\
\llamashort & \textbf{0.76} [0.63, 0.86] & \textbf{0.00} [0.00, 0.07] & \textbf{1.00} [0.93, 1.00] & \textbf{1.00} [0.93, 1.00] \\
\hline
\end{tabular}
\end{table}

\begin{table}[h!]
\raggedright
\renewcommand{\arraystretch}{1.2}
\setlength{\tabcolsep}{6pt}
\begin{tabular}{l|c c c c }
\hline
\multirow{2}{*}{LLM} & \multicolumn{4}{c}{Round $k$ ($k>2$) with $\mathbf{R_{k-1}: \mathbf{RL}}$, mean with Wilson interval, $n=50$} \\
\cline{2-5}
 & R$_{k-2}$: $\LL$ & R$_{k-2}$: $\LR$ & R$_{k-2}$: $\RL$ & R$_{k-2}$: $\RR$ \\
\hline
\claudeshort & \textbf{0.00} [0.00, 0.07] & \textbf{1.00} [0.93, 1.00] & \textbf{0.00} [0.00, 0.07] & \textbf{0.00} [0.00, 0.07] \\
\geminishort & \textbf{0.00} [0.00, 0.07] & \textbf{0.06} [0.02, 0.16] & \textbf{0.00} [0.00, 0.07] & \textbf{0.00} [0.00, 0.07] \\
\gptfourshort & \textbf{0.00} [0.00, 0.07] & \textbf{0.00} [0.00, 0.07] & \textbf{0.00} [0.00, 0.07] & \textbf{0.00} [0.00, 0.07] \\
\gptfiveshort& \textbf{0.64} [0.50, 0.76] & \textbf{0.50} [0.37, 0.63] & \textbf{0.02} [0.00, 0.10] & \textbf{0.16} [0.08, 0.29] \\
\llamashort & \textbf{1.00} [0.93, 1.00] & \textbf{1.00} [0.93, 1.00] & \textbf{0.18} [0.10, 0.31] & \textbf{1.00} [0.93, 1.00] \\
\hline
\end{tabular}
\end{table}

\begin{table}[h!]
\raggedright
\renewcommand{\arraystretch}{1.2}
\setlength{\tabcolsep}{6pt}
\begin{tabular}{l|c c c c }
\hline
\multirow{2}{*}{LLM} & \multicolumn{4}{c}{Round $k$ ($k>2$) with $\mathbf{R_{k-1}: \mathbf{RR}}$, mean with Wilson interval, $n=50$} \\
\cline{2-5}
 & R$_{k-2}$: $\LL$ & R$_{k-2}$: $\LR$ & R$_{k-2}$: $\RL$ & R$_{k-2}$: $\RR$ \\
\hline
\claudeshort & \textbf{1.00} [0.93, 1.00] & \textbf{0.00} [0.00, 0.07] & \textbf{1.00} [0.93, 1.00] & \textbf{1.00} [0.93, 1.00] \\
\geminishort & \textbf{0.58} [0.44, 0.71] & \textbf{0.00} [0.00, 0.07] & \textbf{0.00} [0.00, 0.07] & \textbf{0.00} [0.00, 0.07] \\
\gptfourshort & \textbf{0.08} [0.03, 0.19] & \textbf{0.00} [0.00, 0.07] & \textbf{0.00} [0.00, 0.07] & \textbf{0.00} [0.00, 0.07] \\
\gptfiveshort& \textbf{0.96} [0.87, 0.99] & \textbf{0.32} [0.21, 0.46] & \textbf{0.84} [0.71, 0.92] & \textbf{0.62} [0.48, 0.74] \\
\llamashort & \textbf{1.00} [0.93, 1.00] & \textbf{1.00} [0.93, 1.00] & \textbf{1.00} [0.93, 1.00] & \textbf{1.00} [0.93, 1.00] \\
\hline
\end{tabular}
\caption{Memory-2 experiments: We present cooperation frequency in round one and round two (first table) followed by cooperation frequency in the later rounds (in the four tables that follow). The later round are grouped by the outcome of the previous round, and within each table the columns correspond to the outcome in the second-to-last round. For each scenario, we report the cooperation probability as the mean number of times `L' is played across 50 trials, together with the corresponding Wilson intervals.}
\end{table}
\clearpage
\subsection*{One-shot game strategies}

\begin{table}[h!]
\raggedright
\renewcommand{\arraystretch}{1.2}
\setlength{\tabcolsep}{8pt}
\begin{tabular}{r|c|l}
\hline
LLM & One-shot game cooperation probability  &  Notes\\
\hline
\claudeshort & \textbf{0.00} [0.00, 0.07]& ALLD\\
\geminishort & \textbf{0.16} [0.08, 0.29]& 16\% cooperation\\
\gptfourshort & \textbf{0.00} [0.00, 0.07]& ALLD\\
\gptfiveshort& \textbf{0.00} [0.00, 0.07]& ALLD\\
\llamashort & \textbf{1.00} [0.93, 1.00] & ALLC\\
\hline
\end{tabular}
\vspace{0.5cm}
\caption{One-shot experiment. Cooperation probability in the one-shot game estimated as the frequency of playing `L' in 50 trials of the experiment (in bold). Numbers in bracket represent the Wilson interval for the data. }
\end{table}

\newpage
\subsection{Nash equilibrium strategies, Partners and Rivals among the LLM strategies}

\noindent In this section we analyze whether the LLM strategies estimated from various experiments are Nash equilibrium strategies. In addition we also categorize them into the classes: Partners, Rivals, both or none. We define these terms below.\\

\noindent In these definitions, we use $\pi_{w,\mathrm{g}}(\sigma_1,\sigma_2)$ to denote the expected payoff of the repeated strategy $\sigma_1$ against the repeated game strategy $\sigma_2$ when the repeated game with the stage game $g=(a_\LL,a_\LR,a_\RL,a_\RR)$ has a stopping probability $w$. In Section \ref{Section:Methods} we explain in detail how to compute $\pi$ for memory-1 and memory-2 strategies, including the case $w=0$. \\

\begin{definition}[Nash equilibrium strategy] A repeated game strategy $\sigma$ is a Nash equilibrium for the repeated game with stage game $g$ and stopping probability $w$ if and only if for all strategies $\sigma'$,

\begin{equation}
    \pi_{w,g}(\sigma,\sigma) \geqslant \pi_{w,g}(\sigma',\sigma).
\end{equation}
\label{def:NE}
\end{definition}

\noindent To determine whether an arbitrary memory-$n$ strategy, $\sigma$ is a Nash equilibrium strategy, we use the following useful result. 

\begin{lemma}[Lev{\'\i}nsk{\`y} \textit{et al.} \cite{Levinsky:IJGT:2020}; Sufficient to check just the deviations to deterministic memory-$n$ strategies] A memory-$n$ strategy $\sigma$ is a Nash equilibrium strategy of repeated game with stage game $g$ and stopping probability $w$ if and only if for all strategies $\sigma' \in \mathrm{DetM}n$,

\begin{equation}
    \pi_{w,g}(\sigma,\sigma) \geqslant \pi_{w,g}(\sigma',\sigma).
\end{equation}
\\
\noindent Where $\mathrm{DetM}n$ is the set of all deterministic memory-$n$ strategies.
\label{lemma:levinsky}
\end{lemma}

\noindent Therefore, to check whether a memory-$1$ strategy is a Nash equilibrium, we need to perform $2^5=32$ payoff comparisons and to check whether a memory-$2$ strategy is a Nash equilibrium, we need to perform $2^{21}$ payoff comparisons.

\begin{definition}[Partner in an repeated Prisoner's dilemma, Hilbe \textit{et al.} \cite{Hilbe:NHB:2018}] Consider an repeated prisoner's dilemma with the stage game $g=(a_\LL,a_\LR,a_\RL,a_\RR)$ that satisfies $a_\RL > a_\LL > a_\RR > a_\LR$ and $2 a_\LL > a_\RL + a_\LR$. The repeated game strategy $\sigma$ is called a \textit{partner} in this game if and only if 

\begin{enumerate}
    \item $\sigma$ is a Nash equilibrium strategy of the game.
    \item $\pi_{w,g}(\sigma,\sigma) = a_\LL$.
\end{enumerate}
\label{def:Partner}
\end{definition}

\begin{definition}[Rival in an repeated Prisoner's dilemma, Hilbe \textit{et al.} \cite{Hilbe:NHB:2018}] Consider an repeated prisoner's dilemma with the stage game $g=(a_\LL,a_\LR,a_\RL,a_\RR)$ that satisfies $a_\RL > a_\LL > a_\RR > a_\LR$ and $2 a_\LL > a_\RL + a_\LR$. The strategy $\sigma$ is called a \textit{rival} if and only if for all strategies $\sigma'$

\begin{equation}
    \pi_{w,g}(\sigma,\sigma') \geqslant \pi_{w,g}(\sigma',\sigma).
    \label{equation:rival}
\end{equation}
\\
\noindent Using Lev{\'\i}nsk{\`y} \textit{et al.} \cite{Levinsky:IJGT:2020}, a memory-$n$ strategy $\sigma$ can be labeled ``rival" if it satisfies Eq. (\ref{equation:rival}) for all $\sigma' \in \mathrm{DetM}n$.
\end{definition}

\noindent \textbf{Nash equilibrium, Partners and Rivals among LLM strategies in the original experiment.} We provide two tables for the repeated game with stage-game payoffs $g=(3,0,5,1)$. The first corresponds to the case $w=0$, and the second to $w = 0.01$. In each table, we indicate for every LLM strategy under each framing whether it is a Nash equilibrium strategy (the column N), and whether it qualifies as a Partner (the column P) or a Rival (the column R). A strategy can be both a partner and a rival. \\

\begin{table}[h!]
\centering
\renewcommand{\arraystretch}{1.2}
\setlength{\tabcolsep}{4pt}

\begin{tabular}{r|
ccc|ccc|ccc|ccc|ccc}
\hline
 & \multicolumn{3}{c|}{\claudeshort} 
 & \multicolumn{3}{c|}{\geminishort} 
 & \multicolumn{3}{c|}{\gptfourshort} 
 & \multicolumn{3}{c|}{\gptfiveshort} 
 & \multicolumn{3}{c}{\llamashort} \\
\cline{2-16}
Framing 
 & N & P & R
 & N & P & R
 & N & P & R
 & N & P & R
 & N & P & R \\
\hline

Try to get more points than the other agent.
 & \checkmark &  & \checkmark
 & \checkmark &  & \checkmark
 & \checkmark &  & \checkmark
 & \checkmark &  & \checkmark
 & \checkmark & \checkmark &  \\

Exploit the other agent as much as possible.
 & \checkmark &  & \checkmark
 & \checkmark &  & \checkmark
 & \checkmark &  & \checkmark
 & \checkmark &  & \checkmark
 & \checkmark &  & \checkmark \\

Maximize your own number of points.
 & \checkmark &  & \checkmark
 & \checkmark &  & \checkmark
 & \checkmark &  & \checkmark
 & \checkmark &  \checkmark & 
 & \checkmark & \checkmark & \checkmark \\

Think about winning.
 & \checkmark &  & \checkmark
 &\multicolumn{2}{c}{\makebox[0pt][c]{0.46\%}}&
 & \checkmark &  & \checkmark
 & \checkmark & \checkmark & 
 & \checkmark & \checkmark & \checkmark \\

 &&&
 &&&
 &&&
 &&&
 &&& \\

 \textit{no special instruction}
 & \checkmark & \checkmark & 
 & \checkmark & \checkmark & 
 & \checkmark & \checkmark & \checkmark
 & \checkmark & \checkmark & 
 & \checkmark & \checkmark & \\ 
 
Play like a pro.
 & \checkmark & \checkmark & 
 & \checkmark & \checkmark & 
 & \checkmark & \checkmark & 
 & \checkmark & \checkmark & 
 & \checkmark & \checkmark & \checkmark\\ 

 &&&
 &&&
 &&&
 &&&
 &&& \\
 
Accumulate many points, but do not exploit.
 & \multicolumn{2}{c}{\makebox[0pt][c]{100.0\%}}&
 & \multicolumn{2}{c}{\makebox[0pt][c]{10.44\%}}&
 & \multicolumn{2}{c}{\makebox[0pt][c]{100.0\%}}&
 & \multicolumn{2}{c}{\makebox[0pt][c]{100.0\%}}&
 & \multicolumn{2}{c}{\makebox[0pt][c]{19.59\%}}\\

Maximize your points, keep in mind the other's welfare.
 & \checkmark & \checkmark & 
 & \multicolumn{2}{c}{\makebox[0pt][c]{100.0\%}}&
 &  \multicolumn{2}{c}{\makebox[0pt][c]{100.0\%}}&
 & \checkmark & \checkmark & 
 & \checkmark & \checkmark &  \\

Think about fair outcomes.
 & \checkmark & \checkmark & 
 &  \multicolumn{2}{c}{\makebox[0pt][c]{100.0\%}}&
 & \multicolumn{2}{c}{\makebox[0pt][c]{14.84\%}}&
 & \multicolumn{2}{c}{\makebox[0pt][c]{6.16\%}}&
 & \checkmark & \checkmark &  \\

Be a saint.
 & \multicolumn{2}{c}{\makebox[0pt][c]{100.0\%}}&
 &\multicolumn{2}{c}{\makebox[0pt][c]{100.0\%}}&
 & \multicolumn{2}{c}{\makebox[0pt][c]{100.0\%}}&
 & \multicolumn{2}{c}{\makebox[0pt][c]{100.0\%}}&
 & \checkmark & \checkmark &  \\

\hline
\end{tabular}
\vspace{0.4cm}
\caption{\textbf{Are the memory-1 strategies from the original experiment Nash equilibria, partners, or rivals?}
For each LLM’s memory-1 strategy from the original experiment, we determine whether it is a Nash equilibrium strategy (column N) and whether it acts as a partner (column P) or a rival (column R)in the repeated game with stage-game payoffs $g = (3,0,5,1)$ for $w=0$. A tick means ``yes". Readers will note that a strategy is not a partner if it is not a Nash equilibrium (this is by definition). Additionally, if the strategy is not a Nash equilibrium, we report the percentage of opponents, out of $10^6$ randomly selected memory-1 strategies, that beat it in the its payoff against itself. This gives a sense of how ``close" a strategy is to being a Nash strategy. This number is 0\% for Nash strategies.}
\end{table}

\clearpage

\begin{table}[h!]
\centering
\renewcommand{\arraystretch}{1.2}
\setlength{\tabcolsep}{4pt}

\begin{tabular}{r|
ccc|ccc|ccc|ccc|ccc}
\hline
 & \multicolumn{3}{c|}{\claudeshort} 
 & \multicolumn{3}{c|}{\geminishort} 
 & \multicolumn{3}{c|}{\gptfourshort} 
 & \multicolumn{3}{c|}{\gptfiveshort} 
 & \multicolumn{3}{c}{\llamashort} \\
\cline{2-16}
Framing 
 & N & P & R
 & N & P & R
 & N & P & R
 & N & P & R
 & N & P & R \\
\hline

Try to get more points than the other agent.
 & \checkmark &  & \checkmark
 & \checkmark &  & \checkmark
 & \checkmark &  & \checkmark
 &  \multicolumn{2}{c}{\makebox[0pt][c]{$<$0.01\%}}& 
 & \checkmark & \checkmark &  \\

Exploit the other agent as much as possible.
 & \checkmark &  & \checkmark
 & \checkmark &  & \checkmark
 & \checkmark &  & \checkmark
 &  \multicolumn{2}{c}{\makebox[0pt][c]{$<$0.01\%}}& 
 &  \multicolumn{2}{c}{\makebox[0pt][c]{$<$0.01\%}}& \\

Maximize your own number of points.
 & \checkmark &  & \checkmark
 & \checkmark &  & \checkmark
 & \checkmark &  & \checkmark
 & \checkmark &  \checkmark & 
 & \checkmark & \checkmark &  \\

Think about winning.
 & \checkmark &  & \checkmark
 & \multicolumn{2}{c}{\makebox[0pt][c]{0.44\%}}& 
 & \checkmark &  & \checkmark
 &  \multicolumn{2}{c}{\makebox[0pt][c]{$<$0.01\%}}& 
 & \checkmark & \checkmark & \\

 &&&
 &&&
 &&&
 &&&
 &&& \\

 \textit{no special instruction}
 &  \multicolumn{2}{c}{\makebox[0pt][c]{2.01\%}}&
 &  \multicolumn{2}{c}{\makebox[0pt][c]{$<$0.01\%}}& 
 & \checkmark & \checkmark & 
 &  \multicolumn{2}{c}{\makebox[0pt][c]{0.82\%}}& 
 & \checkmark & \checkmark & \\ 
 
Play like a pro.
 &  \multicolumn{2}{c}{\makebox[0pt][c]{0.41\%}}& 
 &  \multicolumn{2}{c}{\makebox[0pt][c]{0.03\%}}&
 &  \multicolumn{2}{c}{\makebox[0pt][c]{$<$0.01\%}}&
 & \checkmark & \checkmark & 
 & \checkmark & \checkmark & \\ 

 &&&
 &&&
 &&&
 &&&
 &&& \\
 
Accumulate many points, but do not exploit.
 & \multicolumn{2}{c}{\makebox[0pt][c]{100\%}}&
 &  \multicolumn{2}{c}{\makebox[0pt][c]{10.4\%}}&
 & \multicolumn{2}{c}{\makebox[0pt][c]{100\%}}&
 & \multicolumn{2}{c}{\makebox[0pt][c]{100\%}}&
 & \multicolumn{2}{c}{\makebox[0pt][c]{19.59\%}} \\

Maximize your points, keep in mind the other's welfare.
 &  \multicolumn{2}{c}{\makebox[0pt][c]{2.02\%}}& 
 &  \multicolumn{2}{c}{\makebox[0pt][c]{100\%}}&
 &  \multicolumn{2}{c}{\makebox[0pt][c]{100\%}}&
 &  \multicolumn{2}{c}{\makebox[0pt][c]{2.02\%}}& 
 &  \multicolumn{2}{c}{\makebox[0pt][c]{1.36\%}}\\

Think about fair outcomes.
 &  \multicolumn{2}{c}{\makebox[0pt][c]{2.03\%}}& 
 &  \multicolumn{2}{c}{\makebox[0pt][c]{100\%}}& 
 &  \multicolumn{2}{c}{\makebox[0pt][c]{14.96\%}}&  
 &  \multicolumn{2}{c}{\makebox[0pt][c]{6.17\%}}&  
 &  \multicolumn{2}{c}{\makebox[0pt][c]{1.91\%}} \\

Be a saint.
 & \multicolumn{2}{c}{\makebox[0pt][c]{100\%}}&  
 & \multicolumn{2}{c}{\makebox[0pt][c]{100\%}}&
 & \multicolumn{2}{c}{\makebox[0pt][c]{100\%}}& 
 & \multicolumn{2}{c}{\makebox[0pt][c]{100\%}}& 
 & \multicolumn{2}{c}{\makebox[0pt][c]{1.91\%}}\\

\hline
\end{tabular}
\vspace{0.4cm}
\caption{\textbf{Are the memory-1 strategies from the original experiment Nash equilibria, partners, or rivals?}
Recreating the previous table but for $g = (3,0,5,1)$ and $w=0.01$. Readers will note that a strategy is not a partner (by definition) if it is not a Nash equilibrium. Additionally, if the strategy is not a Nash equilibrium, we report the percentage of opponents, out of $10^6$ randomly selected memory-1 strategies, that beat it in the payoff it earns against itself. }
\end{table}

\noindent \textbf{Nash equilibrium, Partners and Rivals among LLM strategies in the stopping probability experiments.} The table below summarizes the properties of the LLM strategies from the stopping-probability experiments for the repeated game with stage-game payoffs $g=(3,0,5,1)$. The four rows correspond to the four stopping-probability treatments, all conducted under the baseline framing with no special instruction. When determining whether an LLM strategy is a Nash equilibrium, and whether it behaves as a partner or a rival, we evaluate it in the repeated game using the corresponding stopping probability.\\

\begin{table}[h!]
\raggedright
\renewcommand{\arraystretch}{1.2}
\setlength{\tabcolsep}{6pt}

\begin{tabular}{r|
ccc|ccc|ccc|ccc|ccc}
\hline
 & \multicolumn{3}{c|}{\claudeshort} 
 & \multicolumn{3}{c|}{\geminishort} 
 & \multicolumn{3}{c|}{\gptfourshort} 
 & \multicolumn{3}{c|}{\gptfiveshort} 
 & \multicolumn{3}{c}{\llamashort} \\
\cline{2-16}
Stopping probability $w$ in prompt
 & N & P & R
 & N & P & R
 & N & P & R
 & N & P & R
 & N & P & R \\
\hline

$0.01$
 &\multicolumn{2}{c}{\makebox[0pt][c]{2.03\%}}&
 &\multicolumn{2}{c}{\makebox[0pt][c]{$<$0.01\%}}& 
 &\checkmark & \checkmark & 
 &\multicolumn{2}{c}{\makebox[0pt][c]{1.48\%}}&
 &\checkmark & \checkmark&  \\

$0.1$
 &\multicolumn{2}{c}{\makebox[0pt][c]{22.33\%}}& 
 &\multicolumn{2}{c}{\makebox[0pt][c]{0.05\%}}& 
 &\checkmark & \checkmark & 
 &\multicolumn{2}{c}{\makebox[0pt][c]{12.95\%}}& 
 &\checkmark & \checkmark&  \\

$0.2$
 &\multicolumn{2}{c}{\makebox[0pt][c]{50.02\%}}& 
 &\multicolumn{2}{c}{\makebox[0pt][c]{0.04\%}}&  
 &\checkmark & \checkmark & 
 &\multicolumn{2}{c}{\makebox[0pt][c]{33.76\%}}&  
 & \checkmark & \checkmark&  \\

$0.5$
 &\multicolumn{2}{c}{\makebox[0pt][c]{100.0\%}}&  
 &\multicolumn{2}{c}{\makebox[0pt][c]{5.44\%}}&  
 &\multicolumn{2}{c}{\makebox[0pt][c]{54.42\%}}&  
 &\multicolumn{2}{c}{\makebox[0pt][c]{73.56\%}}&
 & \checkmark & \checkmark&  \\

\hline
\end{tabular}
\vspace{0.4cm}
\caption{\textbf{Are the memory-1 strategies from the stopping probability experiments Nash equilibria, partners, or rivals?}
For each LLM’s memory-1 strategy from the stopping probability experiments, we determine whether it is a Nash equilibrium strategy (column N) and whether it acts as a partner (column P) or a rival (column R) in the repeated game with stage-game payoffs $g = (3,0,5,1)$ that has the corresponding stopping probability value, $w$. Percentages mean the same as the previous tables.}
\end{table}

\noindent \textbf{Nash equilibrium, Partners and Rivals among LLM strategies in the equal-gains experiments.} The tables below summarize the properties of the LLM strategies from the equal gains experiments. We provide two tables, the first for $w=0$ and the second for $w=0.01$. All experiments were performed for the baseline no instruction case.\\

\begin{table}[h!]
\raggedright
\renewcommand{\arraystretch}{1.2}
\setlength{\tabcolsep}{6pt}
\begin{tabular}{r|
ccc|ccc|ccc|ccc|ccc}
\hline
 & \multicolumn{3}{c|}{\claudeshort} 
 & \multicolumn{3}{c|}{\geminishort} 
 & \multicolumn{3}{c|}{\gptfourshort} 
 & \multicolumn{3}{c|}{\gptfiveshort} 
 & \multicolumn{3}{c}{\llamashort} \\
\cline{2-16}
A. Value of $x$ in $g\!=\!(10,0,10\!+\!x,x)$
 & N & P & R
 & N & P & R
 & N & P & R
 & N & P & R
 & N & P & R \\
\hline

$0$
 & \checkmark & \checkmark &  
 & \checkmark & \checkmark &  
 & \checkmark & \checkmark &  
 & \checkmark & \checkmark &  
 & \checkmark & \checkmark &  
\\

$1$
 & \checkmark & \checkmark &  
 & \checkmark & \checkmark &  
 & \checkmark & \checkmark &  
 & \checkmark & \checkmark &  
 & \checkmark & \checkmark &  
\\

$2$
 & \checkmark & \checkmark &  
 & \checkmark & \checkmark &  
 & \checkmark & \checkmark &  
 & \checkmark & \checkmark &  
 & \checkmark & \checkmark &  
\\

$3$
 & \checkmark & \checkmark &  
 & \checkmark & \checkmark &  
 & \checkmark & \checkmark &  
 & \checkmark & \checkmark &  
 & \multicolumn{2}{c}{\makebox[0pt][c]{9.81\%}}&  
\\

$4$
 & \checkmark & \checkmark &  
 & \multicolumn{2}{c}{\makebox[0pt][c]{$<$0.01\%}}& \checkmark 
 & \checkmark & \checkmark &  
 & \checkmark & \checkmark &  
 & \checkmark & \checkmark &  
\\

$5$
 & \checkmark & \checkmark &  
 & \checkmark & \checkmark &  
 & \checkmark & \checkmark & \checkmark 
 & \checkmark & \checkmark &  
 & \checkmark & \checkmark &  
\\

$6$
 & \multicolumn{2}{c}{\makebox[0pt][c]{33.32\%}}&   
 & \multicolumn{2}{c}{\makebox[0pt][c]{$<$0.01\%}}& \checkmark 
 & \checkmark & \checkmark &  
 & \checkmark & \checkmark &  
 & \multicolumn{2}{c}{\makebox[0pt][c]{15.82\%}}&
\\

$7$
 & \multicolumn{2}{c}{\makebox[0pt][c]{57.13\%}}&  
 & \multicolumn{2}{c}{\makebox[0pt][c]{$<$0.01\%}} & \checkmark 
 & \checkmark & \checkmark & \checkmark 
 & \multicolumn{2}{c}{\makebox[0pt][c]{$<$0.01\%}}& 
 & \multicolumn{2}{c}{\makebox[0pt][c]{95.82\%}}&  
\\

$8$
 & \multicolumn{2}{c}{\makebox[0pt][c]{75.06\%}}&   
 & \multicolumn{2}{c}{\makebox[0pt][c]{$<$0.01\%}} & \checkmark 
 & \checkmark & \checkmark &  
 & \multicolumn{2}{c}{\makebox[0pt][c]{5.13\%}} &  
 & \multicolumn{2}{c}{\makebox[0pt][c]{100.0\%}} &  
\\

$9$
 &  \multicolumn{2}{c}{\makebox[0pt][c]{88.88\%}}&  
 &  \multicolumn{2}{c}{\makebox[0pt][c]{$<$0.01\%}}& \checkmark 
 &  \multicolumn{2}{c}{\makebox[0pt][c]{$<$0.01\%}}& \checkmark 
 &  \multicolumn{2}{c}{\makebox[0pt][c]{12.18\%}}&  
 &  \multicolumn{2}{c}{\makebox[0pt][c]{94.30\%}}&  
\\

$10$
 & \checkmark & \checkmark & \checkmark 
 & \checkmark & \checkmark & \checkmark 
 & \checkmark & \checkmark & \checkmark 
 & \checkmark & \checkmark & \checkmark 
 & \multicolumn{2}{c}{\makebox[0pt][c]{100.0\%}}&  
\\
\hline
\end{tabular}
\vspace{0.2cm}
\end{table}

\begin{table}[h!]
\raggedright
\renewcommand{\arraystretch}{1.2}
\setlength{\tabcolsep}{6pt}
\begin{tabular}{r|
ccc|ccc|ccc|ccc|ccc}
\hline
 & \multicolumn{3}{c|}{\claudeshort} 
 & \multicolumn{3}{c|}{\geminishort} 
 & \multicolumn{3}{c|}{\gptfourshort} 
 & \multicolumn{3}{c|}{\gptfiveshort} 
 & \multicolumn{3}{c}{\llamashort} \\
\cline{2-16}
B. Value of $x$ in $g\!=\!(10,0,10\!+\!x,x)$
 & N & P & R
 & N & P & R
 & N & P & R
 & N & P & R
 & N & P & R \\
\hline
$0$
 & \checkmark & \checkmark &  
 & \checkmark & \checkmark &  
 & \checkmark & \checkmark &  
 & \checkmark & \checkmark &  
 & \checkmark & \checkmark &  
\\

$1$
 & \checkmark & \checkmark &  
 & \multicolumn{2}{c}{\makebox[0pt][c]{$<$0.01\%}} &  
 & \checkmark & \checkmark &  
 & \checkmark & \checkmark &  
 & \checkmark & \checkmark &  
\\

$2$
 & \checkmark & \checkmark &  
 & \multicolumn{2}{c}{\makebox[0pt][c]{$<$0.01\%}} &  
 & \checkmark & \checkmark &  
 & \multicolumn{2}{c}{\makebox[0pt][c]{$<$0.01\%}} &  
 & \checkmark & \checkmark &  
\\

$3$
 & \checkmark & \checkmark &  
 & \checkmark & \checkmark &  
 & \checkmark & \checkmark &  
 & \checkmark & \checkmark &  
 & \multicolumn{2}{c}{\makebox[0pt][c]{11.28\%}} &  
\\

$4$
 & \checkmark & \checkmark &  
 & \multicolumn{2}{c}{\makebox[0pt][c]{0.01\%}} &  
 & \checkmark & \checkmark &  
 & \checkmark & \checkmark &  
 & \checkmark & \checkmark &  
\\

$5$
 & \multicolumn{2}{c}{\makebox[0pt][c]{1.02\%}} &  
 & \multicolumn{2}{c}{\makebox[0pt][c]{0.02\%}} &  
 & \checkmark & \checkmark &  
 & \checkmark & \checkmark &  
 & \checkmark & \checkmark &  
\\

$6$
 & \multicolumn{2}{c}{\makebox[0pt][c]{34.33\%}} &  
 & \multicolumn{2}{c}{\makebox[0pt][c]{$<$0.01\%}} &  
 & \checkmark & \checkmark &  
 & \checkmark & \checkmark &  
 & \multicolumn{2}{c}{\makebox[0pt][c]{18.74\%}} &  
\\

$7$
 & \multicolumn{2}{c}{\makebox[0pt][c]{58.21\%}}&  
 & \multicolumn{2}{c}{\makebox[0pt][c]{$<$0.01\%}}&   
 & \checkmark & \checkmark &  
 & \multicolumn{2}{c}{\makebox[0pt][c]{0.02\%}}&  
 & \multicolumn{2}{c}{\makebox[0pt][c]{97.27\%}}&  
\\

$8$
 & \multicolumn{2}{c}{\makebox[0pt][c]{76.00\%}}&  
 & \multicolumn{2}{c}{\makebox[0pt][c]{$<$0.01\%}}&  
 & \checkmark & \checkmark &  
 & \multicolumn{2}{c}{\makebox[0pt][c]{5.28\%}}&  
 & \multicolumn{2}{c}{\makebox[0pt][c]{100.0\%}}&  
\\

$9$
 & \multicolumn{2}{c}{\makebox[0pt][c]{89.88\%}}& 
 & \multicolumn{2}{c}{\makebox[0pt][c]{$<$0.01\%}}& 
 & \multicolumn{2}{c}{\makebox[0pt][c]{$<$0.01\%}}& 
 & \multicolumn{2}{c}{\makebox[0pt][c]{13.19\%}}&   
 & \multicolumn{2}{c}{\makebox[0pt][c]{95.97\%}}&  
\\

$10$
 &  \multicolumn{2}{c}{\makebox[0pt][c]{100.0\%}}&  
 &  \multicolumn{2}{c}{\makebox[0pt][c]{$<$0.01\%}}&  
 &  \multicolumn{2}{c}{\makebox[0pt][c]{0.17\%}}&  
 &  \multicolumn{2}{c}{\makebox[0pt][c]{$<$0.01\%}}&
 &  \multicolumn{2}{c}{\makebox[0pt][c]{100.0\%}}& 
\\
\hline
\end{tabular}
\vspace{0.2cm}
\caption{\textbf{Are the memory-1 strategies from the equal-gains experiments Nash equilibria, partners, or rivals?}
We determine whether LLM strategies in equal gains from switching experiments are Nash, partners or rivals in $g = (10,0,10\!+\!x,x)$. The first table (A) is for the repeated game with stopping probability $w=0$. The second one (table B) is for the repeated game with $w=0.01$. Percentages mean the same as the previous tables.}
\end{table}
\clearpage
\noindent \textbf{Nash equilibrium, Partners and Rivals among LLM strategies in the memory-2 experiments.} The properties of the LLM strategies from the Memory-2 experiments for the repeated game with stage-game payoffs $g=(3,0,5,1)$ and a stopping probability $w=10^{-10}$ are as follows: claude, gpt-4o and gpt-5 are Nash equilibrium strategies and partners. The remaining two LLM strategies are neither Nash, partners nor rivals.\\

\noindent \textbf{Comparing Nash, Partners and Rivals between original experiment and stopping probability experiments:} For four $w$ values, we show how the strategies from the original experiment (no instruction) and the strategies from the stopping probability experiments perform in being a Nash equilibrium strategy, a partner or a rival. In the table below ``OE" refers to ``Original Experiment" and ``SP" denotes to ``Stopping Probability Experiment".

\begin{table}[h!]
\raggedright
\renewcommand{\arraystretch}{1.2}
\setlength{\tabcolsep}{6pt}

\begin{tabular}{r|r|
ccc|ccc|ccc|ccc|ccc}
\hline
 & &\multicolumn{3}{c|}{\claudeshort} 
 & \multicolumn{3}{c|}{\geminishort} 
 & \multicolumn{3}{c|}{\gptfourshort} 
 & \multicolumn{3}{c|}{\gptfiveshort} 
 & \multicolumn{3}{c}{\llamashort} \\
\cline{3-17}
Stopping probability $w$ & Exp.
 & N & P & R
 & N & P & R
 & N & P & R
 & N & P & R
 & N & P & R \\
\hline

$0$ & OE
 & \checkmark & \checkmark & 
 & \checkmark & \checkmark & 
 & \checkmark & \checkmark & \checkmark
 & \checkmark & \checkmark & 
 & \checkmark & \checkmark & \\ 

&  &  & 
&  &  & 
&  &  & 
&  &  & 
&  &  &  \\

$0.01$ & OE
 &  \multicolumn{2}{c}{\makebox[0pt][c]{2.01\%}}&
 &  \multicolumn{2}{c}{\makebox[0pt][c]{$<$0.01\%}}& 
 & \checkmark & \checkmark & 
 &  \multicolumn{2}{c}{\makebox[0pt][c]{0.82\%}}& 
 & \checkmark & \checkmark & \\ 

&SP
 &\multicolumn{2}{c}{\makebox[0pt][c]{2.03\%}}&
 &\multicolumn{2}{c}{\makebox[0pt][c]{$<$0.01\%}}& 
 &\checkmark & \checkmark & 
 &\multicolumn{2}{c}{\makebox[0pt][c]{1.48\%}}&
 &\checkmark & \checkmark&  \\

&  &  & 
&  &  & 
&  &  & 
&  &  & 
&  &  &  \\

$0.1$ & OE
 &\multicolumn{2}{c}{\makebox[0pt][c]{22.25\%}}&
 &\multicolumn{2}{c}{\makebox[0pt][c]{0.02\%}}&
 &\checkmark & \checkmark & 
 &\multicolumn{2}{c}{\makebox[0pt][c]{9.64\%}}&
 &\checkmark & \checkmark&  \\

&SP
 &\multicolumn{2}{c}{\makebox[0pt][c]{22.33\%}}& 
 &\multicolumn{2}{c}{\makebox[0pt][c]{0.05\%}}& 
 &\checkmark & \checkmark & 
 &\multicolumn{2}{c}{\makebox[0pt][c]{12.95\%}}& 
 &\checkmark & \checkmark&  \\

&  &  & 
&  &  & 
&  &  & 
&  &  & 
&  &  &  \\

$0.2$ & OE
 &\multicolumn{2}{c}{\makebox[0pt][c]{49.97\%}}& 
 &\multicolumn{2}{c}{\makebox[0pt][c]{0.02\%}}&  
 &\checkmark & \checkmark & 
 &\multicolumn{2}{c}{\makebox[0pt][c]{23.48\%}}& 
 &\multicolumn{2}{c}{\makebox[0pt][c]{100.0\%}}&\\

&SP
 &\multicolumn{2}{c}{\makebox[0pt][c]{50.02\%}}& 
 &\multicolumn{2}{c}{\makebox[0pt][c]{0.04\%}}&  
 &\checkmark & \checkmark & 
 &\multicolumn{2}{c}{\makebox[0pt][c]{33.76\%}}&  
 & \checkmark & \checkmark&  \\

&  &  & 
&  &  & 
&  &  & 
&  &  & 
&  &  &  \\

$0.5$ & OE
 &\multicolumn{2}{c}{\makebox[0pt][c]{100.0\%}}& 
 &\multicolumn{2}{c}{\makebox[0pt][c]{31.94\%}}& 
 &\checkmark & \checkmark & 
 &\multicolumn{2}{c}{\makebox[0pt][c]{100.0\%}}& 
 &\multicolumn{2}{c}{\makebox[0pt][c]{100.0\%}}& \\

&SP
 &\multicolumn{2}{c}{\makebox[0pt][c]{100.0\%}}&  
 &\multicolumn{2}{c}{\makebox[0pt][c]{5.44\%}}&  
 &\multicolumn{2}{c}{\makebox[0pt][c]{54.42\%}}&  
 &\multicolumn{2}{c}{\makebox[0pt][c]{73.56\%}}&
 & \checkmark & \checkmark&  \\

&  &  & 
&  &  & 
&  &  & 
&  &  & 
&  &  &  \\
\hline
\end{tabular}
\vspace{0.4cm}
\caption{\textbf{Comparison of strategies from the original experiment and the stopping probability experiment with respect to Nash equilibria, partner, and rival classifications.}  
For each LLM’s memory-1 strategy from the original (no-instruction) experiment and the corresponding strategy from the stopping probability experiment, we determine whether each strategy is a Nash equilibrium (column N), a partner (column P), or a rival (column R) in the repeated game with stage-game payoffs $g = (3,0,5,1)$ and the corresponding stopping probability $w$. Percentages are interpreted as in the previous tables.}
\end{table}

\clearpage
\subsection{Tournaments}

\noindent In this section, we provide tables that summarize the strategies inferred for the LLMs in each experimental treatment and the corresponding outcome of the pairwise tournament conducted between them. In these tournaments, each repeated game has the payoff values $(a_\LL,a_\LR,a_\RL,a_\LR)=(3,0,5,1)$ and the stopping probability $w=0.01$, unless specified otherwise. 

\subsubsection{First ten memory-1 experiments}
\begin{figure}[h!]
\textbf{\textcolor{blue}{Original experiment with no framing}}\\\\\\
\textbf{A \quad Moves of five LLM models in IPD scenarios and their inferred memory-1 strategy.}\\ \\ \\
\raggedright
\renewcommand{\arraystretch}{1.2} 
\setlength{\tabcolsep}{8pt} 
\begin{tabular}{r|c c c c c|c c c c c|r}
\hline
\multirow{3}{*}{LLM model}
 & \multicolumn{5}{c|}{`L' in 50 trials}
 & \multicolumn{5}{c|}{Memory-1 strategy} 
 & \multicolumn{1}{c}{}
\\
\cline{2-11}
 & R1 & LL & LR & RL & RR
 & $p_0$ & $p_1$ & $p_2$ & $p_3$ & $p_4$& Notes \\
 \hline
\claudeshort & 50 & 50 & 0 & 50 & 50 & 1.00 & 1.00 & 0.00 & 1.00 & 1.00 & Forgiver  \\
\geminishort & 41 & 50 & 0 & 1 & 21 & 0.82 & 1.00 & 0.00 & 0.02 & 0.42 & $\sim$WSLS\\
\gptfourshort & 50 & 50 & 0 & 0 & 0 & 1.00 & 1.00 & 0.00 & 0.00 & 0.00 & GRIM \\
\gptfiveshort& 50 & 50 & 0 & 27 & 50 & 1.00 & 1.00 & 0.00 & 0.54 & 1.00 & $\sim$Forgiver\\
\llamashort & 50 & 50 & 13 & 49 & 28 & 1.00 & 1.00 & 0.26 & 0.98 & 0.56 & $\sim$GTFT  \\
\hline
\end{tabular}
\vspace{1.5cm}

\textbf{B \quad Pairwise tournament between the LLM models, represented by their memory-1 strategy.}\newline \newline \newline
\raggedright
\begin{tabular}{r|c c c c c|c|c}
\hline
LLM model& \claudeshort & \geminishort & \gptfourshort & \gptfiveshort& \llamashort&Row sum& Rank\\
\hline
\claudeshort & 3. & 2.9791 & 3. & 3. & 3. & 14.9792 & 2\\
\geminishort & 3.00000085 & 2.96 & 2.59 & 2.99992 & 2.996 & 14.5512 & 5\\
\gptfourshort & 3. & 2.85 & 3. & 3. & 3. & 14.8476 & 4\\
\bf{\gptfiveshort} & \bf{3.} & \bf{2.9792} & \bf{3.} & \bf{3.} & \bf{3.} & \bf{14.9793} & \bf{1}\\
\llamashort & 3. & 2.97 & 3. & 3. & 3. & 14.9675 & 3\\
\hline
\end{tabular}
\vspace{1cm}
\caption{\textbf{Inferred LLM strategies and pairwise tournament for original experiment.} 
\textbf{A)} We summarize the data from the original experiment and present the inferred memory-1 strategy \textbf{B)} Pairwise tournament between the inferred strategies from five models. The stage game payoffs are \((a_\LL, a_\LR, a_\RL, a_\RR) = (3,0,5,1)\) and the stopping probability is $w=0.01$. In this table, we report the payoff obtained by the row-model against each column-model. Rankings are determined by the total payoff accumulated against all five opponents. All numerical values in the table are rounded up to the most relevant decimal place.}
\label{tables:Table_NoInsttruction_AB}
\label{fig:LLM-data}
\end{figure}

\begin{figure}[t!]
\textbf{\textcolor{blue}{``Play like a pro."}}\\\\\\
\textbf{A \quad Moves of five LLM models in IPD scenarios and their inferred memory-1 strategy.}\\ \\ \\
\raggedright
\renewcommand{\arraystretch}{1.2} 
\setlength{\tabcolsep}{8pt} 
\begin{tabular}{r|c c c c c|c c c c c|r}
\hline
\multirow{3}{*}{LLM model}
 & \multicolumn{5}{c|}{`L' in 50 trials}
 & \multicolumn{5}{c|}{Memory-1 strategy} 
 & \multicolumn{1}{c}{}
\\
\cline{2-11}
 & R1 & LL & LR & RL & RR
 & $p_0$ & $p_1$ & $p_2$ & $p_3$ & $p_4$& Notes \\
\hline
\claudeshort & 50 & 50 & 0 & 5 & 50 & 1.00 & 1.00 & 0.00 & 0.10 & 1.00 & $\sim$WSLS  \\
\geminishort & 42 & 50 & 0 & 9 & 16 & 0.84 & 1.00 & 0.00 & 0.18 & 0.32 & $\sim$GRIM  \\
\gptfourshort & 49 & 50 & 0 & 0 & 6 & 0.98 & 1.00 & 0.00 & 0.00 & 0.12 & $\sim$GRIM  \\
\gptfiveshort& 50 & 50 & 0 & 18 & 49 & 1.00 & 1.00 & 0.00 & 0.36 & 0.98 &$\sim$WSLS \\
\llamashort & 50 & 50 & 0 & 0 & 0 & 1.00 & 1.00 & 0.00 & 0.00 & 0.00 & GRIM  \\
\hline
\end{tabular}
\vspace{2cm}

\textbf{B \quad Pairwise tournament between the LLM models, represented by their memory-1 strategy.}\newline \newline \newline
\raggedright
\begin{tabular}{r|c c c c c|c|c}
\hline
LLM model& \claudeshort & \geminishort & \gptfourshort & \gptfiveshort& \llamashort&Row sum& Rank\\
\hline
\claudeshort & 3. & 2.9777 & 2.9928 & 3. & 3. & 14.9705 & 1\\
\geminishort & 2.9977 & 2.9419 & 2.9100 & 2.9978 & 2.6481 & 14.4954 & 5\\
\gptfourshort & 3.00002 & 2.9403 & 2.9685 & 2.99996 & 2.9587 & 14.8674 & 3\\
\gptfiveshort& 3. & 2.9770 & 2.9927 & 3. & 3. & 14.9697 & 2\\
\llamashort & 3. & 2.8317 & 2.9682 & 3. & 3. & 14.7999 & 4\\
\hline
\end{tabular}
\vspace{0.5cm}
\caption{\textbf{LLM models playing the repeated Prisoner's Dilemma (IPD).} 
Recreating Fig. \ref{tables:Table_NoInsttruction_AB} but with the special instruction \textbf{``Play like a pro."}. All other parameters are kept the same.}
\label{tables:7}
\end{figure}

\begin{figure}[t!]
\textbf{\textcolor{blue}{``Try to get more points than the other agent."}}\\\\\\
\textbf{A \quad Moves of five LLM models in IPD scenarios and their inferred memory-1 strategy.}\\ \\ \\
\raggedright
\renewcommand{\arraystretch}{1.2} 
\setlength{\tabcolsep}{8pt} 
\begin{tabular}{r|c c c c c|c c c c c|r}
\hline
\multirow{3}{*}{LLM model}
 & \multicolumn{5}{c|}{`L' in 50 trials}
 & \multicolumn{5}{c|}{Memory-1 strategy} 
 & \multicolumn{1}{c}{}
\\
\cline{2-11}
 & R1 & LL & LR & RL & RR
 & $p_0$ & $p_1$ & $p_2$ & $p_3$ & $p_4$& Notes \\
 \hline
\claudeshort & 0 & 0 & 0 & 0 & 0 & 0.00 & 0.00 & 0.00 & 0.00 & 0.00 & ALLD  \\
\geminishort & 0 & 0 & 0 & 0 & 0 & 0.00 & 0.00 & 0.00 & 0.00 & 0.00 & ALLD  \\
\gptfourshort & 0 & 0 & 0 & 0 & 0 & 0.00 & 0.00 & 0.00 & 0.00 & 0.00 & ALLD  \\
\gptfiveshort& 1 & 1 & 0 & 0 & 0 & 0.02 & 0.02 & 0.00 & 0.00 & 0.00 & $\sim$ALLD  \\
\llamashort & 50 & 50 & 0 & 0 & 1 & 1.00 & 1.00 & 0.00 & 0.00 & 0.02 & $\sim$GRIM  \\
\hline
\end{tabular}
\vspace{2cm}

\textbf{B \quad Pairwise tournament between the LLM models, represented by their memory-1 strategy.}\newline \newline \newline
\raggedright
\begin{tabular}{r|c c c c c|c|c}
\hline
LLM model& \claudeshort & \geminishort & \gptfourshort & \gptfiveshort& \llamashort&Row sum& Rank\\
\hline
\claudeshort & 1. & 1. & 1. & 1.0008 & 1.1169 & 5.1177 & 2\\
\geminishort & 1. & 1. & 1. & 1.0008 & 1.1169 & 5.1177 & 2\\
\gptfourshort & 1. & 1. & 1. & 1.0008 & 1.1169 & 5.1177 & 2\\
\gptfiveshort& 0.9998 & 0.9998 & 0.9998 & 1.0006 & 1.1173 & 5.1173 & 5\\
\llamashort & 0.9708 & 0.9708 & 0.9708 & 0.9712 & 3.& 6.8835 & 1\\
\hline
\end{tabular}
\vspace{0.5cm}
\caption{\textbf{LLM models playing the repeated Prisoner's Dilemma (IPD).} 
Recreating Fig. \ref{tables:Table_NoInsttruction_AB} but with the special instruction \textbf{``Try to get more points than the other agent."}. All other parameters are kept the same.}
\label{tables:1}
\end{figure}

\begin{figure}[t!]
\textbf{\textcolor{blue}{``Exploit the other agent as much as possible."}}\\\\\\
\textbf{A \quad Moves of five LLM models in IPD scenarios and their inferred memory-1 strategy.}\\ \\ \\
\raggedright
\renewcommand{\arraystretch}{1.2} 
\setlength{\tabcolsep}{8pt} 
\begin{tabular}{r|c c c c c|c c c c c|r}
\hline
\multirow{3}{*}{LLM model}
 & \multicolumn{5}{c|}{`L' in 50 trials}
 & \multicolumn{5}{c|}{Memory-1 strategy} 
 & \multicolumn{1}{c}{}
\\
\cline{2-11}
 & R1 & LL & LR & RL & RR
 & $p_0$ & $p_1$ & $p_2$ & $p_3$ & $p_4$& Notes \\
\hline
\claudeshort & 0 & 0 & 0 & 0 & 0 & 0.00 & 0.00 & 0.00 & 0.00 & 0.00 & ALLD  \\
\geminishort & 0 & 0 & 0 & 0 & 0 & 0.00 & 0.00 & 0.00 & 0.00 & 0.00 & ALLD  \\
\gptfourshort & 0 & 0 & 0 & 0 & 0 & 0.00 & 0.00 & 0.00 & 0.00 & 0.00 & ALLD  \\
\gptfiveshort& 7 & 0 & 0 & 0 & 0 & 0.14 & 0.00 & 0.00 & 0.00 & 0.00 & $\sim$ALLD  \\
\llamashort & 11 & 0 & 0 & 0 & 0 & 0.22 & 0.00 & 0.00 & 0.00 & 0.00 & $\sim$ALLD  \\
\hline
\end{tabular}
\vspace{2cm}

\textbf{B \quad Pairwise tournament between the LLM models, represented by their memory-1 strategy.}\newline \newline \newline
\raggedright
\begin{tabular}{r|c c c c c|c|c}
\hline
LLM model& \claudeshort & \geminishort & \gptfourshort & \gptfiveshort& \llamashort&Row sum& Rank\\
\hline
\claudeshort & 1. & 1. & 1. & 1.0056 & 1.0088 & 5.0144 & 1\\
\geminishort & 1. & 1. & 1. & 1.0056 & 1.0088 & 5.0144 & 1\\
\gptfourshort & 1. & 1. & 1. & 1.0056 & 1.0088 & 5.0144 & 1\\
\gptfiveshort& 0.9986 & 0.9986 & 0.9986 & 1.0040 & 1.0071 & 5.0069 & 4\\
\llamashort & 0.9978 & 0.9978 & 0.9978 & 1.0031 & 1.0061 & 5.0026 & 5\\
\hline
\end{tabular}
\vspace{0.5cm}
\caption{\textbf{LLM models playing the repeated Prisoner's Dilemma (IPD).} 
Recreating Fig. \ref{tables:Table_NoInsttruction_AB} but with the special instruction \textbf{``Exploit the other agent as much as possible."}. All other parameters are kept the same.}
\label{tables:5}
\end{figure}

\begin{figure}[t!]
\textbf{\textcolor{blue}{``Maximize your own number of points."}}\\\\\\
\textbf{A \quad Moves of five LLM models in IPD scenarios and their inferred memory-1 strategy.}\\ \\ \\
\raggedright
\renewcommand{\arraystretch}{1.2} 
\setlength{\tabcolsep}{8pt} 
\begin{tabular}{r|c c c c c|c c c c c|r}
\hline
\multirow{3}{*}{LLM model}
 & \multicolumn{5}{c|}{`L' in 50 trials}
 & \multicolumn{5}{c|}{Memory-1 strategy} 
 & \multicolumn{1}{c}{}
\\
\cline{2-11}
 & R1 & LL & LR & RL & RR
 & $p_0$ & $p_1$ & $p_2$ & $p_3$ & $p_4$& Notes \\
\hline
\claudeshort & 0 & 0 & 0 & 0 & 0 & 0.00 & 0.00 & 0.00 & 0.00 & 0.00 & ALLD  \\
\geminishort & 0 & 12 & 0 & 0 & 0 & 0.00 & 0.24 & 0.00 & 0.00 & 0.00 & $\sim$ALLD  \\
\gptfourshort & 0 & 15 & 0 & 0 & 0 & 0.00 & 0.30 & 0.00 & 0.00 & 0.00 & $\sim$ALLD  \\
\gptfiveshort& 50 & 50 & 0 & 16 & 38 & 1.00 & 1.00 & 0.00 & 0.32 & 0.76 & $\sim$WSLS  \\
\llamashort & 50 & 50 & 0 & 0 & 0 & 1.00 & 1.00 & 0.00 & 0.00 & 0.00 & GRIM  \\
\hline
\end{tabular}
\vspace{2cm}

\textbf{B \quad Pairwise tournament between the LLM models, represented by their memory-1 strategy.}\newline \newline \newline
\raggedright
\begin{tabular}{r|c c c c c|c|c}
\hline
LLM model& \claudeshort & \geminishort & \gptfourshort & \gptfiveshort& \llamashort&Row sum& Rank\\
\hline
\claudeshort & 1. & 1. & 1. & 2.74 & 1.04 & 6.7802 & 3\\
\geminishort & 1. & 1. & 1. & 2.74 & 1.04 & 6.7802 & 3\\
\gptfourshort & 1. & 1. & 1. & 2.74 & 1.04 & 6.7802 & 3\\
\gptfiveshort& 0.56 & 0.56 & 0.56 & 3. & 3. & 7.6948 & 2\\
\llamashort & 0.99 & 0.99 & 0.99 & 3. & 3. & 8.9700 & 1\\
\hline

\hline
\end{tabular}
\vspace{0.5cm}
\caption{\textbf{LLM models playing the repeated Prisoner's Dilemma (IPD).} 
Recreating Fig. \ref{tables:Table_NoInsttruction_AB} but with the special instruction \textbf{``Maximize your own number of points."}. All other parameters are kept the same.}
\label{tables:2}
\end{figure}

\begin{figure}[t!]
\textbf{\textcolor{blue}{``Think about winning."}}\\\\\\
\textbf{A \quad Moves of five LLM models in IPD scenarios and their inferred memory-1 strategy.}\\ \\ \\
\raggedright
\renewcommand{\arraystretch}{1.2} 
\setlength{\tabcolsep}{8pt} 
\begin{tabular}{r|c c c c c|c c c c c|r}
\hline
\multirow{3}{*}{LLM model}
 & \multicolumn{5}{c|}{`L' in 50 trials}
 & \multicolumn{5}{c|}{Memory-1 strategy} 
 & \multicolumn{1}{c}{}
\\
\cline{2-11}
 & R1 & LL & LR & RL & RR
 & $p_0$ & $p_1$ & $p_2$ & $p_3$ & $p_4$& Notes \\
\hline
\claudeshort & 0 & 0 & 0 & 0 & 0 & 0.00 & 0.00 & 0.00 & 0.00 & 0.00 & ALLD  \\
\geminishort & 0 & 9 & 0 & 0 & 2 & 0.00 & 0.18 & 0.00 & 0.00 & 0.04 & $\sim$ALLD  \\
\gptfourshort & 0 & 0 & 0 & 0 & 0 & 0.00 & 0.00 & 0.00 & 0.00 & 0.00 & ALLD  \\
\gptfiveshort& 49 & 50 & 0 & 7 & 19 & 0.98 & 1.00 & 0.00 & 0.14 & 0.38 & $\sim$GRIM \\
\llamashort & 50 & 50 & 0 & 0 & 0 & 1.00 & 1.00 & 0.00 & 0.00 & 0.00 & GRIM  \\
\hline
\end{tabular}
\vspace{2cm}

\textbf{B \quad Pairwise tournament between the LLM models, represented by their memory-1 strategy.}\newline \newline \newline
\raggedright
\begin{tabular}{r|c c c c c|c|c}
\hline
LLM model& \claudeshort & \geminishort & \gptfourshort & \gptfiveshort& \llamashort&Row sum& Rank\\
\hline
\claudeshort & 1. & 1.1524 & 1. & 2.1219 & 1.0400 & 6.3143 & 3\\
\geminishort & 0.9619 & 1.1095 & 0.9619 & 2.1042 & 1.0023 & 6.1398 & 5\\
\gptfourshort & 1. & 1.1524 & 1. & 2.1219 & 1.0400 & 6.3143 & 3\\
\gptfiveshort& 0.7195 & 0.8199 & 0.7195 & 2.9942 & 2.9554 & 8.2085 & 2\\
\llamashort & 0.9900 & 1.1408 & 0.9900 & 2.9815 & 3.0000 & 9.1024 & 1\\
\hline
\end{tabular}
\vspace{0.5cm}
\caption{\textbf{LLM models playing the repeated Prisoner's Dilemma (IPD).} 
Recreating Fig. \ref{tables:Table_NoInsttruction_AB} but with the special instruction \textbf{``Think about winning."}. All other parameters are kept the same.}
\label{tables:6}
\end{figure}

\begin{figure}[t!]
\textbf{\textcolor{blue}{``Accumulate many points, but do not exploit."}}\\\\\\
\textbf{A \quad Moves of five LLM models in IPD scenarios and their inferred memory-1 strategy.}\\ \\ \\
\raggedright
\renewcommand{\arraystretch}{1.2} 
\setlength{\tabcolsep}{8pt} 
\begin{tabular}{r|c c c c c|c c c c c|r}
\hline
\multirow{3}{*}{LLM model}
 & \multicolumn{5}{c|}{`L' in 50 trials}
 & \multicolumn{5}{c|}{Memory-1 strategy} 
 & \multicolumn{1}{c}{}
\\
\cline{2-11}
 & R1 & LL & LR & RL & RR
 & $p_0$ & $p_1$ & $p_2$ & $p_3$ & $p_4$& Notes \\
\hline
\claudeshort & 50 & 50 & 50 & 50 & 50 & 1.00 & 1.00 & 1.00 & 1.00 & 1.00 & ALLC  \\
\geminishort & 50 & 50 & 2 & 50 & 50 & 1.00 & 1.00 & 0.04 & 1.00 & 1.00 & $\sim$Forgiver  \\
\gptfourshort & 50 & 50 & 50 & 50 & 50 & 1.00 & 1.00 & 1.00 & 1.00 & 1.00 & ALLC  \\
\gptfiveshort& 50 & 50 & 30 & 50 & 50 & 1.00 & 1.00 & 0.60 & 1.00 & 1.00 &$\sim$Forgiver \\
\llamashort & 50 & 50 & 4 & 50 & 50 & 1.00 & 1.00 & 0.08 & 1.00 & 1.00 & $\sim$Forgiver  \\
\hline
\end{tabular}
\vspace{2cm}

\textbf{B \quad Pairwise tournament between the LLM models, represented by their memory-1 strategy.}\newline \newline \newline
\raggedright
\begin{tabular}{r|c c c c c|c|c}
\hline
LLM model& \claudeshort & \geminishort & \gptfourshort & \gptfiveshort& \llamashort&Row sum& Rank\\
\hline
\claudeshort & 3.0 & 3.0 & 3.0 & 3.0 & 3.0 & 15.0 & 1\\
\geminishort & 3.0 & 3.0 & 3.0 & 3.0 & 3.0 & 15.0 & 1\\
\gptfourshort & 3.0 & 3.0 & 3.0 & 3.0 & 3.0 & 15.0 & 1\\
\gptfiveshort& 3.0 & 3.0 & 3.0 & 3.0 & 3.0 & 15.0 & 1\\
\llamashort & 3.0 & 3.0 & 3.0 & 3.0 & 3.0 & 15.0 & 1\\ 
\hline
\end{tabular}
\vspace{0.5cm}
\caption{\textbf{LLM models playing the repeated Prisoner's Dilemma (IPD).} 
Recreating Fig. \ref{tables:Table_NoInsttruction_AB} but with the special instruction \textbf{``Accumulate many points, but do not exploit."}. All other parameters are kept the same.}
\label{tables:4}
\end{figure}

\begin{figure}[t!]
\textbf{\textcolor{blue}{``Maximize your points, keep in mind the other's welfare."}}\\\\\\
\textbf{A \quad Moves of five LLM models in IPD scenarios and their inferred memory-1 strategy.}\\ \\ \\
\raggedright
\renewcommand{\arraystretch}{1.2} 
\setlength{\tabcolsep}{8pt} 
\begin{tabular}{r|c c c c c|c c c c c|r}
\hline
\multirow{3}{*}{LLM model}
 & \multicolumn{5}{c|}{`L' in 50 trials}
 & \multicolumn{5}{c|}{Memory-1 strategy} 
 & \multicolumn{1}{c}{}
\\
\cline{2-11}
 & R1 & LL & LR & RL & RR
 & $p_0$ & $p_1$ & $p_2$ & $p_3$ & $p_4$& Notes \\
\hline
\claudeshort & 50 & 50 & 0 & 50 & 50 & 1.00 & 1.00 & 0.00 & 1.00 & 1.00 & Forgiver  \\
\geminishort & 50 & 50 & 40 & 50 & 50 & 1.00 & 1.00 & 0.80 & 1.00 & 1.00 & $\sim$ALLC  \\
\gptfourshort & 50 & 50 & 26 & 50 & 50 & 1.00 & 1.00 & 0.52 & 1.00 & 1.00 &$\sim$Forgiver  \\
\gptfiveshort& 50 & 50 & 0 & 50 & 50 & 1.00 & 1.00 & 0.00 & 1.00 & 1.00 & Forgiver  \\
\llamashort & 50 & 50 & 0 & 41 & 50 & 1.00 & 1.00 & 0.00 & 0.82 & 1.00 & $\sim$Forgiver \\
\hline
\end{tabular}
\vspace{2cm}

\textbf{B \quad Pairwise tournament between the LLM models, represented by their memory-1 strategy.}\newline \newline \newline
\raggedright
\begin{tabular}{r|c c c c c|c|c}
\hline
LLM model& \claudeshort & \geminishort & \gptfourshort & \gptfiveshort& \llamashort&Row sum& Rank\\
\hline
\claudeshort & 3. & 3. & 3. & 3. & 3. & 15. & 1\\
\geminishort & 3. & 3. & 3. & 3. & 3. & 15. & 1\\
\gptfourshort & 3. & 3. & 3. & 3. & 3. & 15. & 1\\
\gptfiveshort& 3. & 3. & 3. & 3. & 3. & 15. & 1\\
\llamashort & 3. & 3. & 3. & 3. & 3. & 15. & 1\\
\hline
\end{tabular}
\vspace{0.5cm}
\caption{\textbf{LLM models playing the repeated Prisoner's Dilemma (IPD).} 
Recreating Fig. \ref{tables:Table_NoInsttruction_AB} but with the special instruction \textbf{``Maximize your points, keep in mind the other's welfare."}. All other parameters are kept the same.}
\label{tables:3}
\end{figure}

\begin{figure}[t!]
\textbf{\textcolor{blue}{``Think about fair outcomes."}}\\\\\\
\textbf{A \quad Moves of five LLM models in IPD scenarios and their inferred memory-1 strategy.}\\ \\ \\
\raggedright
\renewcommand{\arraystretch}{1.2} 
\setlength{\tabcolsep}{8pt} 
\begin{tabular}{r|c c c c c|c c c c c|r}
\hline
\multirow{3}{*}{LLM model}
 & \multicolumn{5}{c|}{`L' in 50 trials}
 & \multicolumn{5}{c|}{Memory-1 strategy} 
 & \multicolumn{1}{c}{}
\\
\cline{2-11}
 & R1 & LL & LR & RL & RR
 & $p_0$ & $p_1$ & $p_2$ & $p_3$ & $p_4$& Notes \\
\hline
\claudeshort & 50 & 50 & 0 & 50 & 50 & 1.00 & 1.00 & 0.00 & 1.00 & 1.00 & Forgiver  \\
\geminishort & 50 & 50 & 48 & 50 & 50 & 1.00 & 1.00 & 0.96 & 1.00 & 1.00 & $\sim$ALLC \\
\gptfourshort & 50 & 50 & 3 & 50 & 50 & 1.00 & 1.00 & 0.06 & 1.00 & 1.00 & $\sim$Forgiver  \\
\gptfiveshort& 50 & 50 & 1 & 50 & 50 & 1.00 & 1.00 & 0.02 & 1.00 & 1.00 & $\sim$Forgiver  \\
\llamashort & 50 & 50 & 0 & 49 & 50 & 1.00 & 1.00 & 0.00 & 0.98 & 1.00 & $\sim$Forgiver  \\
\hline
\end{tabular}
\vspace{2cm}

\textbf{B \quad Pairwise tournament between the LLM models, represented by their memory-1 strategy.}\newline \newline \newline
\raggedright
\begin{tabular}{r|c c c c c|c|c}
\hline
LLM model& \claudeshort & \geminishort & \gptfourshort & \gptfiveshort& \llamashort&Row sum& Rank\\
\hline
\claudeshort & 3. & 3. & 3. & 3. & 3. & 15. & 1\\
\geminishort & 3. & 3. & 3. & 3. & 3. & 15. & 1\\
\gptfourshort & 3. & 3. & 3. & 3. & 3. & 15. & 1\\
\gptfiveshort& 3. & 3. & 3. & 3. & 3. & 15. & 1\\
\llamashort & 3. & 3. & 3. & 3. & 3. & 15. & 1\\
\hline
\end{tabular}
\vspace{0.5cm}
\caption{\textbf{LLM models playing the repeated Prisoner's Dilemma (IPD).} 
Recreating Fig. \ref{tables:Table_NoInsttruction_AB} but with the special instruction \textbf{``Think about fair outcomes."}. All other parameters are kept the same.}
\label{tables:8}
\end{figure}

\begin{figure}[t!]
\textbf{\textcolor{blue}{``Be a saint."}}\\\\\\
\textbf{A \quad Moves of five LLM models in IPD scenarios and their inferred memory-1 strategy.}\\ \\ \\
\raggedright
\renewcommand{\arraystretch}{1.2} 
\setlength{\tabcolsep}{8pt} 
\begin{tabular}{r|c c c c c|c c c c c|r}
\hline
\multirow{3}{*}{LLM model}
 & \multicolumn{5}{c|}{`L' in 50 trials}
 & \multicolumn{5}{c|}{Memory-1 strategy} 
 & \multicolumn{1}{c}{}
\\
\cline{2-11}
 & R1 & LL & LR & RL & RR
 & $p_0$ & $p_1$ & $p_2$ & $p_3$ & $p_4$& Notes \\
\hline
\claudeshort & 50 & 50 & 50 & 50 & 50 & 1.00 & 1.00 & 1.00 & 1.00 & 1.00 & ALLC  \\
\geminishort & 50 & 50 & 50 & 50 & 50 & 1.00 & 1.00 & 1.00 & 1.00 & 1.00 & ALLC  \\
\gptfourshort & 50 & 50 & 50 & 50 & 50 & 1.00 & 1.00 & 1.00 & 1.00 & 1.00 & ALLC  \\
\gptfiveshort& 50 & 50 & 50 & 50 & 50 & 1.00 & 1.00 & 1.00 & 1.00 & 1.00 & ALLC  \\
\llamashort & 50 & 50 & 0 & 49 & 50 & 1.00 & 1.00 & 0.00 & 0.98 & 1.00 & $\sim$Forgiver  \\
\hline
\end{tabular}
\vspace{2cm}

\textbf{B \quad Pairwise tournament between the LLM models, represented by their memory-1 strategy.}\newline \newline \newline
\raggedright
\begin{tabular}{r|c c c c c|c|c}
\hline
LLM model& \claudeshort & \geminishort & \gptfourshort & \gptfiveshort& \llamashort&Row sum& Rank\\
\hline
\claudeshort & 3. & 3. & 3. & 3. & 3. & 15. & 1\\
\geminishort & 3. & 3. & 3. & 3. & 3. & 15. & 1\\
\gptfourshort & 3. & 3. & 3. & 3. & 3. & 15. & 1\\
\gptfiveshort& 3. & 3. & 3. & 3. & 3. & 15. & 1\\
\llamashort & 3. & 3. & 3. & 3. & 3. & 15. & 1\\
\hline
\end{tabular}
\vspace{0.5cm}
\caption{\textbf{LLM models playing the repeated Prisoner's Dilemma (IPD).} 
Recreating Fig. \ref{tables:Table_NoInsttruction_AB} but with the special instruction \textbf{``Be a saint."}. All other parameters are kept the same.}
\label{tables:9}
\end{figure}

\clearpage

\begin{figure}
\raggedright
\renewcommand{\arraystretch}{1.2} 
\setlength{\tabcolsep}{4pt} 
\begin{tabular}{r|r r r r r}
Instruction & \claudeshort& \geminishort& \gptfourshort & \gptfiveshort& \llamashort \\
\hline
\textit{no special instruction} & Forgiver& $\sim$WSLS& GRIM& \textbf{$\sim$Forgiver}& $\sim$GTFT\\
Try to get more points than the other agent. & ALLD& ALLD& ALLD& $\sim$ALLD& \textbf{$\sim$GRIM}\\
Exploit the other agent as much as possible. & \textbf{ALLD}& \textbf{ALLD}& \textbf{ALLD}& $\sim$ALLD& $\sim$ALLD\\
Maximize your own number of points. & ALLD& $\sim$ALLD& $\sim$ALLD& $\sim$WSLS& \textbf{GRIM}\\
Think about winning. & ALLD& $\sim$ALLD& ALLD& $\sim$GRIM& \textbf{GRIM}\\
Play like a pro. & \textbf{$\sim$WSLS} & $\sim$GRIM& $\sim$GRIM& $\sim$WSLS& GRIM\\
Accumulate many points, but do not exploit. & \textbf{ALLC}& \textbf{$\sim$Forgiver}& \textbf{ALLC}& \textbf{$\sim$Forgiver}& \textbf{$\sim$Forgiver}\\
Maximize your points, keep in mind the other's welfare. &\textbf{Forgiver} & \textbf{$\sim$ALLC}& \textbf{$\sim$Forgiver}& \textbf{Forgiver}& \textbf{$\sim$Forgiver}\\
Think about fair outcomes. & \textbf{Forgiver}& \textbf{$\sim$ALLC}& \textbf{$\sim$Forgiver}& \textbf{$\sim$Forgiver}& \textbf{$\sim$Forgiver}\\
Be a saint. & \textbf{ALLC} & \textbf{ALLC} & \textbf{ALLC} & \textbf{ALLC} & \textbf{$\sim$Forgiver} \\
\end{tabular}
 \vspace{0.5cm}
 \caption{\textbf{Inferred strategies of the five LLMs across ten experiments. Tournament winners are marked by bolded font.} This summarizes the winner data in \textbf{Fig. S1-10}.}
\end{figure}
\clearpage

\begin{figure}[h!]
\renewcommand{\arraystretch}{1.3}
\setlength{\tabcolsep}{8pt}
\raggedright

% --- Table ---
\textbf{A}\newline \newline 

\begin{tabular}{r|c|ccccc|c}
\hline
 & \multicolumn{5}{c}{All ten tournaments, summed.} & & \\
\hline
Rank & LLM model & \claudeshort & \geminishort & \gptfourshort & \gptfiveshort& \llamashort & Row sum \\
\hline

3& \claudeshort & 22.00 & 22.11 & 21.99 & 24.87 & 22.21 & 113.18 \\
5& \geminishort & 21.96 & 22.01 & 21.47 & 24.85 & 21.81 & 112.10 \\
4& \gptfourshort & 22.00 & 21.94 & 21.97 & 24.87 & 22.16 & 112.94 \\
2& \gptfiveshort& 21.28 & 21.34 & 21.28 & 26.00 & 26.08 & 115.98\\
1& \llamashort & 21.95 & 21.90 & 21.92 & 25.96 & 28.01 & 119.73\\

\hline
\end{tabular}

\vspace{1cm}
\textbf{B} \vspace{0.5cm}
% --- Figure ---

\raggedright
\includegraphics[width=0.8\linewidth]{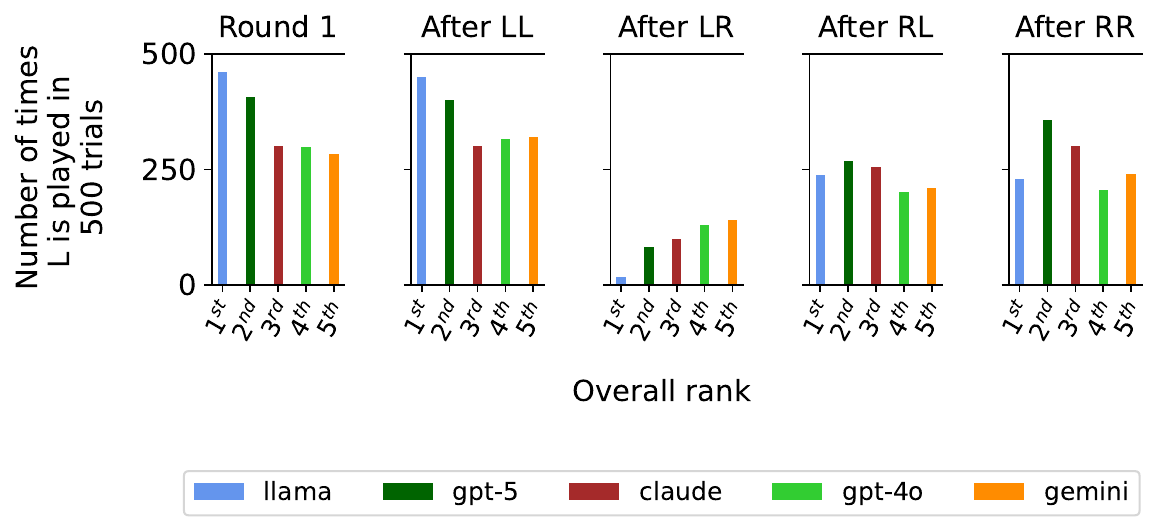}

\vspace{0.5cm}
\caption{\textbf{All ten tournament outcomes, summed.} 
\textbf{A} The winner, once all tournament scores are added, is \llamashort. \textbf{B} We show how often an LLM model choses the move `L', across 10 special instruction cases and in 50 independent trials per instruction.}

\end{figure}

\clearpage

\subsubsection{Stopping probability experiments}

\noindent The following four figures summarize the inferred strategy data from the Stopping probability experiments. Like before we present the outcome of the pairwise tournament. All games are for the payoff values $(a_\LL,a_\LR,a_\RL,a_\RR)\!=\!(3,0,5,1)$ and the stopping probability is the same as for the corresponding experiment.
\\ \\ 

\begin{figure}[h!]
\textbf{\textcolor{blue}{$w=0.01$, $(a_\LL, a_\LR, a_\RL, a_\RR) = (3,0,5,1)$}}\\\\\\
\textbf{A \quad Moves of five LLM models in IPD scenarios and their inferred memory-1 strategy.}\\ \\ \\
\raggedright
\renewcommand{\arraystretch}{1.2}
\setlength{\tabcolsep}{8pt}
\begin{tabular}{r|c c c c c|c c c c c|r}
\hline
\multirow{3}{*}{LLM model} & \multicolumn{5}{c|}{`L' in 50 trials} & \multicolumn{5}{c|}{Memory-1 strategy} & \multicolumn{1}{c}{} \\
\cline{2-11}
 & R1 & LL & LR & RL & RR & $p_0$ & $p_1$ & $p_2$ & $p_3$ & $p_4$ & Notes \\
\hline
\claudeshort & 50 & 50 & 0 & 50 & 50 & 1.00 & 1.00 & 0.00 & 1.00 & 1.00 & Forgiver \\
\geminishort & 49 & 50 & 0 & 0 & 21 & 0.98 & 1.00 & 0.00 & 0.00 & 0.42 & $\sim$GRIM \\
\gptfourshort & 50 & 50 & 0 & 0 & 41 & 1.00 & 1.00 & 0.00 & 0.00 & 0.82 & $\sim$WSLS \\
\gptfiveshort& 50 & 50 & 0 & 43 & 50 & 1.00 & 1.00 & 0.00 & 0.86 & 1.00 & $\sim$Forgiver \\
\llamashort & 50 & 50 & 0 & 0 & 0 & 1.00 & 1.00 & 0.00 & 0.00 & 0.00 & $\sim$GRIM \\
\hline
\end{tabular}
\vspace{1.5cm}
 
\textbf{B \quad Pairwise tournament between the LLM models, represented by their memory-1 strategy.}\newline \newline \newline
\raggedright
\begin{tabular}{r|c c c c c|c|c}
\hline
LLM model& \claudeshort & \geminishort & \gptfourshort & \gptfiveshort& \llamashort&Row sum& Rank\\
\hline
\claudeshort & 3.0000 & 2.9976 & 3.0000 & 3.0000 & 3.0000 & 14.9976 & 1\\
\geminishort & 3.0000 & 2.9951 & 2.9996 & 3.0000 & 2.9549 & 14.9498 & 5\\
\gptfourshort & 3.0000 & 2.9975 & 3.0000 & 3.0000 & 3.0000 & 14.9975 & 3\\
\gptfiveshort& 3.0000 & 2.9976 & 3.0000 & 3.0000 & 3.0000 & 14.9976 & 1\\
\llamashort & 3.0000 & 2.9830 & 3.0000 & 3.0000 & 3.0000 & 14.9830 & 4\\
\hline
\end{tabular}
\vspace{1cm}
\caption{\textbf{Inferred strategies and tournament outcomes for the Stopping probability experiment} \mbox{$w = 0.01$}. The payoff values are (3,0,5,1).}
\end{figure}
\clearpage

\begin{figure}[h!]
\textbf{\textcolor{blue}{$w = 0.1$, $(a_\LL, a_\LR, a_\RL, a_\RR) = (3,0,5,1)$}}\\\\\\
\textbf{A \quad Moves of five LLM models in IPD scenarios and their inferred memory-1 strategy.}\\ \\ \\
\raggedright
\renewcommand{\arraystretch}{1.2}
\setlength{\tabcolsep}{8pt}
\begin{tabular}{r|c c c c c|c c c c c|r}
\hline
\multirow{3}{*}{LLM model} & \multicolumn{5}{c|}{`L' in 50 trials} & \multicolumn{5}{c|}{Memory-1 strategy} & \multicolumn{1}{c}{} \\
\cline{2-11}
 & R1 & LL & LR & RL & RR & $p_0$ & $p_1$ & $p_2$ & $p_3$ & $p_4$ & Notes \\
\hline
\claudeshort & 50 & 50 & 0 & 50 & 50 & 1.00 & 1.00 & 0.00 & 1.00 & 1.00 & Forgiver \\
\geminishort & 37 & 50 & 0 & 0 & 17 & 0.74 & 1.00 & 0.00 & 0.00 & 0.34 & $\sim$S-(1,0,0,$x$) \\
\gptfourshort & 50 & 50 & 0 & 0 & 13 & 1.00 & 1.00 & 0.00 & 0.00 & 0.26 & $\sim$GRIM \\
\gptfiveshort& 50 & 50 & 0 & 42 & 49 & 1.00 & 1.00 & 0.00 & 0.84 & 0.98 & $\sim$Forgiver \\
\llamashort & 50 & 50 & 0 & 0 & 2 & 1.00 & 1.00 & 0.00 & 0.00 & 0.04 & $\sim$GRIM \\
\hline
\end{tabular}
\vspace{1.5cm}
 
\textbf{B \quad Pairwise tournament between the LLM models, represented by their memory-1 strategy.}\newline \newline \newline
\raggedright
\begin{tabular}{r|c c c c c|c|c}
\hline
LLM model& \claudeshort & \geminishort & \gptfourshort & \gptfiveshort& \llamashort&Row sum& Rank\\
\hline
\claudeshort & 3.0000 & 2.7318 & 3.0000 & 3.0000 & 3.0000 & 14.7318 & 1\\
\geminishort & 3.0011 & 2.6078 & 2.7971 & 3.0089 & 2.5914 & 14.0166 & 5\\
\gptfourshort & 3.0000 & 2.7058 & 3.0000 & 3.0000 & 3.0000 & 14.7058 & 3\\
\gptfiveshort& 3.0000 & 2.7309 & 3.0000 & 3.0000 & 3.0000 & 14.7309 & 2\\
\llamashort & 3.0000 & 2.6808 & 3.0000 & 3.0000 & 3.0000 & 14.6808 & 4\\
\hline
\end{tabular}
\vspace{1cm}
\caption{\textbf{Inferred strategies and tournament outcomes for the Stopping probability experiment} \mbox{$w = 0.1$}. The payoff values are (3,0,5,1).}
\end{figure}
\clearpage

\begin{figure}[h!]
\textbf{\textcolor{blue}{$w = 0.2$, $(a_\LL, a_\LR, a_\RL, a_\RR) = (3,0,5,1)$}}\\\\\\
\textbf{A \quad Moves of five LLM models in IPD scenarios and their inferred memory-1 strategy.}\\ \\ \\
\raggedright
\renewcommand{\arraystretch}{1.2}
\setlength{\tabcolsep}{8pt}
\begin{tabular}{r|c c c c c|c c c c c|r}
\hline
\multirow{3}{*}{LLM model} & \multicolumn{5}{c|}{`L' in 50 trials} & \multicolumn{5}{c|}{Memory-1 strategy} & \multicolumn{1}{c}{} \\
\cline{2-11}
 & R1 & LL & LR & RL & RR & $p_0$ & $p_1$ & $p_2$ & $p_3$ & $p_4$ & Notes \\
\hline
\claudeshort & 50 & 50 & 0 & 50 & 50 & 1.00 & 1.00 & 0.00 & 1.00 & 1.00 & Forgiver \\
\geminishort & 25 & 50 & 0 & 0 & 14 & 0.50 & 1.00 & 0.00 & 0.00 & 0.28 & S-(1,0,0,$x$) \\
\gptfourshort & 50 & 50 & 0 & 0 & 27 & 1.00 & 1.00 & 0.00 & 0.00 & 0.54 & $\sim$WSLS \\
\gptfiveshort& 50 & 50 & 0 & 39 & 50 & 1.00 & 1.00 & 0.00 & 0.78 & 1.00 & $\sim$Forgiver \\
\llamashort & 50 & 50 & 0 & 0 & 0 & 1.00 & 1.00 & 0.00 & 0.00 & 0.00 & GRIM \\
\hline
\end{tabular}
\vspace{1.5cm}
 
\textbf{B \quad Pairwise tournament between the LLM models, represented by their memory-1 strategy.}\newline \newline \newline
\raggedright
\begin{tabular}{r|c c c c c|c|c}
\hline
LLM model& \claudeshort & \geminishort & \gptfourshort & \gptfiveshort& \llamashort&Row sum& Rank\\
\hline
\claudeshort & 3.0000 & 2.1468 & 3.0000 & 3.0000 & 3.0000 & 14.1468 & 3\\
\geminishort & 3.0741 & 2.0913 & 2.8657 & 3.0741 & 2.3267 & 13.4322 & 5\\
\gptfourshort & 3.0000 & 2.1596 & 3.0000 & 3.0000 & 3.0000 & 14.1596 & 2\\
\gptfiveshort& 3.0000 & 2.1468 & 3.0000 & 3.0000 & 3.0000 & 14.1468 & 3\\
\llamashort & 3.0000 & 2.1928 & 3.0000 & 3.0000 & 3.0000 & 14.1928 & 1\\
\hline
\end{tabular}
\vspace{1cm}
\caption{\textbf{Inferred strategies and tournament outcomes for the Stopping probability experiment} \mbox{$w = 0.2$}. The payoff values are (3,0,5,1).}
\end{figure}
\clearpage

\begin{figure}[h!]
\textbf{\textcolor{blue}{$w = 0.5$, $(a_\LL, a_\LR, a_\RL, a_\RR) = (3,0,5,1)$}}\\\\\\
\textbf{A \quad Moves of five LLM models in IPD scenarios and their inferred memory-1 strategy.}\\ \\ \\
\raggedright
\renewcommand{\arraystretch}{1.2}
\setlength{\tabcolsep}{8pt}
\begin{tabular}{r|c c c c c|c c c c c|r}
\hline
\multirow{3}{*}{LLM model} & \multicolumn{5}{c|}{`L' in 50 trials} & \multicolumn{5}{c|}{Memory-1 strategy} & \multicolumn{1}{c}{} \\
\cline{2-11}
 & R1 & LL & LR & RL & RR & $p_0$ & $p_1$ & $p_2$ & $p_3$ & $p_4$ & Notes \\
\hline
\claudeshort & 50 & 50 & 0 & 50 & 50 & 1.00 & 1.00 & 0.00 & 1.00 & 1.00 & Forgiver \\
\geminishort & 23 & 44 & 0 & 0 & 3 & 0.46 & 0.88 & 0.00 & 0.00 & 0.06 & $\sim$DGRIM \\
\gptfourshort & 50 & 50 & 0 & 0 & 28 & 1.00 & 1.00 & 0.00 & 0.00 & 0.56 & $\sim$WSLS \\
\gptfiveshort& 49 & 50 & 0 & 9 & 32 & 0.98 & 1.00 & 0.00 & 0.18 & 0.64 & $\sim$WSLS \\
\llamashort & 50 & 50 & 0 & 0 & 0 & 1.00 & 1.00 & 0.00 & 0.00 & 0.00 & GRIM \\
\hline
\end{tabular}
\vspace{1.5cm}
 
\textbf{B \quad Pairwise tournament between the LLM models, represented by their memory-1 strategy.}\newline \newline \newline
\raggedright
\begin{tabular}{r|c c c c c|c|c}
\hline
LLM model& \claudeshort & \geminishort & \gptfourshort & \gptfiveshort& \llamashort&Row sum& Rank\\
\hline
\claudeshort & 3.0000 & 1.4561 & 3.0000 & 2.9587 & 3.0000 & 13.4148 & 5\\
\geminishort & 3.3859 & 1.7933 & 3.2492 & 3.2437 & 2.9914 & 14.6637 & 1\\
\gptfourshort & 3.0000 & 1.4925 & 3.0000 & 2.9604 & 3.0000 & 13.4529 & 4\\
\gptfiveshort& 3.0103 & 1.4909 & 3.0045 & 2.9661 & 2.9969 & 13.4689 & 3\\
\llamashort & 3.0000 & 1.5611 & 3.0000 & 2.9624 & 3.0000 & 13.5235 & 2\\
\hline
\end{tabular}
\vspace{1cm}
\caption{\textbf{Inferred strategies and tournament outcomes for the Stopping probability experiment} \mbox{$w = 0.5$}. The payoff values are (3,0,5,1).}
\end{figure}

\clearpage

\clearpage

\noindent Next, we sum up the tournament outcomes from all four continuation treatment tournaments and present the aggregate outcomes. 
\vspace{1cm}
\begin{figure}[h!]
\textbf{Aggregated results from five tournaments, each conduced for $(a_\LL,a_\LR,a_\RL,a_\RR)=(3,0,5,1)$.}\newline \newline \newline
\raggedright
\begin{tabular}{r|c c c c c|c|c}
\hline
LLM model& \claudeshort & \geminishort & \gptfourshort & \gptfiveshort& \llamashort&Row sum& Rank\\
\hline
\claudeshort & 12.0000 & 8.9608 & 12.0000 & 11.9587 & 12.0000 & 56.9195 & 5\\
\geminishort & 12.6058 & 9.4136 & 12.0195 & 12.4214 & 11.1868 & 57.6471 & 1\\
\gptfourshort & 12.0000 & 9.0766 & 12.0000 & 11.9614 & 12.0000 & 57.0380 & 3\\
\gptfiveshort& 12.0103 & 9.0111 & 12.0008 & 11.9661 & 11.9969 & 56.9853 & 4\\
\llamashort & 12.0000 & 9.1664 & 12.0000 & 11.9624 & 12.0000 & 57.1288 & 2\\
\hline
\end{tabular}
\vspace{1cm}
\caption{We sum up the tournament matrices from all four continuation treatment tournaments, each conducted for $(a_\LL,a_\LR,a_\RL,a_\RR)=(3,0,5,1)$, and present an aggregate tournament matrix with rankings.}
\end{figure}

\clearpage
\subsubsection{Equal gains from switching experiments}

We do the same for the equal gains from switching experiments. The eleven figures that follow summarize the data, inferred strategy and outcome of the tournaments conducted for strategies from each tournament. The games are conducted for the stopping probability $w\!=\!0.01$. The payoff values for each experiment is $(a_\LL,a_\LR,a_\RL,a_\RR)\!=\!(10,0,10\!+\!x,x)$.

\vspace{1cm}

\begin{figure}[h!]
\textbf{\textcolor{blue}{$x = 0$}}\\\\
\textbf{A \quad Moves of five LLM models in IPD scenarios and their inferred memory-1 strategy.}\\ \\ \\
\raggedright
\renewcommand{\arraystretch}{1.2}
\setlength{\tabcolsep}{8pt}
\begin{tabular}{r|c c c c c|c c c c c|r}
\hline
\multirow{3}{*}{LLM model} & \multicolumn{5}{c|}{`L' in 50 trials} & \multicolumn{5}{c|}{Memory-1 strategy} & \multicolumn{1}{c}{} \\
\cline{2-11}
 & R1 & LL & LR & RL & RR & $p_0$ & $p_1$ & $p_2$ & $p_3$ & $p_4$ & Notes \\
\hline
\claudeshort & 50 & 50 & 0 & 50 & 50 & 1.00 & 1.00 & 0.00 & 1.00 & 1.00 & Forgiver \\
\geminishort & 50 & 50 & 2 & 4 & 50 & 1.00 & 1.00 & 0.04 & 0.08 & 1.00 & $\sim$WSLS \\
\gptfourshort & 50 & 50 & 0 & 0 & 50 & 1.00 & 1.00 & 0.00 & 0.00 & 1.00 & WSLS \\
\gptfiveshort& 50 & 50 & 4 & 50 & 50 & 1.00 & 1.00 & 0.08 & 1.00 & 1.00 & $\sim$Forgiver \\
\llamashort & 50 & 50 & 50 & 49 & 50 & 1.00 & 1.00 & 1.00 & 0.98 & 1.00 & $\sim$ALLC \\
\hline
\end{tabular}
\vspace{1.5cm}
 
\textbf{B \quad Pairwise tournament between the LLM models, represented by their memory-1 strategy. $\delta=0.99$}\newline \newline \newline
\raggedright
\begin{tabular}{r|c c c c c|c|c}
\hline
LLM model& \claudeshort & \geminishort & \gptfourshort & \gptfiveshort& \llamashort&Row sum& Rank\\
\hline
\claudeshort & 10.0000 & 10.0000 & 10.0000 & 10.0000 & 10.0000 & 50.0000 & 1\\
\geminishort & 10.0000 & 10.0000 & 10.0000 & 10.0000 & 10.0000 & 50.0000 & 1\\
\gptfourshort & 10.0000 & 10.0000 & 10.0000 & 10.0000 & 10.0000 & 50.0000 & 1\\
\gptfiveshort& 10.0000 & 10.0000 & 10.0000 & 10.0000 & 10.0000 & 50.0000 & 1\\
\llamashort & 10.0000 & 10.0000 & 10.0000 & 10.0000 & 10.0000 & 50.0000 & 1\\
\hline
\end{tabular}
\vspace{1cm}
\caption{\textbf{Inferred strategies and tournament outcomes for `equal gains from switching' experiments.} For $x=0$.}
\end{figure}
\clearpage

\begin{figure}[h!]
\textbf{\textcolor{blue}{$x = 1$}}\\ \\
\textbf{A \quad Moves of five LLM models in IPD scenarios and their inferred memory-1 strategy.}\\ \\ \\
\raggedright
\renewcommand{\arraystretch}{1.2}
\setlength{\tabcolsep}{8pt}
\begin{tabular}{r|c c c c c|c c c c c|r}
\hline
\multirow{3}{*}{LLM model} & \multicolumn{5}{c|}{`L' in 50 trials} & \multicolumn{5}{c|}{Memory-1 strategy} & \multicolumn{1}{c}{} \\
\cline{2-11}
 & R1 & LL & LR & RL & RR & $p_0$ & $p_1$ & $p_2$ & $p_3$ & $p_4$ & Notes \\
\hline
\claudeshort & 50 & 50 & 0 & 50 & 50 & 1.00 & 1.00 & 0.00 & 1.00 & 1.00 & Forgiver \\
\geminishort & 46 & 50 & 0 & 0 & 36 & 0.92 & 1.00 & 0.00 & 0.00 & 0.72 & $\sim$WSLS \\
\gptfourshort & 50 & 50 & 0 & 0 & 47 & 1.00 & 1.00 & 0.00 & 0.00 & 0.94 & $\sim$WSLS \\
\gptfiveshort& 50 & 50 & 0 & 32 & 49 & 1.00 & 1.00 & 0.00 & 0.64 & 0.98 & $\sim$Forgiver \\
\llamashort & 50 & 50 & 43 & 49 & 48 & 1.00 & 1.00 & 0.86 & 0.98 & 0.96 & $\sim$ALLC \\
\hline
\end{tabular}
\vspace{1.5cm}
 
\textbf{B \quad Pairwise tournament between the LLM models, represented by their memory-1 strategy. $\delta=0.99$}\newline \newline \newline
\raggedright
\begin{tabular}{r|c c c c c|c|c}
\hline
LLM model& \claudeshort & \geminishort & \gptfourshort & \gptfiveshort& \llamashort&Row sum& Rank\\
\hline
\claudeshort & 10.0000 & 9.9792 & 10.0000 & 10.0000 & 10.0000 & 49.9792 & 1\\
\geminishort & 9.9913 & 9.9622 & 9.9902 & 9.9909 & 9.9977 & 49.9324 & 4\\
\gptfourshort & 10.0000 & 9.9786 & 10.0000 & 10.0000 & 10.0000 & 49.9786 & 3\\
\gptfiveshort& 10.0000 & 9.9789 & 10.0000 & 10.0000 & 10.0000 & 49.9789 & 2\\
\llamashort & 10.0000 & 9.9161 & 10.0000 & 10.0000 & 10.0000 & 49.9161 & 5\\
\hline
\end{tabular}
\vspace{1cm}
\caption{$x=1$}
\end{figure}

\begin{figure}[h!]
\textbf{\textcolor{blue}{$x = 2$}}\\\\
\textbf{A \quad Moves of five LLM models in IPD scenarios and their inferred memory-1 strategy.}\\ \\ \\
\raggedright
\renewcommand{\arraystretch}{1.2}
\setlength{\tabcolsep}{8pt}
\begin{tabular}{r|c c c c c|c c c c c|r}
\hline
\multirow{3}{*}{LLM model} & \multicolumn{5}{c|}{`L' in 50 trials} & \multicolumn{5}{c|}{Memory-1 strategy} & \multicolumn{1}{c}{} \\
\cline{2-11}
 & R1 & LL & LR & RL & RR & $p_0$ & $p_1$ & $p_2$ & $p_3$ & $p_4$ & Notes \\
\hline
\claudeshort & 50 & 50 & 0 & 50 & 50 & 1.00 & 1.00 & 0.00 & 1.00 & 1.00 & Forgiver \\
\geminishort & 47 & 50 & 0 & 1 & 17 & 0.94 & 1.00 & 0.00 & 0.02 & 0.34 & $\sim$GRIM \\
\gptfourshort & 50 & 50 & 0 & 0 & 5 & 1.00 & 1.00 & 0.00 & 0.00 & 0.10 & $\sim$GRIM \\
\gptfiveshort& 49 & 50 & 0 & 30 & 47 & 0.98 & 1.00 & 0.00 & 0.60 & 0.94 & $\sim$Forgiver \\
\llamashort & 50 & 50 & 13 & 50 & 50 & 1.00 & 1.00 & 0.26 & 1.00 & 1.00 & $\sim$Forgiver \\
\hline
\end{tabular}
\vspace{1.5cm}
 
\textbf{B \quad Pairwise tournament between the LLM models, represented by their memory-1 strategy. $\delta=0.99$}\newline \newline \newline
\raggedright
\begin{tabular}{r|c c c c c|c|c}
\hline
LLM model& \claudeshort & \geminishort & \gptfourshort & \gptfiveshort& \llamashort&Row sum& Rank\\
\hline
\claudeshort & 10.0000 & 9.9693 & 10.0000 & 9.9938 & 10.0000 & 49.9631 & 1\\
\geminishort & 9.9897 & 9.9077 & 9.8642 & 9.9829 & 9.9912 & 49.7357 & 5\\
\gptfourshort & 10.0000 & 9.8931 & 10.0000 & 9.9870 & 10.0000 & 49.8801 & 4\\
\gptfiveshort& 9.9963 & 9.9600 & 9.9703 & 9.9922 & 9.9979 & 49.9166 & 3\\
\llamashort & 10.0000 & 9.9643 & 10.0000 & 9.9948 & 10.0000 & 49.9591 & 2\\
\hline
\end{tabular}
\vspace{1cm}
\caption{$x=2$.}
\end{figure}

\begin{figure}[h!]
\textbf{\textcolor{blue}{$x = 3$}}\\\\
\textbf{A \quad Moves of five LLM models in IPD scenarios and their inferred memory-1 strategy.}\\ \\ \\
\raggedright
\renewcommand{\arraystretch}{1.2}
\setlength{\tabcolsep}{8pt}
\begin{tabular}{r|c c c c c|c c c c c|r}
\hline
\multirow{3}{*}{LLM model} & \multicolumn{5}{c|}{`L' in 50 trials} & \multicolumn{5}{c|}{Memory-1 strategy} & \multicolumn{1}{c}{} \\
\cline{2-11}
 & R1 & LL & LR & RL & RR & $p_0$ & $p_1$ & $p_2$ & $p_3$ & $p_4$ & Notes \\
\hline
\claudeshort & 50 & 50 & 0 & 50 & 50 & 1.00 & 1.00 & 0.00 & 1.00 & 1.00 & Forgiver \\
\geminishort & 50 & 50 & 0 & 3 & 12 & 1.00 & 1.00 & 0.00 & 0.06 & 0.24 & $\sim$GRIM \\
\gptfourshort & 50 & 50 & 0 & 0 & 2 & 1.00 & 1.00 & 0.00 & 0.00 & 0.04 & $\sim$GRIM \\
\gptfiveshort& 50 & 50 & 0 & 25 & 48 & 1.00 & 1.00 & 0.00 & 0.50 & 0.96 & $\sim$Forgiver \\
\llamashort & 50 & 50 & 30 & 39 & 50 & 1.00 & 1.00 & 0.60 & 0.78 & 1.00 & $\sim$ALLC\\
\hline
\end{tabular}
\vspace{1.5cm}
 
\textbf{B \quad Pairwise tournament between the LLM models, represented by their memory-1 strategy. $\delta=0.99$}\newline \newline \newline
\raggedright
\begin{tabular}{r|c c c c c|c|c}
\hline
LLM model& \claudeshort & \geminishort & \gptfourshort & \gptfiveshort& \llamashort&Row sum& Rank\\
\hline
\claudeshort & 10.0000 & 10.0000 & 10.0000 & 10.0000 & 10.0000 & 50.0000 & 1\\
\geminishort & 10.0000 & 10.0000 & 10.0000 & 10.0000 & 10.0000 & 50.0000 & 1\\
\gptfourshort & 10.0000 & 10.0000 & 10.0000 & 10.0000 & 10.0000 & 50.0000 & 1\\
\gptfiveshort& 10.0000 & 10.0000 & 10.0000 & 10.0000 & 10.0000 & 50.0000 & 1\\
\llamashort & 10.0000 & 10.0000 & 10.0000 & 10.0000 & 10.0000 & 50.0000 & 1\\
\hline
\end{tabular}
\vspace{1cm}
\caption{$x=3$}
\end{figure}

\begin{figure}[h!]
\textbf{\textcolor{blue}{$x = 4$}}\\\\
\textbf{A \quad Moves of five LLM models in IPD scenarios and their inferred memory-1 strategy.}\\ \\ \\
\raggedright
\renewcommand{\arraystretch}{1.2}
\setlength{\tabcolsep}{8pt}
\begin{tabular}{r|c c c c c|c c c c c|r}
\hline
\multirow{3}{*}{LLM model} & \multicolumn{5}{c|}{`L' in 50 trials} & \multicolumn{5}{c|}{Memory-1 strategy} & \multicolumn{1}{c}{} \\
\cline{2-11}
 & R1 & LL & LR & RL & RR & $p_0$ & $p_1$ & $p_2$ & $p_3$ & $p_4$ & Notes \\
\hline
\claudeshort & 50 & 50 & 0 & 50 & 50 & 1.00 & 1.00 & 0.00 & 1.00 & 1.00 & Forgiver \\
\geminishort & 37 & 50 & 0 & 0 & 0 & 0.74 & 1.00 & 0.00 & 0.00 & 0.00 & $\sim$GRIM \\
\gptfourshort & 50 & 50 & 0 & 0 & 5 & 1.00 & 1.00 & 0.00 & 0.00 & 0.10 & $\sim$GRIM \\
\gptfiveshort& 50 & 50 & 0 & 19 & 36 & 1.00 & 1.00 & 0.00 & 0.38 & 0.72 & $\sim$WSLS \\
\llamashort & 50 & 50 & 17 & 48 & 36 & 1.00 & 1.00 & 0.34 & 0.96 & 0.72 & $\sim$Forgiver \\
\hline
\end{tabular}
\vspace{1.5cm}
 
\textbf{B \quad Pairwise tournament between the LLM models, represented by their memory-1 strategy. $\delta=0.99$}\newline \newline \newline
\raggedright
\begin{tabular}{r|c c c c c|c|c}
\hline
LLM model& \claudeshort & \geminishort & \gptfourshort & \gptfiveshort& \llamashort&Row sum& Rank\\
\hline
\claudeshort & 10.0000 & 7.9174 & 10.0000 & 10.0000 & 10.0000 & 47.9174 & 3\\
\geminishort & 9.7465 & 7.2971 & 8.6979 & 9.5372 & 9.8056 & 45.0843 & 5\\
\gptfourshort & 10.0000 & 8.3369 & 10.0000 & 10.0000 & 10.0000 & 48.3369 & 1\\
\gptfiveshort& 10.0000 & 8.0011 & 10.0000 & 10.0000 & 10.0000 & 48.0011 & 2\\
\llamashort & 10.0000 & 7.8938 & 10.0000 & 10.0000 & 10.0000 & 47.8938 & 4\\
\hline
\end{tabular}
\vspace{1cm}
\caption{$x=4$}
\end{figure}

\begin{figure}[h!]
\textbf{\textcolor{blue}{$x = 5$}}\\\\
\textbf{A \quad Moves of five LLM models in IPD scenarios and their inferred memory-1 strategy.}\\ \\ \\
\raggedright
\renewcommand{\arraystretch}{1.2}
\setlength{\tabcolsep}{8pt}
\begin{tabular}{r|c c c c c|c c c c c|r}
\hline
\multirow{3}{*}{LLM model} & \multicolumn{5}{c|}{`L' in 50 trials} & \multicolumn{5}{c|}{Memory-1 strategy} & \multicolumn{1}{c}{} \\
\cline{2-11}
 & R1 & LL & LR & RL & RR & $p_0$ & $p_1$ & $p_2$ & $p_3$ & $p_4$ & Notes \\
\hline
\claudeshort & 50 & 50 & 0 & 50 & 50 & 1.00 & 1.00 & 0.00 & 1.00 & 1.00 & Forgiver \\
\geminishort & 40 & 50 & 0 & 0 & 1 & 0.80 & 1.00 & 0.00 & 0.00 & 0.02 & $\sim$GRIM \\
\gptfourshort & 50 & 50 & 0 & 0 & 0 & 1.00 & 1.00 & 0.00 & 0.00 & 0.00 & GRIM \\
\gptfiveshort& 50 & 50 & 0 & 16 & 37 & 1.00 & 1.00 & 0.00 & 0.32 & 0.74 & $\sim$WSLS\\
\llamashort & 50 & 50 & 5 & 50 & 43 & 1.00 & 1.00 & 0.10 & 1.00 & 0.86 & $\sim$Forgiver \\
\hline
\end{tabular}
\vspace{1.5cm}
 
\textbf{B \quad Pairwise tournament between the LLM models, represented by their memory-1 strategy. $\delta=0.99$}\newline \newline \newline
\raggedright
\begin{tabular}{r|c c c c c|c|c}
\hline
LLM model& \claudeshort & \geminishort & \gptfourshort & \gptfiveshort& \llamashort&Row sum& Rank\\
\hline
\claudeshort & 10.0000 & 9.2431 & 10.0000 & 10.0000 & 10.0000 & 49.2431 & 1\\
\geminishort & 10.0025 & 8.3056 & 9.0008 & 9.9198 & 9.9903 & 47.2190 & 5\\
\gptfourshort & 10.0000 & 9.0284 & 10.0000 & 10.0000 & 10.0000 & 49.0284 & 4\\
\gptfiveshort& 10.0000 & 9.2246 & 10.0000 & 10.0000 & 10.0000 & 49.2246 & 2\\
\llamashort & 10.0000 & 9.2009 & 10.0000 & 10.0000 & 10.0000 & 49.2009 & 3\\
\hline
\end{tabular}
\vspace{1cm}
\caption{$x=5$}
\end{figure}

\begin{figure}[h!]
\textbf{\textcolor{blue}{$x = 6$}}\\\\
\textbf{A \quad Moves of five LLM models in IPD scenarios and their inferred memory-1 strategy.}\\ \\ \\
\raggedright
\renewcommand{\arraystretch}{1.2}
\setlength{\tabcolsep}{8pt}
\begin{tabular}{r|c c c c c|c c c c c|r}
\hline
\multirow{3}{*}{LLM model} & \multicolumn{5}{c|}{`L' in 50 trials} & \multicolumn{5}{c|}{Memory-1 strategy} & \multicolumn{1}{c}{} \\
\cline{2-11}
 & R1 & LL & LR & RL & RR & $p_0$ & $p_1$ & $p_2$ & $p_3$ & $p_4$ & Notes \\
\hline
\claudeshort & 50 & 50 & 0 & 50 & 50 & 1.00 & 1.00 & 0.00 & 1.00 & 1.00 & Forgiver \\
\geminishort & 31 & 50 & 0 & 0 & 0 & 0.62 & 1.00 & 0.00 & 0.00 & 0.00 & $\sim$GRIM \\
\gptfourshort & 50 & 50 & 0 & 0 & 1 & 1.00 & 1.00 & 0.00 & 0.00 & 0.02 & $\sim$GRIM \\
\gptfiveshort& 50 & 50 & 0 & 16 & 24 & 1.00 & 1.00 & 0.00 & 0.32 & 0.48 & $\sim$GRIM \\
\llamashort & 50 & 50 & 17 & 37 & 28 & 1.00 & 1.00 & 0.34 & 0.74 & 0.58 & $\sim$Forgiver \\
\hline
\end{tabular}
\vspace{1.5cm}
 
\textbf{B \quad Pairwise tournament between the LLM models, represented by their memory-1 strategy. $\delta=0.99$}\newline \newline \newline
\raggedright
\begin{tabular}{r|c c c c c|c|c}
\hline
LLM model& \claudeshort & \geminishort & \gptfourshort & \gptfiveshort& \llamashort&Row sum& Rank\\
\hline
\claudeshort & 10.0000 & 7.3343 & 10.0000 & 10.0000 & 10.0000 & 47.3343 & 4\\
\geminishort & 10.3895 & 7.5470 & 8.5910 & 9.7298 & 10.2738 & 46.5312 & 5\\
\gptfourshort & 10.0000 & 8.4134 & 10.0000 & 10.0000 & 10.0000 & 48.4134 & 1\\
\gptfiveshort& 10.0000 & 7.7301 & 10.0000 & 10.0000 & 10.0000 & 47.7301 & 2\\
\llamashort & 10.0000 & 7.4037 & 10.0000 & 10.0000 & 10.0000 & 47.4037 & 3\\
\hline
\end{tabular}
\vspace{1cm}
\caption{$x=6$}
\end{figure}

\begin{figure}[h!]
\textbf{\textcolor{blue}{$x = 7$}}\\\\
\textbf{A \quad Moves of five LLM models in IPD scenarios and their inferred memory-1 strategy.}\\ \\ \\
\raggedright
\renewcommand{\arraystretch}{1.2}
\setlength{\tabcolsep}{8pt}
\begin{tabular}{r|c c c c c|c c c c c|r}
\hline
\multirow{3}{*}{LLM model} & \multicolumn{5}{c|}{`L' in 50 trials} & \multicolumn{5}{c|}{Memory-1 strategy} & \multicolumn{1}{c}{} \\
\cline{2-11}
 & R1 & LL & LR & RL & RR & $p_0$ & $p_1$ & $p_2$ & $p_3$ & $p_4$ & Notes \\
\hline
\claudeshort & 50 & 50 & 0 & 50 & 50 & 1.00 & 1.00 & 0.00 & 1.00 & 1.00 & Forgiver \\
\geminishort & 28 & 50 & 0 & 0 & 0 & 0.58 & 1.00 & 0.00 & 0.00 & 0.00 & $\sim$GRIM \\
\gptfourshort & 50 & 50 & 0 & 0 & 0 & 1.00 & 1.00 & 0.00 & 0.00 & 0.00 & GRIM \\
\gptfiveshort& 49 & 50 & 0 & 6 & 22 & 0.98 & 1.00 & 0.00 & 0.12 & 0.44 & $\sim$GRIM \\
\llamashort & 50 & 50 & 18 & 43 & 36 & 1.00 & 1.00 & 0.36 & 0.86 & 0.72 & $\sim$Forgiver \\
\hline
\end{tabular}
\vspace{1.5cm}
 
\textbf{B \quad Pairwise tournament between the LLM models, represented by their memory-1 strategy. $\delta=0.99$}\newline \newline \newline
\raggedright
\begin{tabular}{r|c c c c c|c|c}
\hline
LLM model& \claudeshort & \geminishort & \gptfourshort & \gptfiveshort& \llamashort&Row sum& Rank\\
\hline
\claudeshort & 10.0000 & 7.2626 & 10.0000 & 9.9941 & 10.0000 & 47.2567 & 4\\
\geminishort & 10.8506 & 8.0165 & 8.7820 & 10.0424 & 10.9781 & 48.6696 & 2\\
\gptfourshort & 10.0000 & 8.7106 & 10.0000 & 9.9988 & 10.0000 & 48.7094 & 1\\
\gptfiveshort& 10.0016 & 7.7698 & 9.8998 & 9.9921 & 10.0024 & 47.6657 & 3\\
\llamashort & 10.0000 & 7.1733 & 10.0000 & 9.9921 & 10.0000 & 47.1654 & 5\\
\hline
\end{tabular}
\vspace{1cm}
\caption{$x=7$}
\end{figure}

\begin{figure}[h!]
\textbf{\textcolor{blue}{$x = 8$}}\\\\
\textbf{A \quad Moves of five LLM models in IPD scenarios and their inferred memory-1 strategy.}\\ \\ \\
\raggedright
\renewcommand{\arraystretch}{1.2}
\setlength{\tabcolsep}{8pt}
\begin{tabular}{r|c c c c c|c c c c c|r}
\hline
\multirow{3}{*}{LLM model} & \multicolumn{5}{c|}{`L' in 50 trials} & \multicolumn{5}{c|}{Memory-1 strategy} & \multicolumn{1}{c}{} \\
\cline{2-11}
 & R1 & LL & LR & RL & RR & $p_0$ & $p_1$ & $p_2$ & $p_3$ & $p_4$ & Notes \\
\hline
\claudeshort & 50 & 50 & 0 & 50 & 50 & 1.00 & 1.00 & 0.00 & 1.00 & 1.00 & Forgiver \\
\geminishort & 24 & 50 & 0 & 0 & 0 & 0.48 & 1.00 & 0.00 & 0.00 & 0.00 & $\sim$DGRIM \\
\gptfourshort & 50 & 50 & 0 & 0 & 1 & 1.00 & 1.00 & 0.00 & 0.00 & 0.02 & $\sim$GRIM \\
\gptfiveshort& 49 & 49 & 0 & 13 & 19 & 0.98 & 0.98 & 0.00 & 0.26 & 0.38 & $\sim$Forgiver \\
\llamashort & 50 & 50 & 22 & 49 & 28 & 1.00 & 1.00 & 0.44 & 0.98 & 0.58 & $\sim$Forgiver \\
\hline
\end{tabular}
\vspace{1.5cm}
 
\textbf{B \quad Pairwise tournament between the LLM models, represented by their memory-1 strategy. $\delta=0.99$}\newline \newline \newline
\raggedright
\begin{tabular}{r|c c c c c|c|c}
\hline
LLM model& \claudeshort & \geminishort & \gptfourshort & \gptfiveshort& \llamashort&Row sum& Rank\\
\hline
\claudeshort & 10.0000 & 6.8695 & 10.0000 & 9.4383 & 10.0000 & 46.3078 & 3\\
\geminishort & 11.5731 & 8.4658 & 9.1120 & 10.6263 & 11.6280 & 51.4051 & 1\\
\gptfourshort & 10.0000 & 8.8384 & 10.0000 & 10.2743 & 10.0000 & 49.1128 & 2\\
\gptfiveshort& 10.2365 & 6.4788 & 7.7803 & 9.5181 & 10.3158 & 44.3296 & 5\\
\llamashort & 10.0000 & 6.8256 & 10.0000 & 9.3584 & 10.0000 & 46.1840 & 4\\
\hline
\end{tabular}
\vspace{1cm}
\caption{$x=8$}
\end{figure}

\begin{figure}[h!]
\textbf{\textcolor{blue}{$x = 9$}}\\\\
\textbf{A \quad Moves of five LLM models in IPD scenarios and their inferred memory-1 strategy.}\\ \\ \\
\raggedright
\renewcommand{\arraystretch}{1.2}
\setlength{\tabcolsep}{8pt}
\begin{tabular}{r|c c c c c|c c c c c|r}
\hline
\multirow{3}{*}{LLM model} & \multicolumn{5}{c|}{`L' in 50 trials} & \multicolumn{5}{c|}{Memory-1 strategy} & \multicolumn{1}{c}{} \\
\cline{2-11}
 & R1 & LL & LR & RL & RR & $p_0$ & $p_1$ & $p_2$ & $p_3$ & $p_4$ & Notes \\
\hline
\claudeshort & 50 & 50 & 0 & 50 & 50 & 1.00 & 1.00 & 0.00 & 1.00 & 1.00 & TFT \\
\geminishort & 5 & 50 & 0 & 0 & 0 & 0.10 & 1.00 & 0.00 & 0.00 & 0.00 & $\sim$DGRIM \\
\gptfourshort & 48 & 50 & 0 & 0 & 0 & 0.96 & 1.00 & 0.00 & 0.00 & 0.00 & $\sim$GRIM\\
\gptfiveshort& 48 & 50 & 0 & 11 & 24 & 0.96 & 1.00 & 0.00 & 0.22 & 0.48 & $\sim$GRIM \\
\llamashort & 50 & 50 & 17 & 33 & 19 & 1.00 & 1.00 & 0.34 & 0.66 & 0.38 & $\sim$GTFT \\
\hline
\end{tabular}
\vspace{1.5cm}
 
\textbf{B \quad Pairwise tournament between the LLM models, represented by their memory-1 strategy. $\delta=0.99$}\newline \newline \newline
\raggedright
\begin{tabular}{r|c c c c c|c|c}
\hline
LLM model& \claudeshort & \geminishort & \gptfourshort & \gptfiveshort& \llamashort&Row sum& Rank\\
\hline
\claudeshort & 10.0000 & 5.0296 & 9.7791 & 9.9909 & 10.0000 & 44.7996 & 5\\
\geminishort & 13.6226 & 9.0109 & 9.1820 & 12.0662 & 12.4434 & 56.3251 & 1\\
\gptfourshort & 10.1610 & 9.0186 & 9.9220 & 10.1726 & 10.1086 & 49.3828 & 2\\
\gptfiveshort& 10.0063 & 6.4229 & 9.6964 & 9.9953 & 10.0056 & 46.1265 & 3\\
\llamashort & 10.0000 & 6.0909 & 9.8263 & 9.9895 & 10.0000 & 45.9067 & 4\\
\hline
\end{tabular}
\vspace{1cm}
\caption{$x=9$}
\end{figure}

\begin{figure}[h!]
\textbf{\textcolor{blue}{$x = 10$}}\\\\
\textbf{A \quad Moves of five LLM models in IPD scenarios and their inferred memory-1 strategy.}\\ \\ \\
\raggedright
\renewcommand{\arraystretch}{1.2}
\setlength{\tabcolsep}{8pt}
\begin{tabular}{r|c c c c c|c c c c c|r}
\hline
\multirow{3}{*}{LLM model} & \multicolumn{5}{c|}{`L' in 50 trials} & \multicolumn{5}{c|}{Memory-1 strategy} & \multicolumn{1}{c}{} \\
\cline{2-11}
 & R1 & LL & LR & RL & RR & $p_0$ & $p_1$ & $p_2$ & $p_3$ & $p_4$ & Notes \\
\hline
\claudeshort & 50 & 50 & 0 & 50 & 0 & 1.00 & 1.00 & 0.00 & 1.00 & 0.00 & Forgiver \\
\geminishort & 14 & 46 & 0 & 0 & 0 & 0.28 & 0.92 & 0.00 & 0.00 & 0.00 & $\sim$DGRIM \\
\gptfourshort & 50 & 50 & 0 & 0 & 0 & 1.00 & 1.00 & 0.00 & 0.00 & 0.00 & GRIM \\
\gptfiveshort& 8 & 19 & 0 & 1 & 0 & 0.16 & 0.38 & 0.00 & 0.02 & 0.00 & $\sim$ALLD \\
\llamashort & 50 & 50 & 47 & 36 & 13 & 1.00 & 1.00 & 0.94 & 0.72 & 0.26 & $(11110)$ \\
\hline
\end{tabular}
\vspace{1.5cm}
 
\textbf{B \quad Pairwise tournament between the LLM models, represented by their memory-1 strategy. $\delta=0.99$}\newline \newline \newline
\raggedright
\begin{tabular}{r|c c c c c|c|c}
\hline
LLM model& \claudeshort & \geminishort & \gptfourshort & \gptfiveshort& \llamashort&Row sum& Rank\\
\hline
\claudeshort & 10.0000 & 9.9031 & 10.0000 & 9.9003 & 10.0000 & 49.8034 & 4\\
\geminishort & 10.0969 & 10.0000 & 10.0969 & 9.9849 & 17.9255 & 58.1041 & 2\\
\gptfourshort & 10.0000 & 9.9031 & 10.0000 & 9.9022 & 10.0000 & 49.8054 & 3\\
\gptfiveshort& 10.0997 & 10.0151 & 10.0978 & 10.0000 & 17.9703 & 58.1829 & 1\\
\llamashort & 10.0000 & 2.0745 & 10.0000 & 2.0297 & 10.0000 & 34.1042 & 5\\
\hline
\end{tabular}
\vspace{1cm}
\caption{$x=10$}
\end{figure}

\clearpage

\noindent We sum up the results from all eleven equal gains from switching tournaments ($x$ from 0 to 10) and present the aggregate outcomes. 
\vspace{1cm}
\begin{figure}[h!]
\textbf{Aggregated results from eleven tournaments, $\delta=0.99$}\newline \newline \newline
\raggedright
\begin{tabular}{r|c c c c c|c|c}
\hline
LLM model& \claudeshort & \geminishort & \gptfourshort & \gptfiveshort& \llamashort&Row sum& Rank\\
\hline
\claudeshort & 110.0000 & 93.5081 & 109.7791 & 109.3174 & 110.0000 & 532.6046 & 4\\
\geminishort & 116.2627 & 98.5129 & 103.3170 & 111.8803 & 123.0336 & 553.0065 & 1\\
\gptfourshort & 110.1610 & 102.1212 & 109.9220 & 110.3350 & 110.1086 & 542.6478 & 2\\
\gptfiveshort& 110.3405 & 95.5813 & 107.4446 & 109.4977 & 118.2920 & 541.1561 & 3\\
\llamashort & 110.0000 & 86.5432 & 109.8263 & 101.3645 & 110.0000 & 517.7340 & 5\\
\hline
\end{tabular}
\vspace{1cm}
\caption{All eleven tournament matrices corresponding to the eleven equal gains from switching treatment, summed up. Tournaments we conducted for $w=0.01$.}
\end{figure}

\clearpage

\section{Method: Computing payoffs between repeated games strategies}
\label{Section:Methods}

\noindent Because of Lemma~\ref{lemma:levinsky}, and since we classify only memory-1 and memory-2 strategies as Nash, Partners, or Rivals, we compute payoffs only for memory-1 vs.\ memory-1 and memory-2 vs.\ memory-2 strategy pairs. Below we outline how these computations are carried out. For memory-1 vs.\ memory-1 strategies, we describe the procedure both for the case $w=0$ and for $w>0$. For memory-2 vs.\ memory-2 strategies, we provide the method for the $w>0$ case.

\subsection*{Memory-1 vs memory-1}

Consider the the pair of memory-1 strategies, $\q = (q_0,q_1,q_2,q_3,q_4)$ and $\p = (p_0,p_1,p_2,p_3,p_4)$, playing against each other in the repeated game with the stage game $g=(a_\LL,a_\LR,a_\RL,a_\RR)$. Here the player using $\q$ is denoted as player 1 and the player using $\p$ is denoted as player 2 (meaning the outcome $\LR$ is read as the situation where the player using $\q$ played L and the player using $\p$ played R). Here $p_{0}$ and $q_{0}$ denote the probabilities of cooperating in the first round.  The quantities $p_{1}, p_{2}, p_{3}, p_{4}$ are the probabilities with which strategy $\p$ cooperates after the outcomes $\LL$, $\LR$, $\RL$, and $\RR$, respectively. The corresponding values for the strategy $\q$ represent the probability to cooperate after $\LL, \RL, \LR$ and $\RR$ respectively.
\\

\noindent \textbf{The} $w\mathbf{>0}$ \textbf{case:} The probabilities that the first round finds itself in the states $\LL$, $\LR, \RL$ or $\RR$ are given by the vector $\mathbf{v}_0 = \left(p_0q_0, p_0(1-q_0), (1-p_0)q_0, (1-p_0)(1-q_0)\right)$. The transition matrix $\mathbf{M}$ defined in the following manner gives the conditional probabilities that the next state of the game is $j$ (in column) if the current state is $i$ (in row):

\begin{equation}
\mathbf{M} \;=\;
\begin{pmatrix}
q_{1}p_{1} & q_{1}(1-p_{1}) & (1-q_{1})p_{1} & (1-q_{1})(1-p_{1}) \\[6pt]
q_{2}p_{3} & q_{2}(1-p_{3}) & (1-q_{2})p_{3} & (1-q_{2})(1-p_{3}) \\[6pt]
q_{3}p_{2} & q_{3}(1-p_{2}) & (1-q_{3})p_{2} & (1-q_{3})(1-p_{2}) \\[6pt]
q_{4}p_{4} & q_{4}(1-p_{4}) & (1-q_{4})p_{4} & (1-q_{4})(1-p_{4})
\end{pmatrix}
\label{Eq:matrix}
\end{equation}
\\ \\ 
\noindent The states are ordered $\LL, \LR, \RL$ and $\RR$ in the matrix. The expected per-round payoff of player 1 from the whole game (that lasts on expectation for $w^{-1}$ rounds) is given by 

\begin{equation}
   \pi_{w,g}(\q,\p) = \left(w\right)\left(\pi_1(\q,\p) + (1-w) \pi_2(\q,\p) + (1-w)^2 \pi_3(\q,\p) + ...\right) 
   \label{equation:geometric-sum}
\end{equation}
\\ 
\noindent Where $\pi_t(\q,\p)$ is the expected payoff that strategy $\q$ earns against strategy $\p$ in round $t$. Payoffs at round $t$ are taken in expectation because players may use stochastic strategies. It is possible to simplify the above expression to 

\begin{equation}
    \pi_{w,g}(\q,\p) = \langle w \mathbf{v}_0 (\mathbf{1}-(1-w)\mathbf{M})^{-1}, g \rangle
    \label{equation:payoff-formula-inverse}
\end{equation}
\\
\noindent Where $\mathbf{1}$ is the identity matrix of size 4. For more details please see Sigmund \cite{Sigmund:Book:2010}.\\ 

\noindent \textbf{The} $w=0$ \textbf{case:} In the case $w\!=\!0$, the geometric-series expression in Eq.~(\ref{equation:geometric-sum}) no longer applies. Instead, payoffs are defined using the limit-of-means approach (see the Supplementary Information of Glynatsi \textit{et al.}~\cite{Glytnasi:PNAS:2024}, for example). In their analysis, however, strategies are defined solely by their continuation plans (i.e., memory-1 strategies are defined by $p_1,p_2,p_3,p_4$ only, leaving $p_0$ unspecified). For compatibility, they assume that strategies are subject to random execution error. Under a small execution-error assumption, play eventually moves between all absorbing sets of the original Markov chain (the one without errors). Expected payoffs are obtained by computing the stationary distribution over the states $\LL$, $\LR$, $\RL$, and $\RR$ (a unique stationary distribution exists because the chain is irreducible due to errors). Therefore, it is possible to compute payoffs without explicitly specifying how the game begins. For example, two Tit-for-Tat continuation plans with $(p_1,p_2,p_3,p_4)=(1,0,1,0)$ may end in the $\LL$ cycle, the $\RR$ cycle, or alternate between $\LR$ and $\RL$, depending on the initial move. With execution errors, however, the process drifts among these terminal sets, and the limit-of-means reflects the time spent in each of them. This is the approach in Glynatsi \textit{et al.}~\cite{Glytnasi:PNAS:2024} and in many preceding papers.\\

\noindent In our case, however, we assume no execution error. Instead, the strategy explicitly specifies the first-round cooperation probability. We use this, together with $\mathbf{M}$ (computed from the continuation plans as in Eq.~(\ref{Eq:matrix})), to determine the probabilities with which the play enters each absorbing set. In our setting, once the play reaches an absorbing set, it cannot leave. We then weight the per-round payoff earned at each absorbing set by the probability of entering it to compute the expected per-round payoff of the strategies.
For example, consider two Tit-for-Tat strategies: one cooperates with probability 0.4 in round 1, and the other with probability 0.6. The probability with which the play ends up in the $\LL$ cycle is $0.4 \times 0.6\!=\!0.24$, the probability with which it ends up in the $\RR$ cycle is $0.4 \times 0.6\!=\!0.24$ and the probability with which it enters the $\RL$-$\LR$ cycle is $1\!-\!0.48\!=\!0.52$. The payoff of the first player is thus $0.24(a_\LL) + 0.24(a_\RR) + 0.52(a_\RL+a_\LR)/2$.\\ \\

\subsection*{Memory-2 vs memory-2}

\noindent Here we describe the method of computing payoffs in a game between two memory-2 strategies, $\p$ and $\q$. Our method is outlined for just the $w>0$ case. The approach is similar to the memory-1 vs memory-1 case. A memory-2 strategy $\p$ is defined as follows:

\begin{equation}
    \p = (p_0; p_\LL, p_\LR, p_\RL, p_\RR; \{p_{\mathrm{o_{-1}},\mathrm{o_{-2}}}\}_{o_{-1}, o_{-2} \in \{\LL,\LR,\RL,\RR\}})
\end{equation}

\noindent where $p_0$ is the probability to cooperate in round 1, $p_\LL,p_\LR,p_\RL,p_\RR$ are probabilities to cooperate in round 2 based on outcome of round 1. Finally $p_{\mathrm{o_{-1}},\mathrm{o_{-2}}}$ is the probability to cooperate in a round if the outcome of the last round and the round before that are $o_{-1}$ and $o_{-2}$ respectively. Here $o_{-1}, o_{-2} \in \{\LL,\LR,\RL,\RR\}$. Thus a memory-2 strategy is vector with 21 entries, each a real number in $[0,1]$. Later on, for brevity, we denote $p_{\mathrm{o_{-1}},\mathrm{o_{-2}}}$ with $p_k$ where $k=4(i-1)+j$, is an integer and $i$ and $j$ are the indices of $o_{-1}$ and $o_{-2}$ respectively in the list $\{\LL,\LR,\RL,\RR\}$.\\

\noindent The expected payoff that strategy $\p$ earns against $\q$ in the repeated game with stage game $g=(R,S,T,P)$ and stopping probability $w$ is given by

\begin{equation}
        \pi_{w,g}(\p,\q) = \langle w \mathbf{v}_0 (\mathbf{1}-(1-w)\mathbf{M})^{-1}, g \rangle
\end{equation}

\noindent Unlike Eq. (\ref{equation:payoff-formula-inverse}), here the matrix $\mathbf{M}$ and the vector $\mathbf{v}_0$ are of shapes $16\times16$ and $1 \times 16$ respectively. They are computed in the following manner. The vector $\mathbf{v}_0$ is given as

\begin{equation}
    \mathbf{v}_0 = 
    x1 \odot x2,
\end{equation}

\noindent where $x_1$ and $x_2$ are two vectors with 16 entries and $\odot$ represents element-wise multiplication between them. The vectors $x_1$ and $x_2$ are given as

\begin{align*}
    x_1 = \Big(&p_{0} q_{0}, \quad p_{0}(1-q_{0}), \quad (1-p_{0}) q_{0}, \quad (1-p_{0})(1-q_{0}),\\
    &p_{0} q_{0}, \quad p_{0}(1-q_{0}), \quad (1-p_{0}) q_{0}, \quad (1-p_{0})(1-q_{0}),\\
    &p_{0} q_{0}, \quad p_{0}(1-q_{0}), \quad (1-p_{0}) q_{0}, \quad (1-p_{0})(1-q_{0}),\\
    &p_{0} q_{0}, \quad p_{0}(1-q_{0}), \quad (1-p_{0}) q_{0}, \quad (1-p_{0})(1-q_{0}) \Big)
\end{align*}

\begin{align*}
    x_2 = \Big(
&p_{\LL} q_{\LL}, \quad p_{\LR} q_{\RR}, \quad p_{\RL} q_{\LR}, \quad p_{\RR} q_{\RR}, \\
&p_{\LL}(1-q_{\LL}), \quad p_{\LR}(1-q_{\RR}), \quad p_{\RL}(1-q_{\LR}), \quad p_{\RR}(1-q_{\RR}), \\
&(1-p_{\LL}) q_{\LL}, \quad (1-p_{\LR}) q_{\RR}, \quad (1-p_{\RL}) q_{\LR}, \quad (1-p_{\RR}) q_{\RR}, \\
&(1-p_{\LL})(1-q_{\LL}), \quad (1-p_{\LR})(1-q_{\RR}), \quad (1-p_{\RL})(1-q_{\LR}), \quad (1-p_{\RR})(1-q_{\RR})
    \Big)
\end{align*}\\

\noindent The $16\times16$ matrix $\mathbf{M}$ is defined below. The element in row $a$ and column $b$ in matrix $\mathbf{M}$ is given by the following piecewise function.

\[
\mathbf{M}_{a,b} =
\begin{cases}
p_a\, q_a, & \text{if } 1\leqslant b\leqslant4, \quad a = 4(b-1) + l, \quad \forall l \in [0,1,2,3], \\[2mm]
p_{a}\, (1 - q_{a}), & \text{if } 5\leqslant b\leqslant8, \quad a = 4(b-5) + l, \quad \forall l \in [0,1,2,3], \\[1mm]
(1 - p_{a})\, q_{a}, & \text{if } 9\leqslant b\leqslant12, \quad a = 4(b-9) + l, \quad \forall l \in [0,1,2,3], \\[1mm]
(1 - p_{a})\, (1 - q_{a}), & \text{if } 13\leqslant b\leqslant16, \quad a = 4(b-13)+l, \quad \forall l \in [0,1,2,3], \\[1mm]
0, & \text{otherwise.}
\end{cases}
\]
\vspace{1cm}

\clearpage

\section{Extended Figures}
\begin{figure}[h!]
    \centering
    \includegraphics[width=0.95\linewidth]{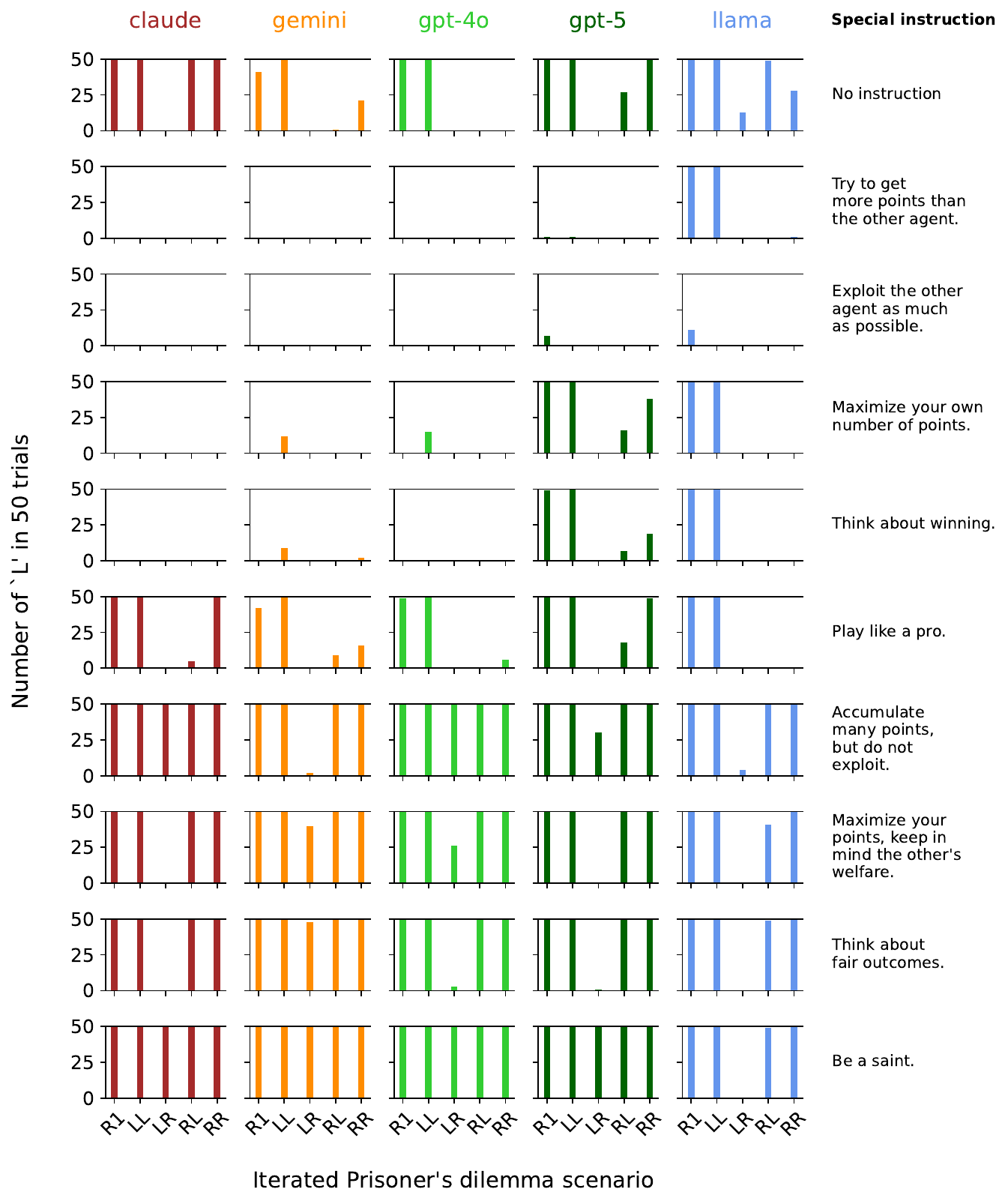}
    \vspace{0.3cm}
    \caption{\textbf{The data from the original experiment with no special instruction and from the nine framing treatments.} We plot the number of times the five LLMs choose $L$ in five scenarios of the repeated Prisoner’s Dilemma (IPD), each conducted over 50 trials. The treatments are arranged in rows and the LLMs in columns. The five scenarios considered are: Round 1 (R1), and cases where the last round is $\LL$, $\LR$, $\RL$, or $\RR$.} 
    \label{fig:Fig-diamonds-099}
\end{figure}

\begin{figure}[h!]
    \centering
    \includegraphics[width=0.85\linewidth]{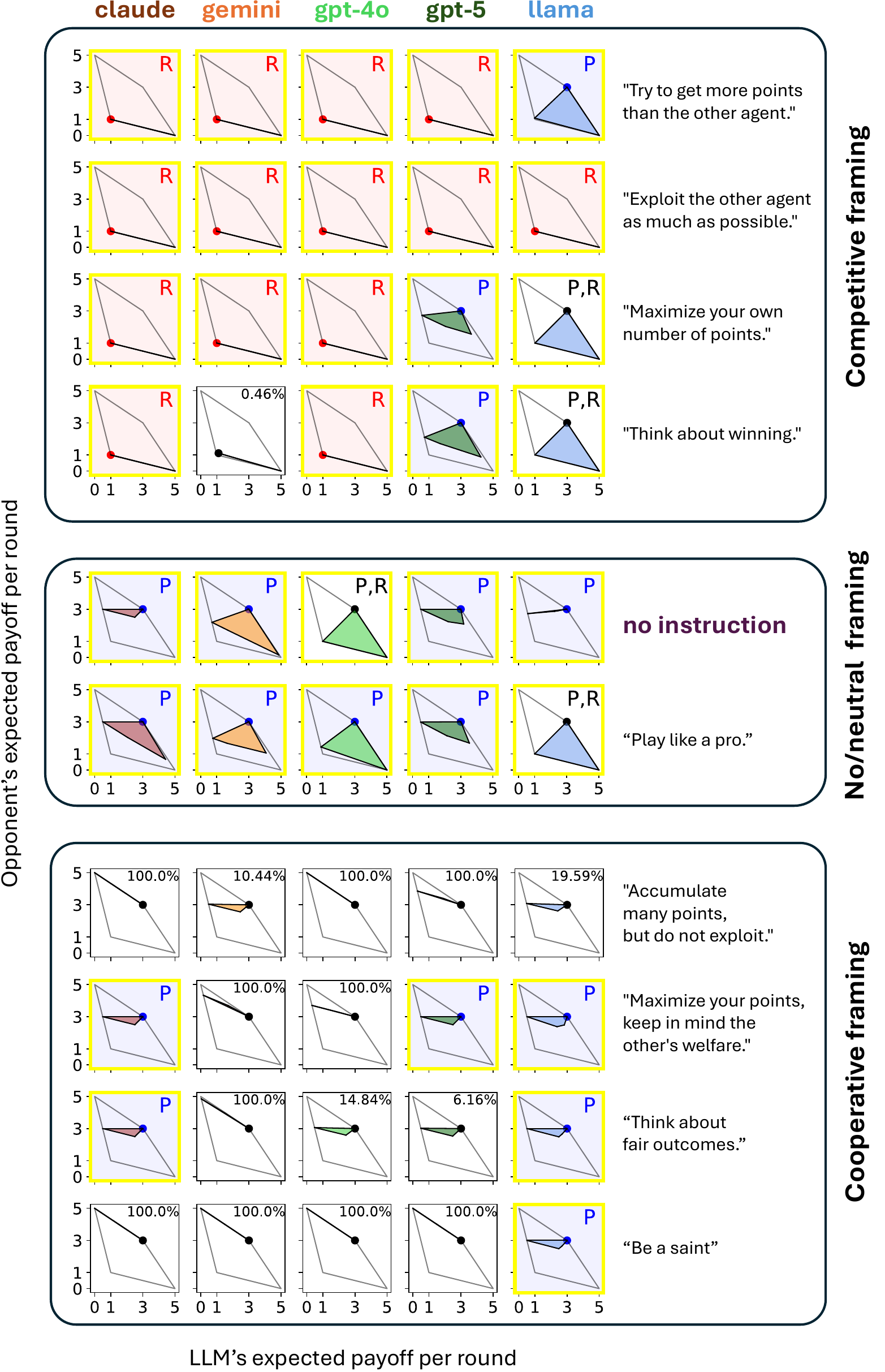}
    \vspace{0.3cm}
    \caption{\textbf{How do opponents fare against LLM strategies in the repeated Prisoner's Dilemma? Partners and Rivals among LLM strategies.} Here we recreate the payoff diamonds from the main text,  for the original experiment and all nine framings in game where the stopping probability $w=0$. } 
    \label{fig:Fig-diamonds-1}
\end{figure}

\begin{figure}[h!]
    \centering
    \includegraphics[width=0.85\linewidth]{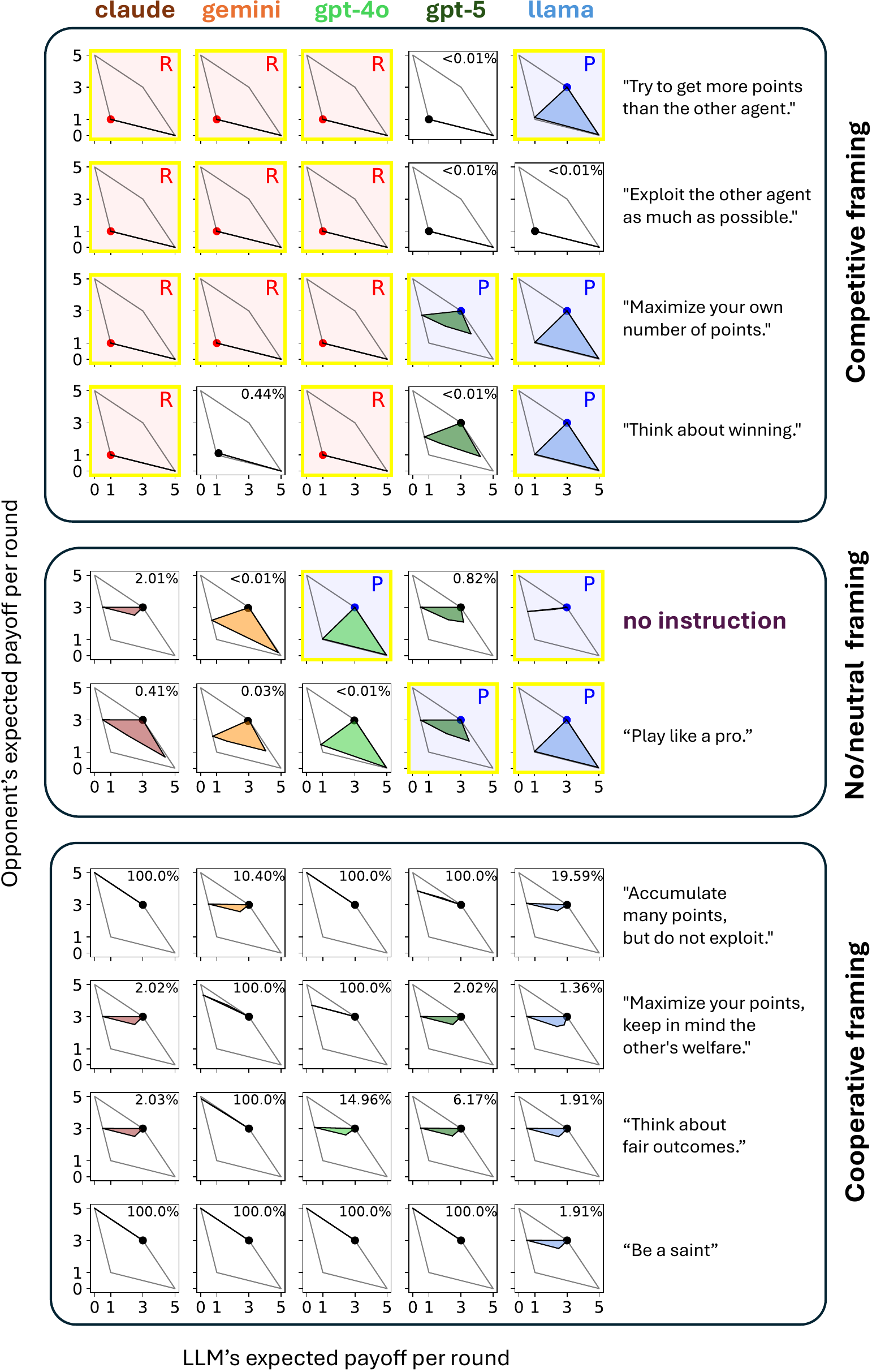}
    \vspace{0.3cm}
    \caption{\textbf{How do opponents fare against LLM strategies in the repeated Prisoner's Dilemma? Partners and Rivals among LLM strategies.} Here we recreate the payoff diamonds from the main text,  for the original experiment and all nine framings but for the game in which the stopping probability $w=0.01$. } 
    \label{fig:Fig-diamonds-099}
\end{figure}

\clearpage

\clearpage

% \begin{equation}
% v_{2} = 
% \begin{pmatrix}

% \end{pmatrix}
% \end{equation}

% \[
% x = x_{1} \odot x_{2},
% \]

% \section*{Supplementary Figures}

\clearpage

\bibliographystyle{naturemag}
\bibliography{bibliography}